\documentclass[useAMS,usenatbib]{mn2e}

\usepackage{graphicx}
\usepackage{graphics,float}                
\usepackage{epsf}

\usepackage{amsmath}                                
\usepackage{amssymb}
\usepackage{picinpar}                               
\usepackage{relsize,setspace,subfigure}
\usepackage{tabularx}
\usepackage[innercaption]{sidecap}                  
\usepackage[figuresright]{rotating}
\usepackage{url}

\usepackage{natbib}      

\def\etal{{\it et al}.}

\def\hmpc{h^{-1} {\rm Mpc}}

\def\textindent#1{\indent{#1\enspace}\ignorespaces}
\def\itemitem{\par\indent \hangindent2\parindent \textindent}

\newcommand{\cgal}{\texttt{CGAL}\ }

\def\Mpch{~h^{-1} {\rm Mpc}}

\title[A Cosmic Watershed: the WVF Void Detection Technique]{A Cosmic Watershed: the WVF Void Detection Technique}

\author[Platen, van de Weygaert \& Jones]{Erwin Platen\thanks{E-mail: platen@astro.rug.nl (EP)}, 
Rien van de Weygaert \& Bernard J.T. Jones$^{1}$\\
  $^{1}$Kapteyn Astronomical Institute, University of Groningen, P.O.
  Box 800, 9700 AV, Groningen, The Netherlands.}

\begin{document}

\date{Accepted 2007 June 15; Received ...; in original form ...}

\pagerange{\pageref{firstpage}--\pageref{lastpage}} \pubyear{2007}

\maketitle

\label{firstpage}

\begin{abstract}On megaparsec scales the Universe is permeated by an intricate filigree of 
clusters, filaments, sheets and voids, the Cosmic Web. For the understanding of its dynamical 
and hierarchical history it is crucial to identify objectively its complex morphological 
components. One of the most characteristic aspects is that of the dominant underdense 
Voids, the product of a hierarchical process driven by the collapse of minor voids in addition to 
the merging of large ones. 

In this study we present an objective void finder technique which involves 
a minimum of assumptions about the scale, structure and shape of voids. Our void finding method, 
the Watershed Void Finder (WVF), is based upon the Watershed Transform, a well-known technique 
for the segmentation of images. Importantly, the technique has the potential to trace the existing 
manifestations of a void hierarchy. The basic watershed transform is augmented by a variety of 
correction procedures to remove spurious structure resulting from sampling noise. 

This study contains a detailed description of the WVF. We demonstrate how it is 
able to trace and identify, relatively parameter free, voids and their surrounding 
(filamentary and planar) boundaries. We test the technique on a set of 
Kinematic Voronoi models, heuristic spatial models for a cellular 
distribution of matter. Comparison of the WVF segmentations of low noise and high noise 
Voronoi models with the quantitatively known spatial characteristics 
of the intrinsic Voronoi tessellation shows that the size and shape 
of the voids are succesfully retrieved. WVF manages to even reproduce 
the full void size distribution function.\end{abstract}


\section{Introduction}
Voids form a prominent aspect of the distribution of galaxies and matter 
on megaparsec scales. They are enormous regions with sizes in the range
of $20-50h^{-1}$ Mpc that are practically devoid of any galaxy and
usually roundish in shape. Forming an essential ingredient of the {\it
Cosmic Web} \citep{BonKof}, they are surrounded by elongated
filaments, sheetlike walls and dense compact clusters.  Together they
define the salient weblike pattern of galaxies and matter which
pervades the observable Universe.

Voids have been known as a feature of galaxy surveys since the first
surveys were compiled \citep{ChiRoo,GreTho,EinJoe}. Following the
discovery by \citep{KirOem,KirOem2} of the most dramatic specimen, the
Bo\"otes void, a hint of their central position within a weblike
arrangement came with the first CfA redshift slice
\citep{LapGel}. This view has recently been expanded dramatically as
maps of the spatial distribution of hundreds of thousands of galaxies
in the 2dFGRS \citep{Col} and SDSS \citep{Abz} have become available.

Voids are a manifestation of the cosmic structure formation process as
it reaches a non-linear stage of evolution. Structure
forms by gravitational instability from a primordial Gaussian field of
small amplitude density perturbations, where voids emerge out of the depressions
\citep[e.g.][]{Ick,WeyKam}. They mark the transition scale at which
perturbations have decoupled from the Hubble flow and organized themselves 
into recognizable structural features.  Early theoretical models of
void formation \citep{HofSha,Ick,Ber,BluDac} were followed and
generalized by the first numerical simulations of void centered
universes \citep{RegGel,WeyKam, DubDac,MarWas}.

In recent years the huge increase in computational resources has
enabled N-body simulations to resolve in detail the intricate
substructure of voids within the context of hierarchical cosmological 
structure formation scenarios \citep{MatWhi,GotKly,HoeGus,ArbMul,GolVog,ColShe,PadCec}. 
They confirm the theoretical expectation of voids having a rich substructure as a
result of their hierarchical buildup. Theoretically this evolution has
been succesfully embedded in the extended Press-Schechter description
\citep{PreSch,BonCol,She}. \cite{SheWey} showed how this can be
described by a two-barrier excursion set formalism \citep[also
see][]{FurPir}. The two barriers refer to the two processes dictating
the evolution of voids: their merging into ever larger voids as well
as the collapse and disappearance of small ones embedded in overdense
regions \citep[see][]{WeyShe}.

Besides representing a key constituent of the cosmic matter
distribution voids are interesting and important for a variety of
reasons. First, they are a prominent feature of the megaparsec
Universe. A proper and full understanding of the formation and
dynamics of the Cosmic Web is not possible without understanding the
structure and evolution of voids \citep{SheWey}. Secondly, they are a
probe of cosmological parameters.  The outflow from the voids depends
on the matter density parameter $\Omega_m$, the Hubble parameter
$H(t)$ and possibly on the cosmological constant $\Lambda$ \citep[see
e.g.][]{BerWey,DekRee,MarWas,FliTri}. These parameters also dictate
their redshift space distortions \citep{RydMel,SchRyd} while their
intrinsic structure and shape is sensitive to various aspects of the
power spectrum of density fluctuations \citep{ParLee}. A third point
of interest concerns the galaxies in voids.  Voids provide a unique and
still largely pristine environment for studying the evolution of
galaxies \citep{HofSil,LitWei,Peebles}. The recent interest in
environmental influences on galaxy formation has stimulated substantial
activity in this direction \citep{SzoGor,GroGel,MatWhi,FriPir,BenHoy,GotKly,
HoeGus,FurPir,HoyVog,RojVog,Pati1,CecPad}.

Despite the considerable interest in voids a fairly basic yet highly
significant issue remains: identifying voids and tracing their outline
within the complex spatial geometry of the Cosmic Web. There is not an 
unequivocal definition of what a void is and as
a result there is considerable disagreement on the precise outline of
such a region \citep[see e.g.][]{ShaFel}. Because of the vague and
diverse definitions, and the diverse interests in voids,
there is a plethora of void identification procedures
\citep{KauFai,ElaPir,AikMae,HoyVog,ArbMul,PliBas,Pati2,
ColShe,ShaFel,HahPor,Ney}. 

The ``sphere-based'' voidfinder algorithm of \cite{ElaPir} has been at the 
basis of most voidfinding methods. However, this succesful approach will not 
be able to analyze complex spatial configurations in which voids may have arbitrary
shapes and contain a range and variety of substructures. A somewhat 
related and tessellation based voidfinding technique that still 
is under development is ZOBOV \citep{Ney}. It is the voidfinder 
equivalent to the VOBOZ halofinder method \citep{neyrinck2005}. 

Here we introduce and test a new and objective voidfinding formalism
that has been specifically designed to dissect the multiscale character of 
the void network and the weblike features marking its boundaries. Our 
{\it Watershed Void Finder} (WVF) is based on the watershed algorithm 
\citep{BeuLan,BeuMey}. It stems from the field of mathematical morphology 
and image analysis. 

The WVF is defined with respect to the DTFE density field
of a discrete point distribution \citep{SchWey}. This assures an
optimal sensitivity to the morphology of spatial structures and yields 
an unbiased probe of substructure in the mass
distribution \citep[see e.g][]{Oka,SchWey}. Because the WVF void finder 
does not impose a priori constraints on the size, morphology and shape of a 
voids it provides a basis for analyzing the intricacies of an evolving void
hierarchy. Indeed, this has been a major incentive towards its
development.

This study is the first in a series. Here we will define 
and describe the Watershed Void Finder and investigate its performance with 
respect to a test model of spatial weblike distributions, Voronoi 
kinematic models. Having assured the success of WVF to trace and measure 
the spatial characteristics of these models the follow-up study will 
address the application of WVF on a number of GIF N-body simulations 
of structure formation \citep{kauffmann1999}. Amongst others, WVF will be 
directed towards characterizing the hierarchical structure of the 
megaparsec void population \citep{SheWey}. For a comparison of the 
WVF with other void finder methods we refer to the extensive 
study of \cite{colpear2007}.

In the following sections we will first describe how the fundamental concepts of mathematical 
morphology have been translated into a tool for the analysis of cosmological density 
fields inferred from a discrete N-body simulation or galaxy redshift survey point 
distribution (sect. 2 \& 3). To test our method we have applied it 
to a set of heuristic and flexible models of a cellular spatial distribution of 
points, Voronoi clustering models. These are described in section 4. In section 5 we 
present the quantitative analysis of our test results and a comparison with 
the known intrinsic properties of the test models. In section 6 we evaluate 
our findings and discuss the prospects for the analysis of cosmological N-body
simulations.

\begin{figure*}
     \includegraphics[width=0.33\textwidth]{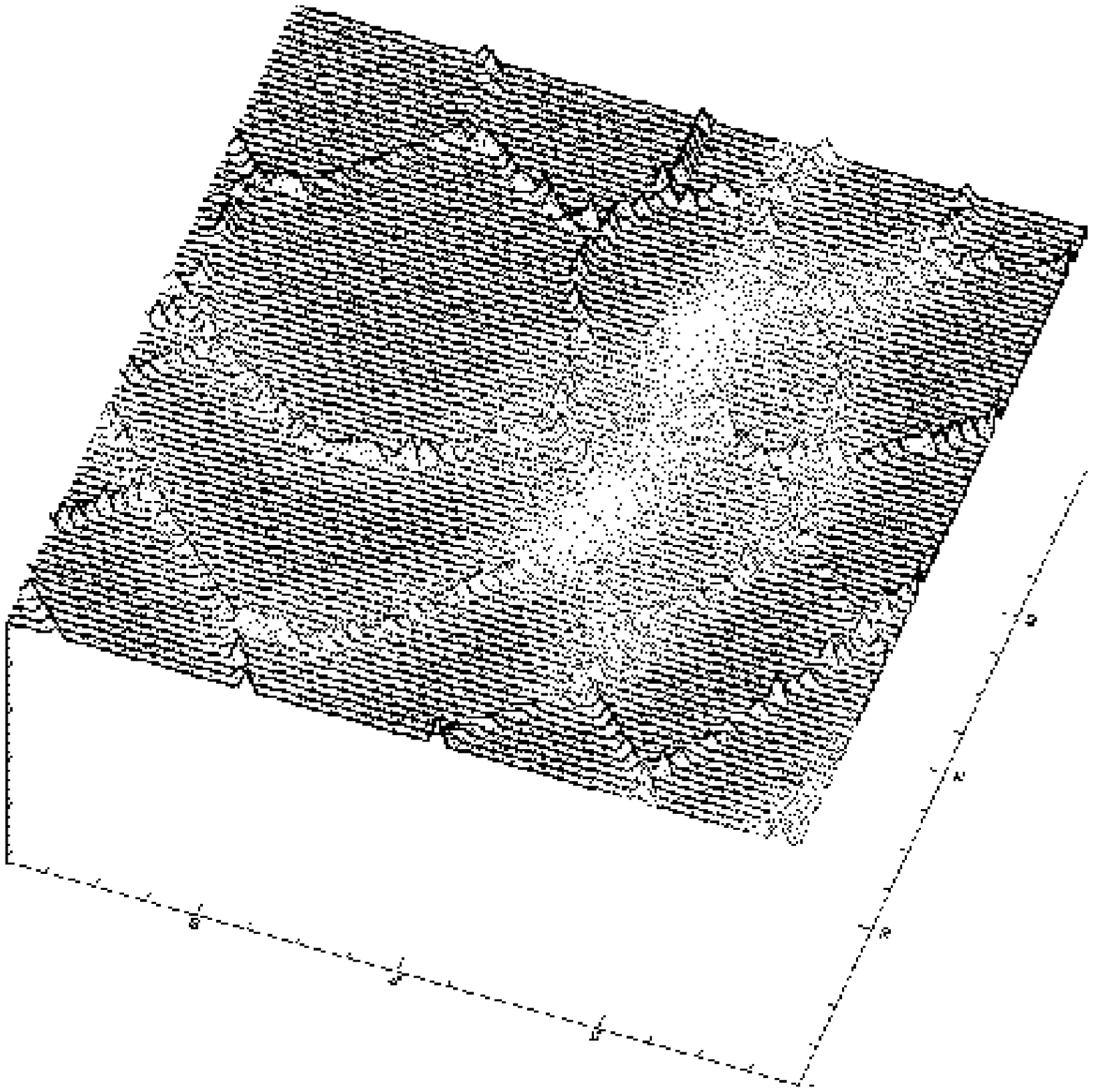}
     \includegraphics[width=0.33\textwidth]{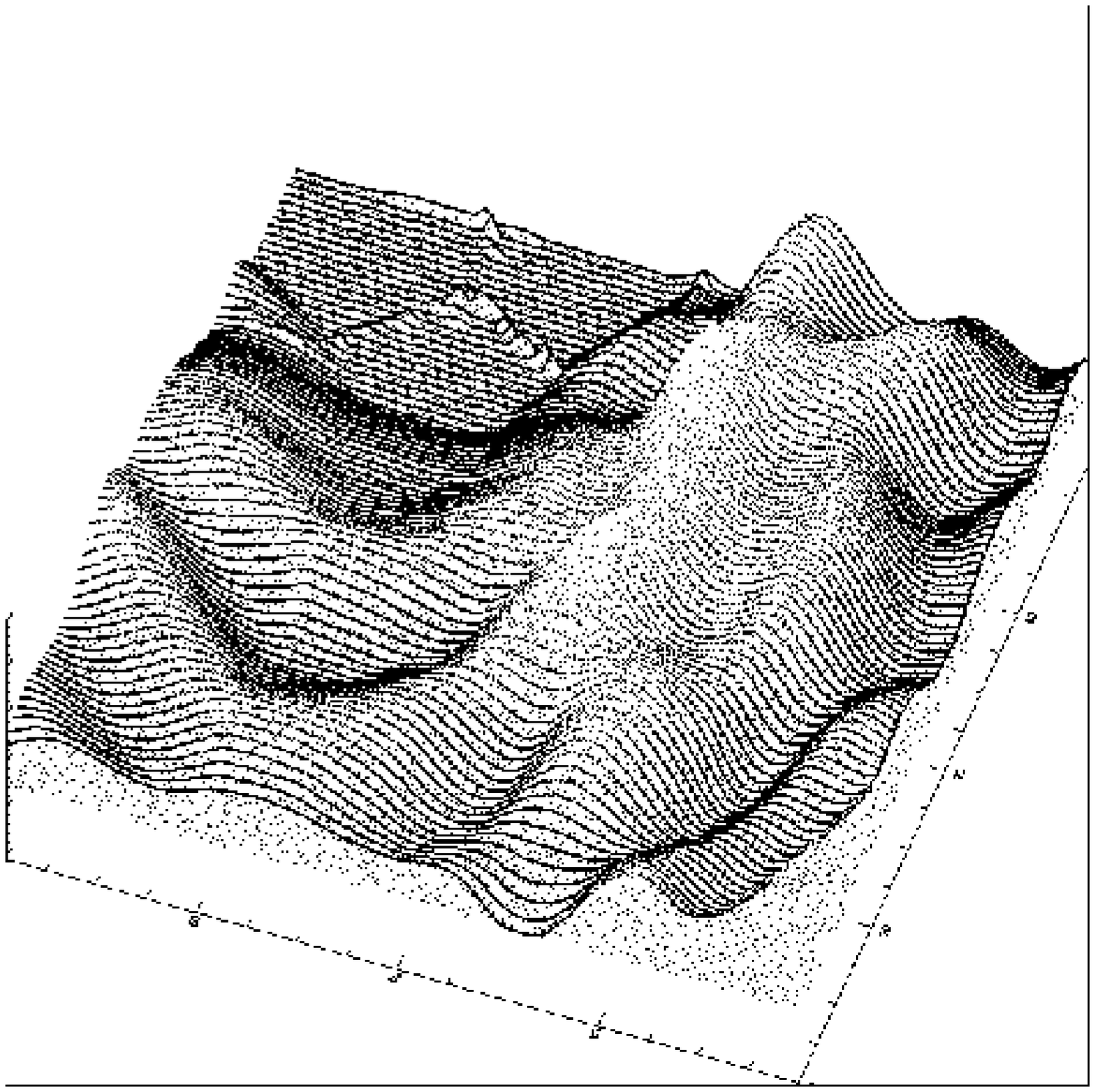}
     \includegraphics[width=0.33\textwidth]{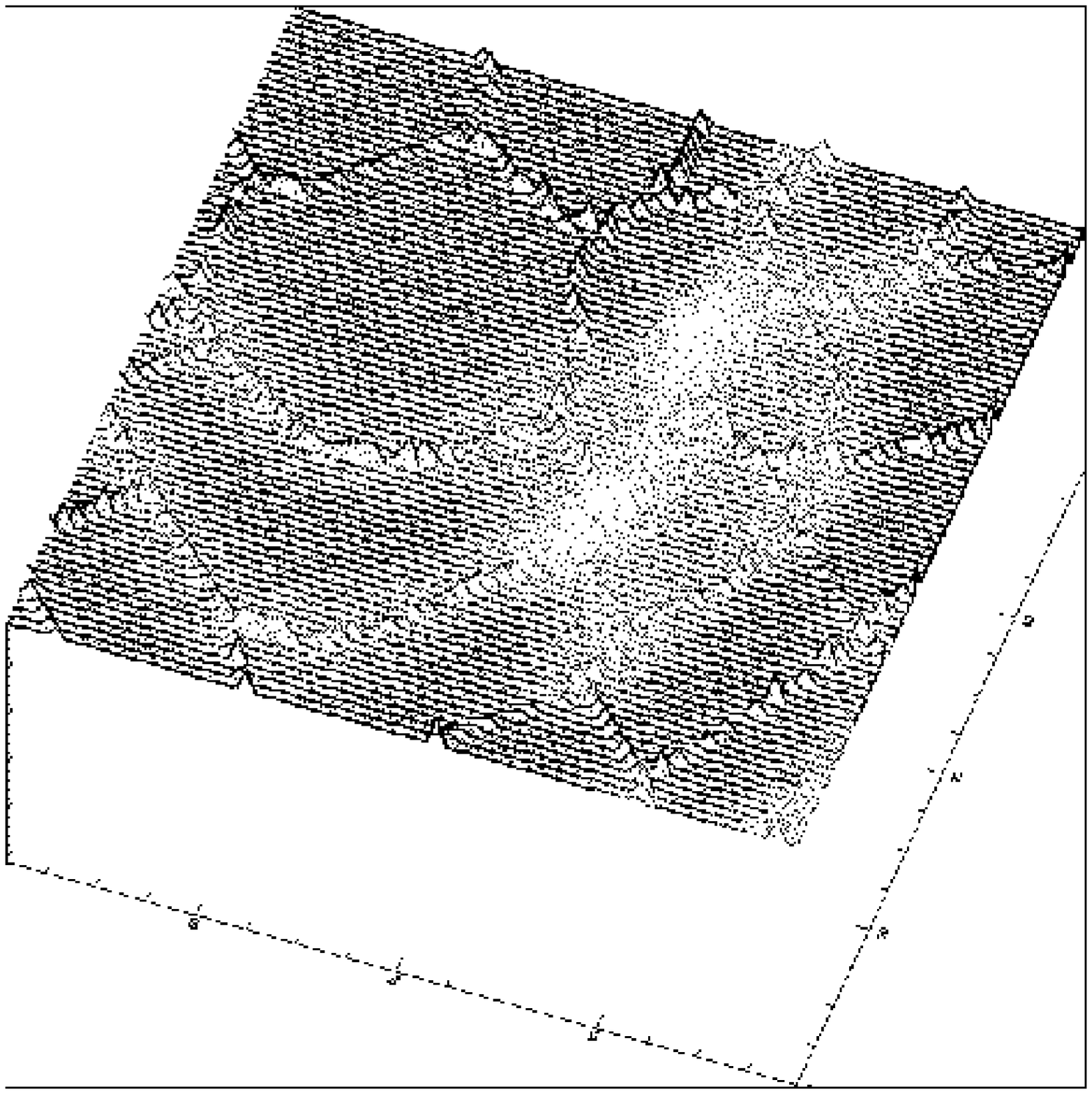}
     \caption{Three frames illustrating the principle of the watershed
       transform. The lefthand frame shows the surface to be segmented. 
       Starting from the local minima the surrounding basins of the surface 
       start to flood as the water level continues to rise (dotted plane 
       initially below the surface). Where two basins meet up near a ridge 
       of the density surface, a ``dam'' is erected (central frame). Ultimately, 
       the entire surface is flooded, leaving a network of dams 
       defines a segmented volume and delineates the corresponding 
       cosmic web (righthand frame).}
     \label{fig:proc}
\end{figure*}

\section{the Watershed Void Finder}
\label{sec:method}
The new void finding algorithm which we introduce here is based on the
{\it watershed transform} of \cite{BeuLan} and \cite{BeuMey}. A more extensive and 
technical description of the basic concepts of mathematical morphology 
and the basic watershed algorithm in terms of homotopy transformations on 
lattices \citep{kresch1998} is provided in appendix~\ref{app:mathmorph} 
and ~\ref{app:wshedimpl}. 

\subsection{the Watershed Transform (WST)}
The watershed transform is used for segmenting images into distinct
regions and objects. The Watershed Transform (WST) is a concept
defined within the context of mathematical morphology, and was first
introduced by \cite{BeuLan}. The basic idea behind the WST finds its 
origin in geophysics. The WST delineates the boundaries of the
separate domains, i.e. the {\it basins}, into which yields of, 
for example, rainfall will collect.

The word {\it watershed} refers to the analogy of a 
landscape being flooded by a rising level of
water. Suppose we have a surface in the shape of a landscape (first
image of Fig.~\ref{fig:proc}). The surface is pierced at the location
of each of the minima. As the water-level rises a growing fraction of
the landscape will be flooded by the water in the expanding
basins. Ultimately basins will meet at the ridges corresponding to
saddle-points in the density field. This intermediate step is
plotted in the second image of Fig.~\ref{fig:proc}. The ridges define the
boundaries of the basins, enforced by means of a sufficiently high
dam. The final result (see last in Fig.~\ref{fig:proc}) of the
completely immersed landscape is a division of the landscape into
individual cells, separated by the {\it ridge dams}. In the 
remainder of this study we will use the word {\it ``segment''} to describe 
the watershed's cells. 

\subsection{Watershed segments: qualities}
The watershed algorithm holds several advantages with respect to other 
voidfinders:

\begin{itemize}
\item Within an ideal smooth density field (i.e. without
noise) it will identify voids in a parameter free way. No
predefined values have to be introduced. In less ideal, 
and realistic, circumstances a few parameters have to be 
set for filtering out discreteness noise. Their values are 
guided by the properties of the data.
\item The watershed works directly on the topology of the field and 
does not reply on a predefined geometry/shape. By implication the 
identified voids may have any shape. 
\item The watershed naturally places the {\it divide lines} on the
crests of a field. The void boundary will be detected even 
when its boundary is distorted. 
\item The transform naturally produces closed contours. As long 
as minima are well chosen the watershed transform will not be
sensitive to local protrusions between two adjacent voids.
\end{itemize}

\noindent Obviously we can only extract structural information to the extent that 
the point distribution reflects the underlying structure. Undersampling 
and shotnoise always conspire to obfiscate the results, but we believe 
the present methodology provides an excellent way of handling this. 

\subsection{Voids and watersheds}
The Watershed Void Finder (WVF) is an implementation of the watershed
transform within a cosmological context. The watershed method is
perfectly suited to study the holes and boundaries in the distribution
of galaxies, and holds the specific promise of being able to recognize
the void hierarchy that has been the incentive for our study.

The analogy of the WST with the cosmological context is straightforward: 
{\it voids} are to be identified with the {\it basins}, while the {\it filaments} 
and {\it walls} of the cosmic web are the ridges separating the voids from 
each other.

\subsection{The Watershed Void Finder: Outline}
An outline of the steps of the watershed procedure within its
cosmological context is as follows:
\begin{itemize}
\item {\bf DTFE}: Given a point distribution (N-body, redshift
survey), the Delaunay Tessellation Field Estimator
\citep[DTFE,][]{SchWey} is used to define a continuous density
field throughout the sample volume. This guarantees a density field
which retains the morphological character of the underlying point
distribution, i.e. the hierarchical nature, the web-like morphology
dominated by filaments and walls, and the presence voids is
warranted.\\

\item {\bf Grid Sampling}: 
For practical processing purposes the DTFE field is sampled 
on a grid. The optimal grid size has to assure the resolution of  
all morphological structures while minimizing the number 
of needed gridcells. This criterion suggests a grid 
with gridcells whose size is in the order of the interparticle 
separation.\\ 

\item {\bf Rank-Ordered Filtering}: The DTFE density field is
adaptively smoothed by means of {\it Natural Neighbour Maxmin and
Median} filtering. This involves the computation of the median,
minimum or maximum of densities within the {\it contiguous Voronoi
cell}, the region defined by a point and its {\it natural neighbours
}.\\

\item {\bf Contour Levels}: The image is transformed into a discrete
set of density levels. The levels are defined by a uniform
partitioning of the cumulative density distribution.\\

\item {\bf Pixel Noise}: With an opening and closing (operation to 
be defined in appendix.~\ref{app:mathmorph}) of 2 pixel radius
we further reduce pixel by pixel fluctuations.\\

\item {\bf Field Minima}: The minima in the smoothed density field are
identified as the pixels (grid cells) which are exclusively surrounded 
by neighbouring grid-cells with a higher density value.\\

\item {\bf Flooding}: The {\it flooding procedure} starts at the
location of the minima. At successively increasing flood levels the
surrounding region with a density lower than the corresponding
density threshold is added to the {\it basin} of a particular
minimum. The flooding is illustrated in Fig.~\ref{fig:proc}.\\

\item {\bf Segmentation}: Once a pixel is reached by two distinct
basins it is identified as belonging to their segmentation
boundary. By continuing this procedure up to the maximum density
level the whole region has been segmented into distinct {\it void
patches}.\\

\item {\bf Hierarchy Correction}: A correction is
necessary to deal with effects related to the intrinsic hierarchical
nature of the void distribution. The correction involves the removal
of segmentation boundaries whose density is lower than some
density threshold. The natural threshold value would be the 
typical void underdensity $\Delta=-0.8$ (see sect.~\ref{sec:threshold}). 
Alternatively, dependent on the application, one may chose to take a 
user-defined value.\\
\end{itemize}

\begin{figure*}
  \begin{minipage}[b]{\linewidth}
    \includegraphics[angle=0,width=.49\textwidth]{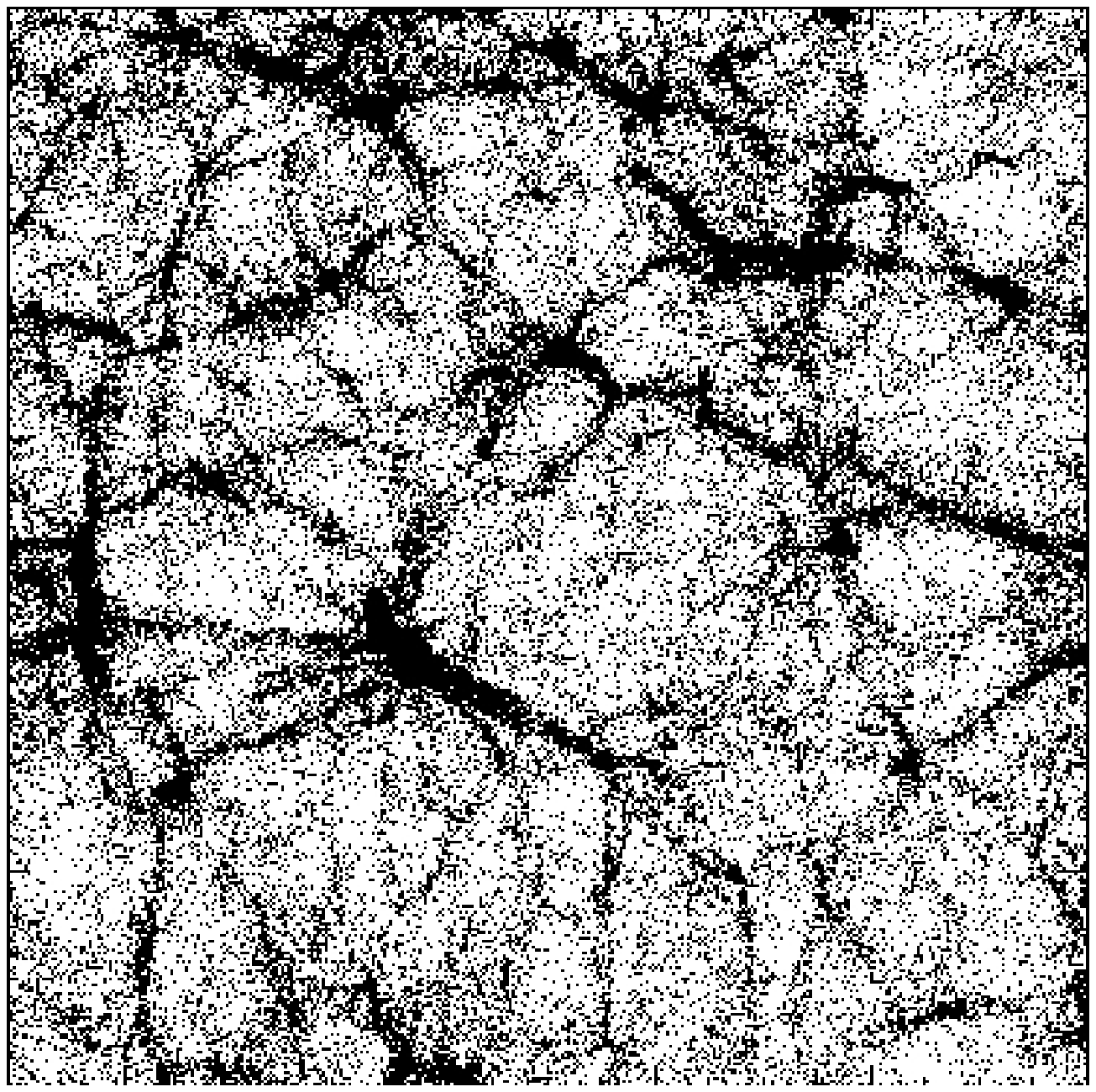}
    \hspace{0.3cm}
    \includegraphics[angle=0,width=.49\textwidth]{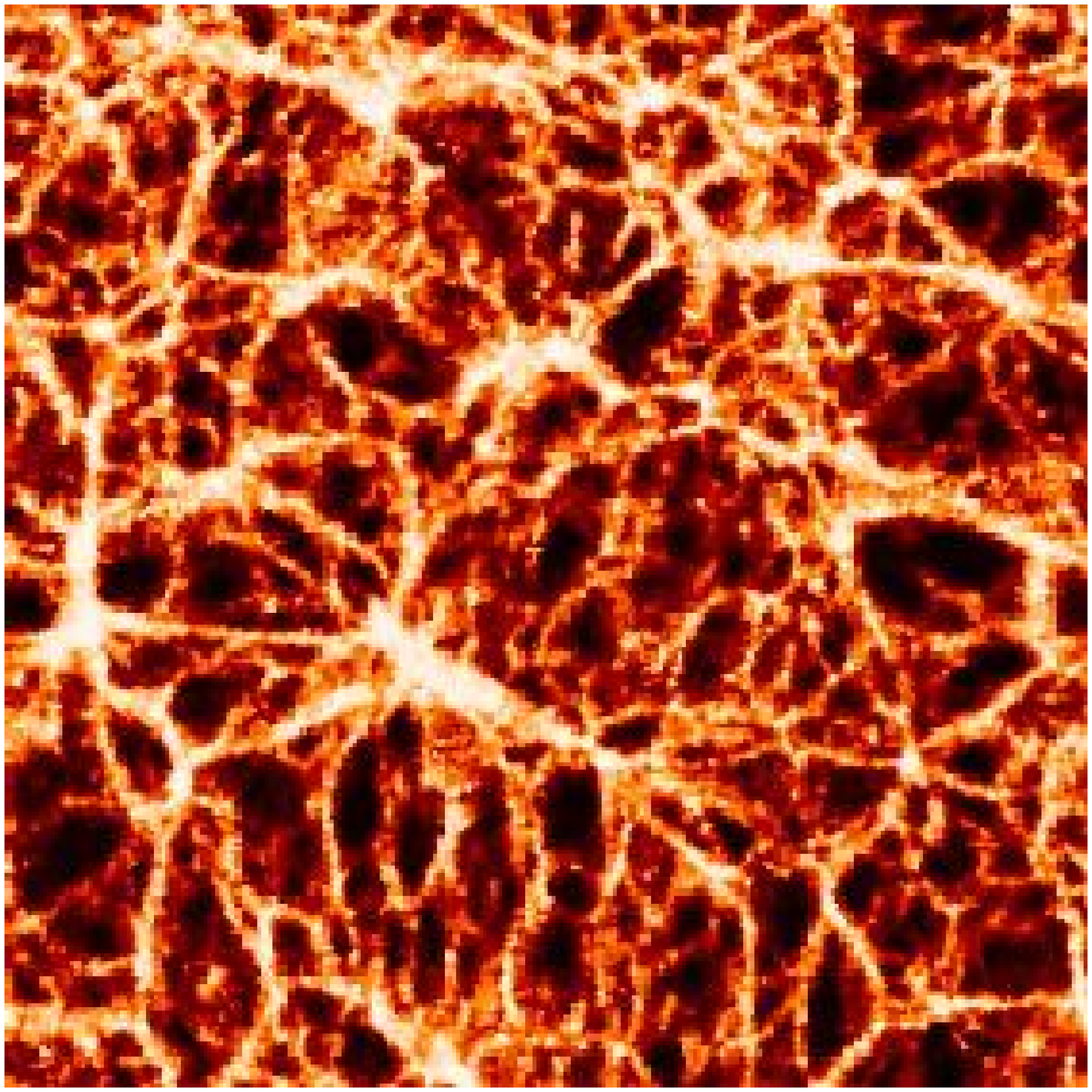}
  \end{minipage}
  \vskip 0.4truecm
  \begin{minipage}[b]{\linewidth}
    \includegraphics[angle=0,width=.49\textwidth]{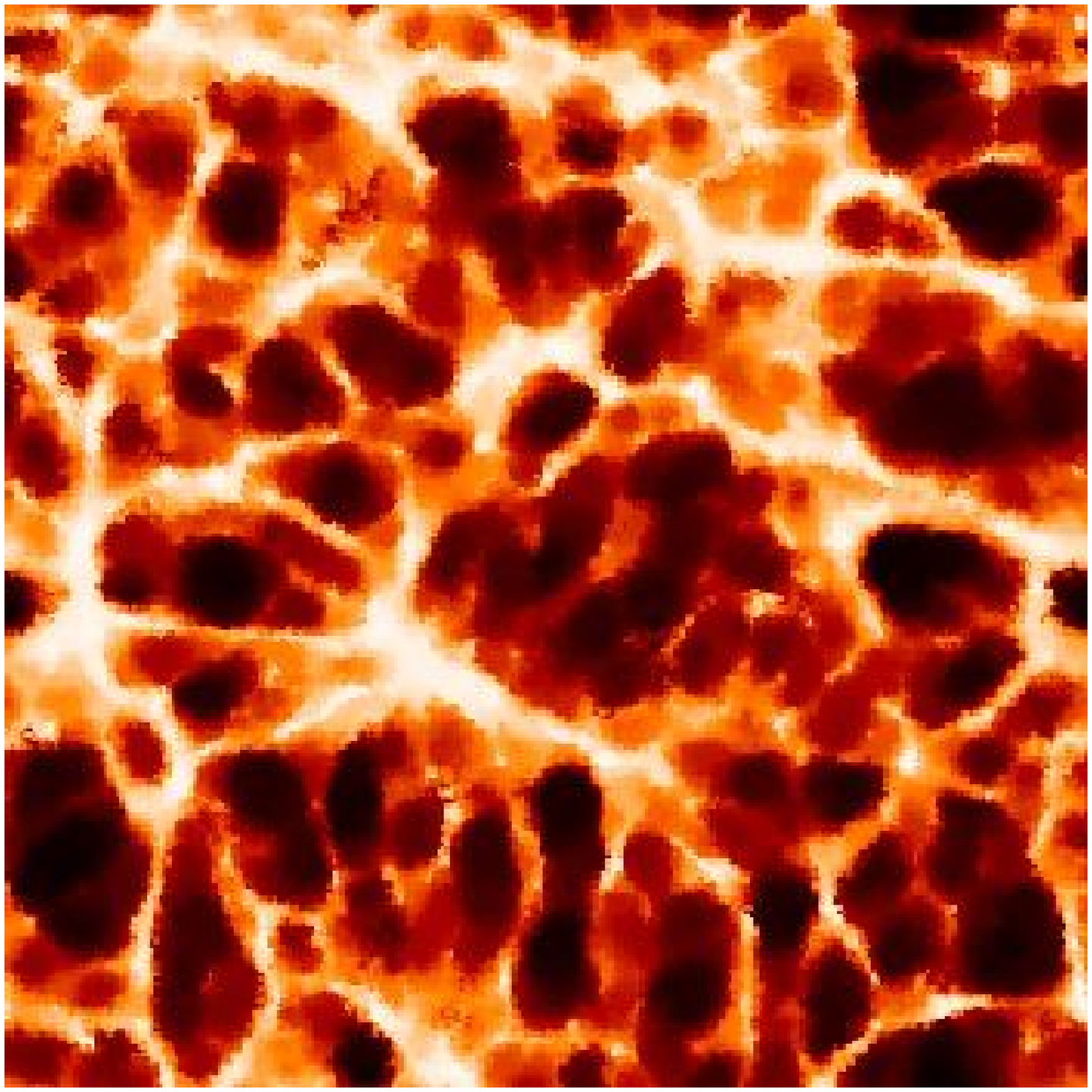}
    \hspace{0.3cm}
    \includegraphics[angle=0,width=.49\textwidth]{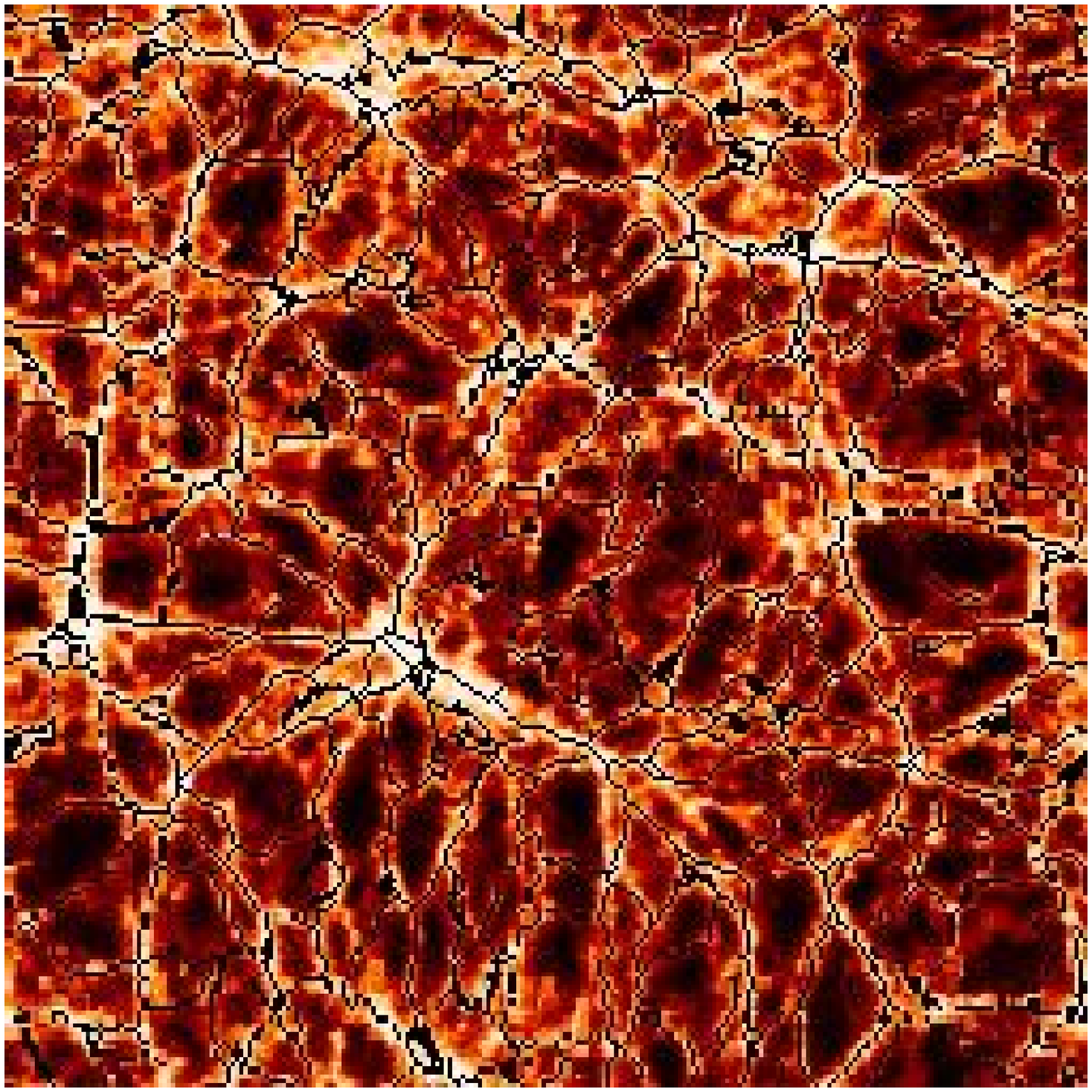}
  \end{minipage}
  \caption{A visualization of several intermediate steps of the
  Watershed VoidFinding method. The top lefthand frame shows the
  particles of a slice in the LCDM GIF simulation. The corresponding 
  DTFE density field is shown in the top righthand frame. The 
  next, bottom lefthand, frame shows the resulting 5th order 
  median-filtered image. Bottom righthand frame: the resulting WVF segmentation, 
  computed on the basis of the median filtered image. The image shows the 
  superposition of WVF ridges (black) on the original density field.} 
  \label{fig:wvf}
\end{figure*}

\subsection{WVF by example:\\ \ \ \ \ \ \ \ Voids in a $\Lambda$CDM simulation}
A direct impression of the watershed voidfinding method is most
readily obtained via the illustration of a representative example. In
Fig.~\ref{fig:wvf} the watershed procedure has been applied to the
cosmological GIF2 simulation \citep{kauffmann1999}.

The N-body particle distribution (lefthand Fig.~\ref{fig:wvf}) is
translated into a density field using the DTFE method. The application
of the DTFE method is described in section~\ref{sec:dtfe}, the details
of the DTFE procedure are specified in Appendix~\ref{app:dtfe}.

The DTFE density field is sampled and interpolated on a $256^3$ grid,
the result of which is shown in the top righthand frame of
Fig.~\ref{fig:wvf}. The gray-scales are fixed by uniformly sampling 
the cumulative density distribution, ensuring that all grayscale values 
have the same amount of volume.

The DTFE density field is smoothed by means of the adaptive {\it
Natural Neighbour Median} filtering described in
sect.~\ref{sec:natnghb}. This procedure determines the filtered
density values at the location of the particles. Subsequently, these
are interpolated onto a grid. This field is translated into a
grayscale image following the same procedure as that for the raw DTFE
image (bottom lefthand panel).

The minima in the smoothed density field are identified and marked as
the flooding centres for the watershed transform.  The resulting WVF
segmentation is shown in the bottom righthand frame of
Fig.~\ref{fig:wvf}.

The correspondence between the Cosmic Web, its voids and
the watershed segmentation is striking. There is an almost
perfect one-to-one correspondence between the segmentation and the
void regions in the underlying density field. The WVF method does not depend on any
predefined shape. As a result, the recovered voids do follow their
natural shape. A qualitative assessment of the whole simulation cube
reveals that voids are very elongated and have a preferential
orientation within the cosmic web, perhaps dictated by the megaparsec
tidal force field \citep[see e.g.][]{ParLee}.

Clearly, the Watershed Void Finder is able to extract substructure at
any level present in the density distribution. While this is an
advantage with respect to tracing the presence of substructure within
voids it does turn into a disadvantage when seeking to trace the
outline of large scale voids or when dealing with noise in the
dataset. While the noise-induced artificial segments are suppresed by
means of the full machinery of Markers (sect.~\ref{sec:marker}),
Void Patch Merging (sect.~\ref{sec:merging}) and 
Natural Neighbour Rank filtering (sect.~\ref{sec:natnghb}), it
are the latter two which may deal with intrinsic void hierarchy.

The follow-up study \citep{PlaWey07} will involve a detailed
quantitative analyze of volume and shapes of the voids in the GIF2
mass distribution for a sequence of timesteps.

\section{Method: detailed description}
In order to appreciate the various steps of the Watershed Void Finder 
outlined in the previous section we need to describe a few of the 
essential steps in more detail. 

To process a point sample into a spatial density field we use 
DTFE. To detect voids of a particular scale it is necessary to 
remove statistically insignificant voids generated by the 
shotnoise of the discrete point sample as well as physically 
significant subvoids. In order to retain only the statistically 
significicant voids we introduce and apply Natural Neighbour 
Rank-Order filtering. Hierarchy Merging is used for the removal 
of subvoids which one would wish to exclude from a specific 
void study.

\subsection{The DTFE density field}
\label{sec:dtfe}
The input samples for our analysis are mostly samples of galaxy
positions obtained by galaxy redshift surveys or the positions of a
large number of particles produced by N-body simulations of cosmic
structure formation. In order to define a proper continuous field from
a discrete distribution of points -- computer particles or galaxies --
we translate the spatial point sample into a continuous density field
by means of the Delaunay Tessellation Field Estimator
\citep[DTFE,][]{SchWey}.

\subsubsection{DTFE}
The DTFE technique \citep{SchWey} recovers fully volume-covering and
volume-weighted continuous fields from a discrete set of sampled field
values. The method has been developed by \cite{SchWey} and forms an
elaboration of the velocity interpolation scheme introduced by
\cite{BerWey}. It is based upon the use of the Voronoi and Delaunay
tessellations of a given spatial point distribution to form the basis
of a natural, fully self-adaptive filter in which the
Delaunay tessellations are used as multidimensional interpolation
intervals. A typical example of a DTFE processed field is the one
shown in the top row of Fig.~\ref{fig:wvf}: the particles of a GIF
N-body simulation \citep{kauffmann1999} are translated into the continuous
density field in the righthand frame.

\begin{figure}
  \begin{center}
  \includegraphics[angle=0,width=.4\textwidth]{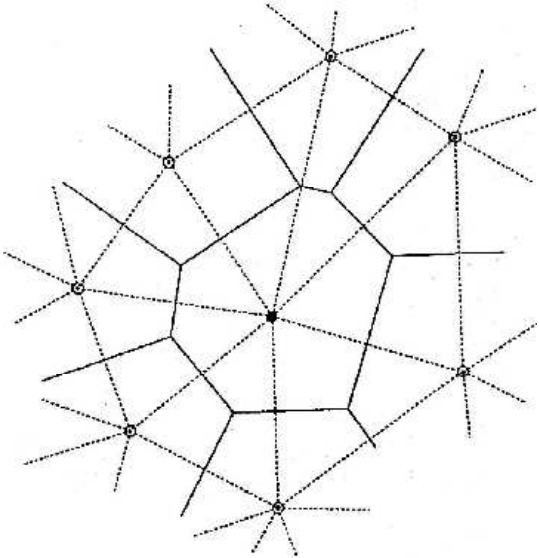}
  \end{center}
  \caption{Natural Neighbours of a point. The black dot represents 
the central point, the open circles its {\it Natural Neighbours}. The 
solid edges mark the Voronoi cell surrounding the central point, 
along with the connecting Voronoi edges. The dashed lines delineate 
the corresponding Delaunay triangles. The central Voronoi cell is 
surrounded by its related Delaunay triangles, defining the 
Natural Neighbours. The image is an illustration of the dual 
relationship between Voronoi and Delaunay tessellations.}
  \label{fig:natnbrs}
\end{figure}

The primary ingredient of the DTFE method is the Delaunay tessellation
of the particle distribution.  The Delaunay tessellation of a point
set is the uniquely defined and volume-covering tessellation of
mutually disjunct Delaunay tetrahedra (triangles in 2D). Each is
defined by the set of four points whose circumscribing sphere does not
contain any of the other points in the generating set \citep{Del}. The
Delaunay tessellation and the Voronoi tessellation of the point set
are each others {\it dual}. The Voronoi tessellation is the division
of space into mutually disjunct polyhedra, each polyhedron consisting
of the part of space closer to the defining point than any of the
other points \citep{Vor,Oka}

DTFE exploits three properties of Voronoi and Delaunay tessellations
\citep{Sch,SchWey07}. The tessellations are very sensitive to the
local point density. DTFE uses this to define a local estimate of the
density on the basis of the inverse of the volume of the tessellation
cells. Equally important is their sensitivity to the local geometry of
the point distribution. This allows them to trace anisotropic features
such as encountered in the cosmic web. Finally, DTFE exploits the
adaptive and minimum triangulation properties of Delaunay
tessellations in using them as adaptive spatial interpolation
intervals for irregular point distributions. In this way it is the first
order version of the {\it Natural Neighbour method}
\citep{BraSam,Suk,Wat}.

Within the cosmological context a major -- and crucial --
characteristic of a processed DTFE density field is that it is capable
of delineating three fundamental characteristics of the spatial
structure of the megaparsec cosmic matter distribution. It outlines
the full hierarchy of substructures present in the sampling
point distribution, relating to the standard view of structure in the
Universe having arisen through the gradual hierarchical buildup of
matter concentrations. DTFE also reproduces any anisotropic
patterns in the density distribution without diluting their intrinsic
geometrical properties. This is particularly important when analyzing
the the prominent filamentary and planar features marking the 
Cosmic Web. A third important aspect of DTFE is that it outlines the
presence and shape of voidlike regions. Because of the interpolation
definition of the DTFE field reconstruction voids are rendered as
regions of slowly varying and moderately low density values.

\begin{figure}
     \includegraphics[width=0.155\textwidth]{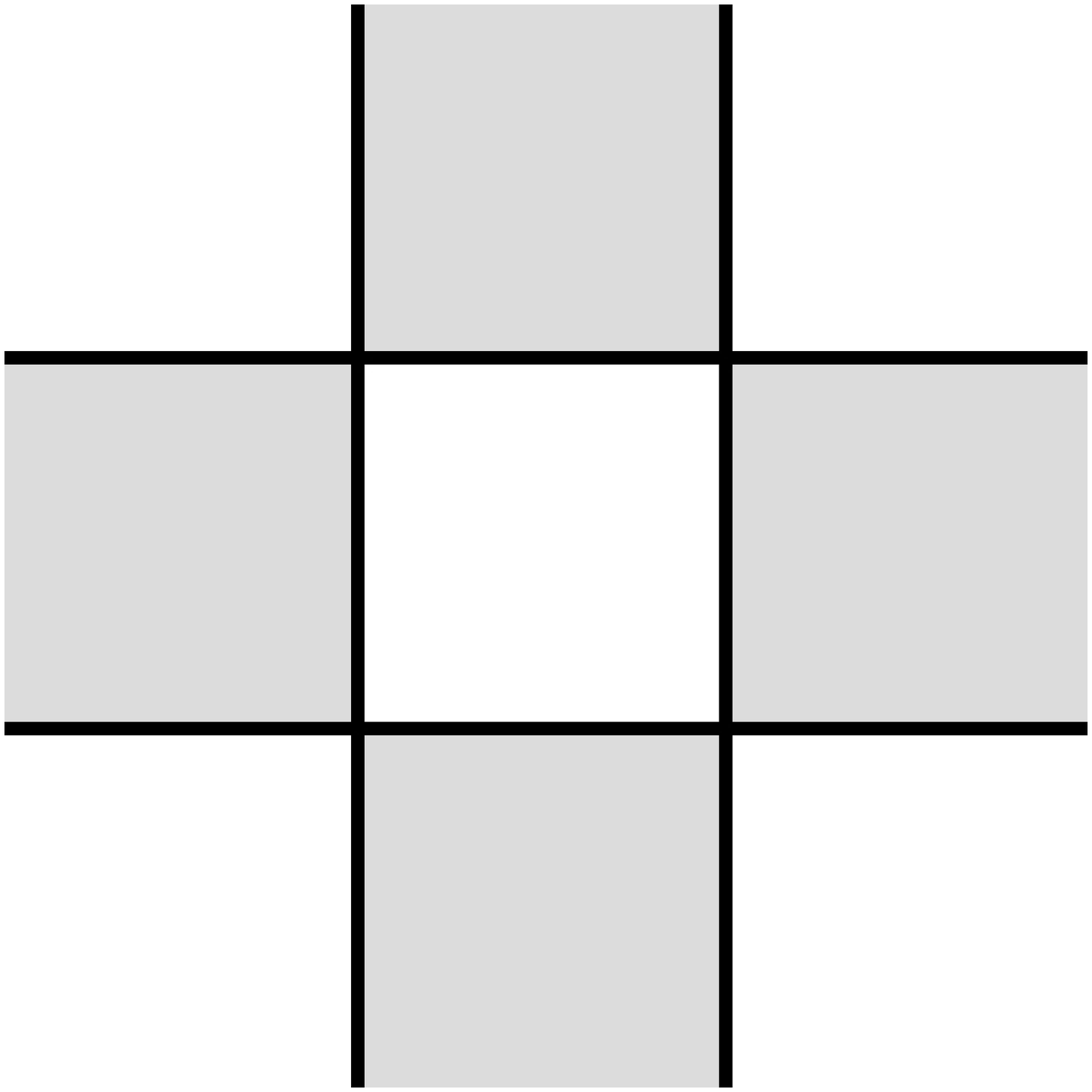}
     \includegraphics[width=0.155\textwidth]{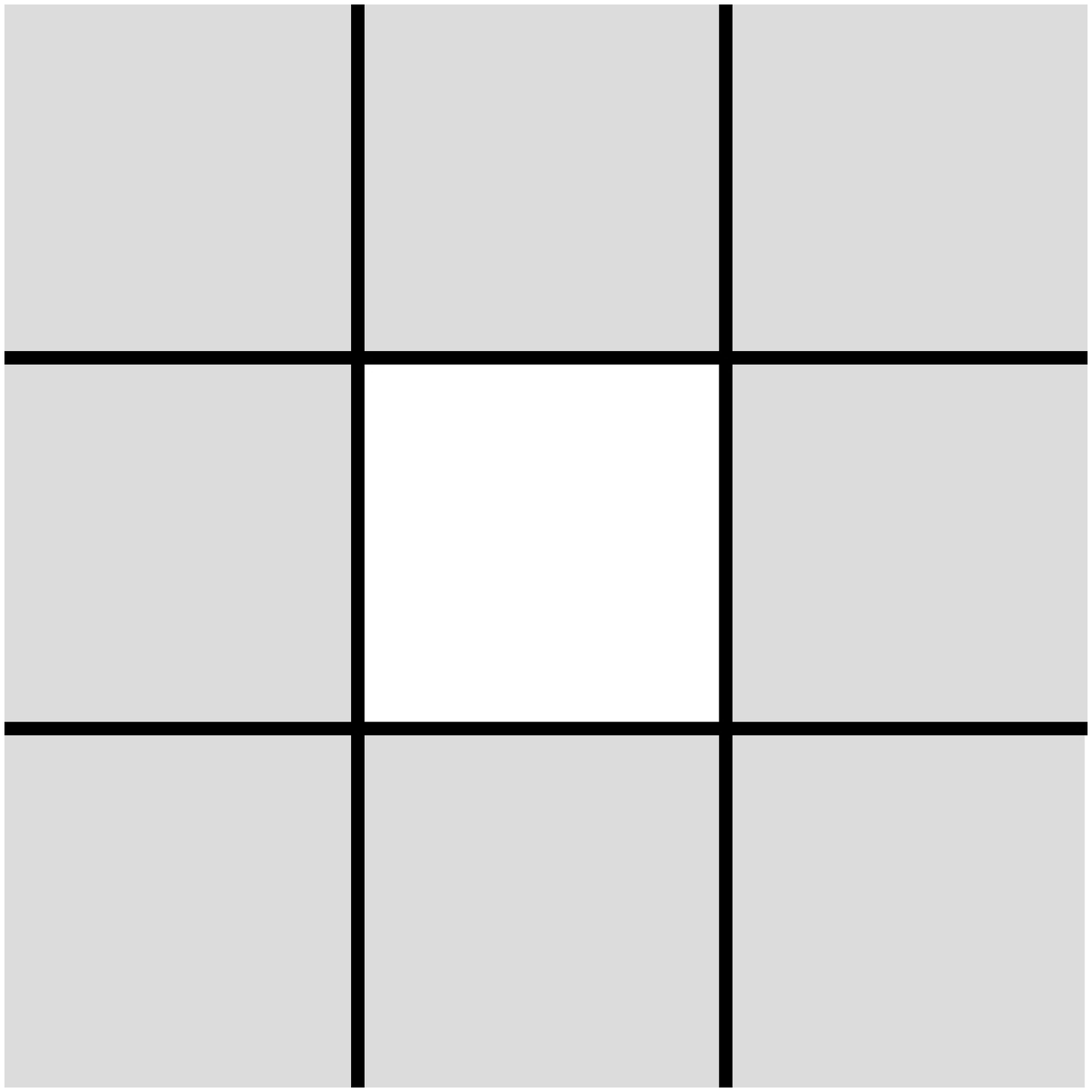}
     \includegraphics[width=0.155\textwidth]{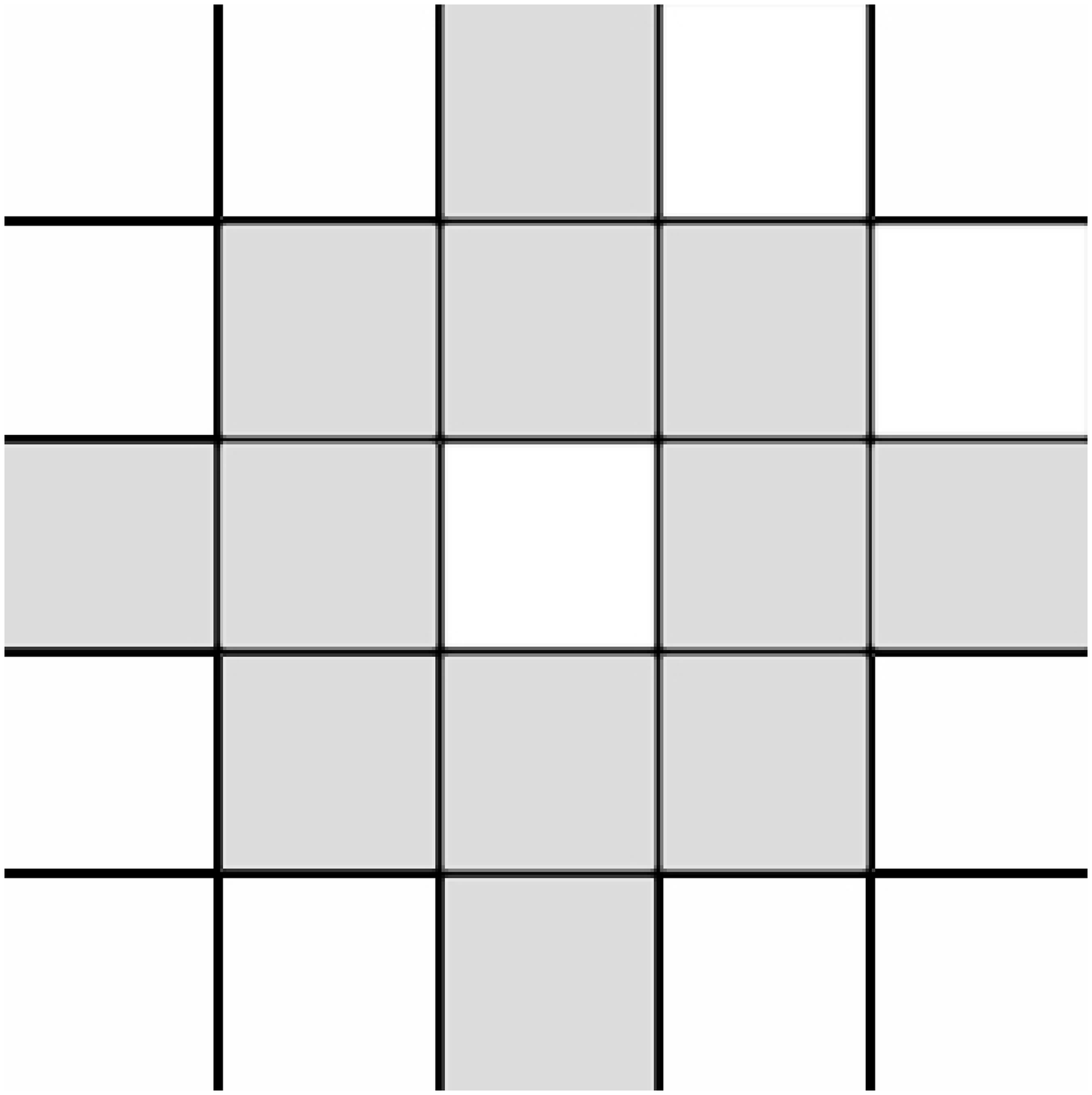}
     \caption{Examples of 2-D grid connectivities. By default the central 
square is white. Cells connected to the centre are represented by gray squares. 
Lefthand frame: a 4-connectivity. Centre frame: a 8-connectivity. Righthand 
frame: a structure element representing a ball of 2 pixels.}
     \label{fig:connect}
\end{figure}

A more detailed outline of the DTFE reconstruction procedure can be found in 
appendix~\ref{app:dtfe}. 

\subsubsection{DTFE grid}
\label{sec:dtfegrid}
DTFE involves the estimate of a continuous field throughout the 
complete sample volume. To process the DTFE field through the WVF 
machinery we sample the field on a grid. It is important to 
choose a grid which is optimally suited for the void finding 
purpose of the WVF method. On the one hand, the grid values should 
represent all physically significant structural features (voids) 
in the sample volume. On the other hand, the grid needs to be 
as coarse as possible in order to suppress the detection of 
spurious and insignificant features. The latter is also beneficial from 
a viewpoint of computational efficiency. This is achieved by adopting 
a gridsize in the order of the mean interparticle distance.

\begin{figure*}
  \includegraphics[width=0.47\textwidth,clip]{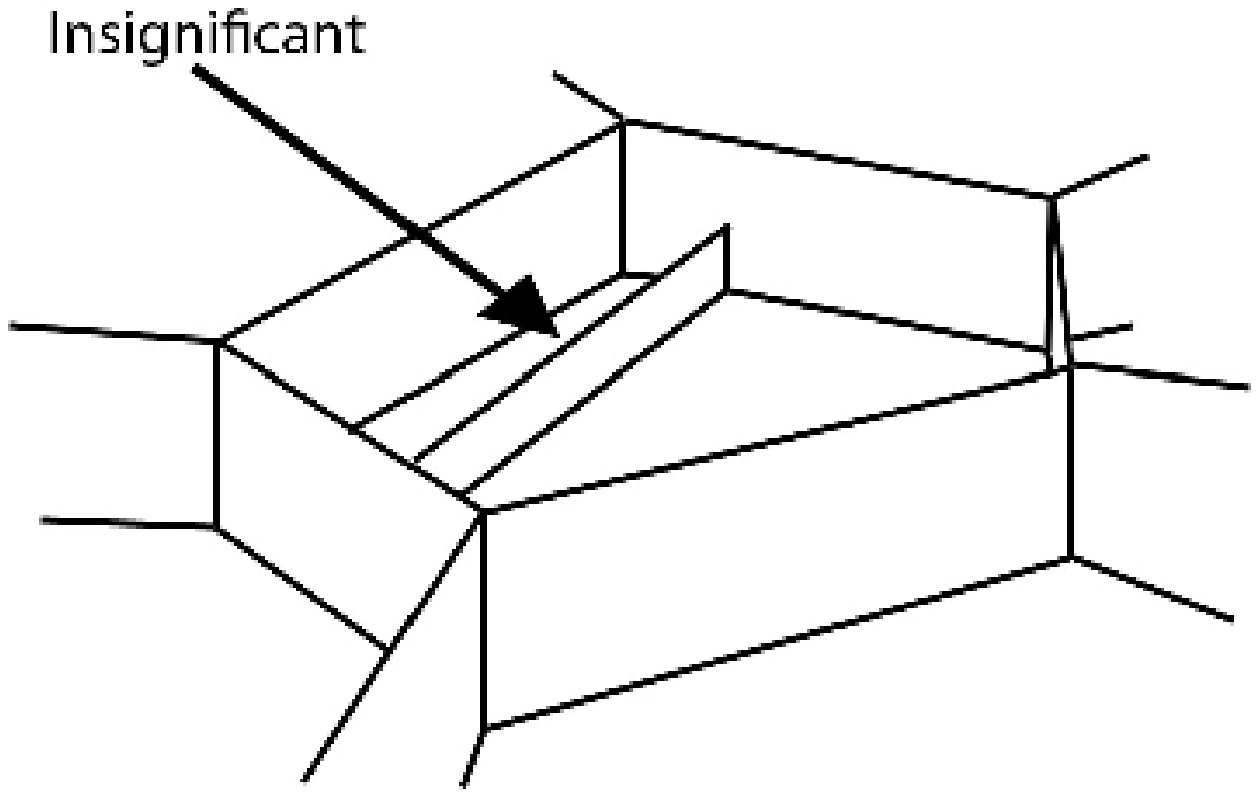}(a)~~
  \includegraphics[angle=0,width=0.47\textwidth]{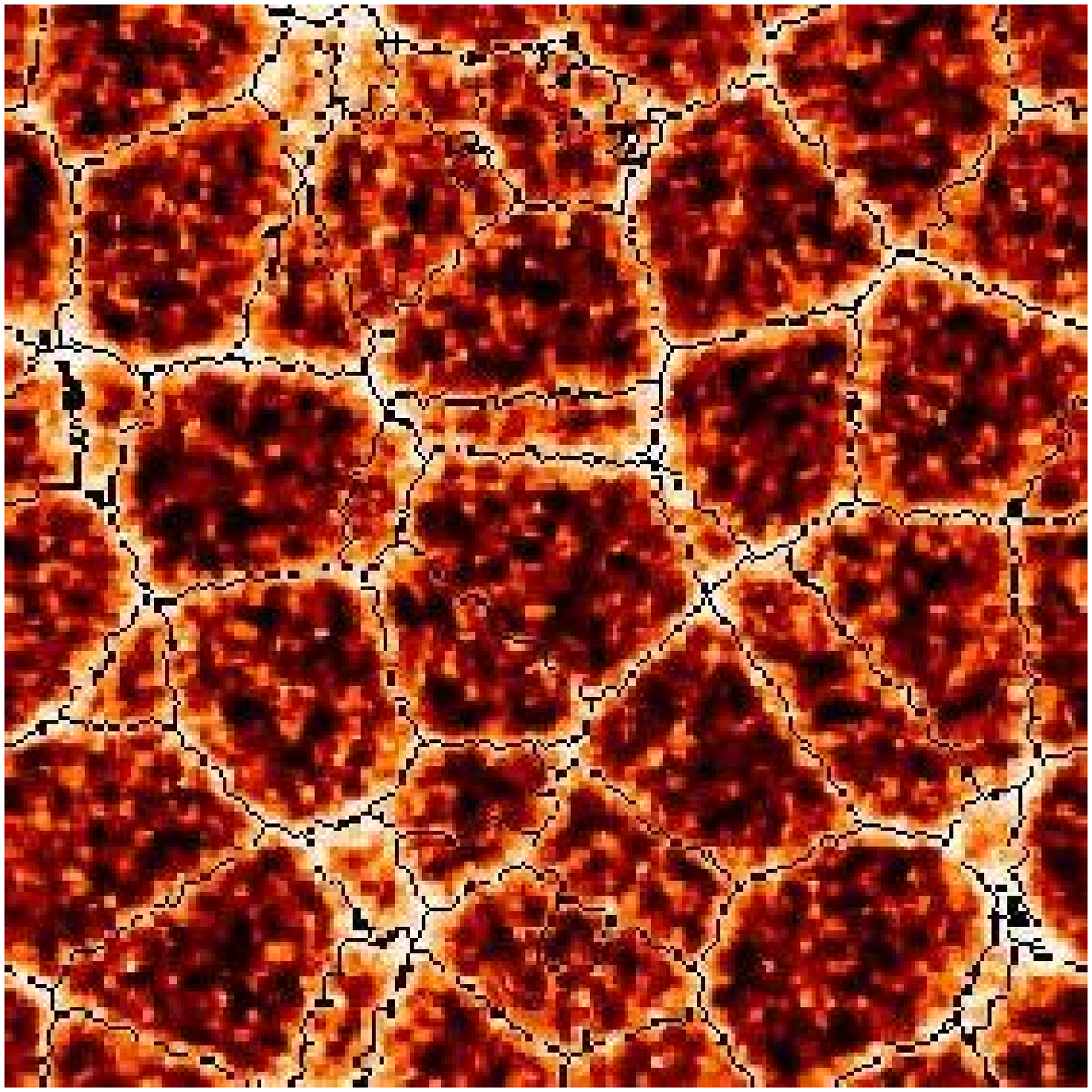}~~(b)
  \caption{The concept of hierarchical watershed. Not all divide lines
  produced by the watershed may be relevant. They are removed if they
  do not fulfil a particular criterium (e.g. if they have a contrast
  lower than some threshold). Only the significant watershed segments
  survive. The segmentation after 5 iterative density smoothings and
  removal of boundaries below a contrast of 0.8.}
  \label{fig:pyrvor152}
\end{figure*}

The DTFE grid sampling is accomplished through Monte Carlo sampling 
within each grid cell. Within each gridcell the DTFE density value 
is measured at 10 randomly distributed sample points. The grid value 
is taken to be their average.

\subsection{Natural Neighbour Rank-Ordered filtering}
\label{sec:natnghb}
A major and novel ingredient of our WVF method intended to eliminate 
shot noise in the DTFE density field reconstructions is that of a
natural non-linear filtering extension: the {\it Natural Neighbour 
Rank-Ordered filtering}

We invoke two kinds of non-linear adaptive smoothing techniques, {\it
Median Filtering} and {\it Max/Min Filtering}, the latter originating
in mathematical morphology (MM). Both filters are rank order filters,
and both have well known behaviour. They have a few important
properties relevant for our purposes. Median filtering is very
effective in removing shot noise while preserving the locations of
edges. The max/min filters are designed to remove morphological
features arising from shot noise (see appendix~\ref{app:mathmorph}).

The filters are defined over neighbourhoods. These are often named
connectivity or, alternatively, structure elements. Image analysis
usually deals with regular two-dimensional image grids. The most
common situation for such grids are straightforward 
4-connectivities or 8-connectivities (see Fig.~\ref{fig:connect}). 
When a more arbitrary shape is used one usually refers to it as a 
structure element.

In the situation of our interest we deal with irregularly spaced data,
rendering it impossible to use any of the above neighbourhoods. It is
the Delaunay triangulation which defines a natural neighbourhood for
these situations. For any point it consists of its {\it Natural
Neighbours}, i.e. all points to which it is connected via an edge of
the Delaunay triangulation (see Fig.~\ref{fig:natnbrs}). This may be
extended to any higher order natural neighbourhood: e.g. a second
order neighbourhood would include the natural neighbours of the (first
order) natural neighbours.

The advantages of following this approach are the same as those for
the DTFE procedure: the Natural Neighbour filtering -- shortly named
NN-{\it median filtering} or NN-{\it min/max filtering} -- forms a
natural extension to our DTFE based formalism. It shares in the major
advantage of being an entirely natural and self-adaptive
procedure. The smoothing kernel is compact in regions of high point
concentrations, while it is extended in regions of low density.

\subsubsection{Implementation NN Rank-Order filtering}
Implementing the min/max and median Natural Neighbour filters within
the DTFE method is straightforward. The procedure starts with the DTFE
density value at each of the (original) sample points. These may be
the particles in an N-body simulation or the galaxies in a redshift
survey. For each point in the sample the next step consists of the
determination of the median, maximum or minimum value over the set of
density values made up by that of the point itself and those of its
natural neighbours. The new ``filtered'' density values are assigned
to the points as the first-order filter value. This process is
continued for a number of iterative steps, each step yielding a higher
order filtering step.

The number of iterative steps of the natural neighbour smoothing is 
dependent on the size of the structure to be resolved and the 
sampling density within its realm. Testing has shown that a reasonable 
order of magnitude estimate is the mean number of sample points along the 
diameter of the structure. As an illustration of this criterion one may want 
to consult the low noise and high noise Voronoi models in fig.~\ref{fig:segorg}. 
While the void cells of the low noise models contain on average 6 points per 
cell diameter, the void cells of the high noise model contain around 16. Fifth-order 
filtering sufficed for the low noise model, 20-th order for the 
high noise model (fig.~\ref{fig:vor11} and fig.~\ref{fig:vor15})

In the final step, following the specified order of the filtering process,
the filtered density values -- determined at the particle positions -- 
are interpolated onto a regular grid for practical processing purposes 
(see sec.~\ref{sec:dtfegrid}). 
\begin{table*}
\caption{\small Parameters of the Voronoi kinematic model realizations: number of cells, 
cell filling factor, percentages of galaxies within each of the morphological components (clusters, filaments, 
walls, field) and the Gaussian width of clusters, filaments and walls.}

\begin{center}
\begin{tabular}{||c||c|c||c||c|c||c|c||c|c||}
\hline 
\hline
&&&&&&&&&\\
Model & M & cell & field & wall & $R_w$ & filament & $R_f$ & cluster & $R_c$ \\
& & filling & & & & & & &    \\
& & factor& & & ($\hmpc$) & & ($\hmpc$) & & ($\hmpc$)   \\
&{\hskip 1.5cm} & {\hskip 1.2cm} & {\hskip 1.2cm} &{\hskip 1.0cm} &{\hskip 1.0cm} &{\hskip 1.0cm} &{\hskip 1.0cm} &{\hskip 1.0cm} &\\
\hline 
&&&&&&&&&\\
 High noise & 180 & 0.500 & 50.0 & 38.3 & 1.0 & 10.6 & 1.0 & 1.1 & 0.5 \\
&&&&&&&&&\\
 Low noise & 180 & 0.025 & 2.5 & 16.4 & 1.0 & 40.6 & 1.0 & 40.5 & 0.5 \\
&&&&&&&&&\\
\hline
\hline
\end{tabular}
\end{center}
\label{tab:vorkinmpar}
\end{table*}

An example of a fifth-order median filtering process is shown in the
bottom lefthand frame of Fig.~\ref{fig:wvf}. The comparison with the
original DTFE field (top righthand frame, Fig.~\ref{fig:wvf}) reveals
the adaptive nature of the filtering process, suppressing noise in the
low-density areas while retaining the overall topology of the density
field. Figs.~\ref{fig:vor11}b and ~\ref{fig:vor15}b show it in the presence of 
controlled noise. 

\subsection{Markers and False Segment Removal}
\label{sec:marker}
Following the NN-median smoothing of the DTFE density field, and some
minor pixel noise removals, the WVF proceeds by identifying the
significant minima of the density field. These are the {\it
Markers} for the watershed transform. In the case of a cosmological
density field the markers are the central deep minima in the
(smoothed) density field.

Almost without exception the Markers do not involve all minima
in a raw unfiltered density field. The minima originating from shot
noise need to be eliminated. In the unfiltered field each regional
minimum would correspond to a catchment basin, producing
over-segmented results: significant watershed basins would tend to get
subdivided into an overabundance of smaller insignificant
patches. While most of these segments are not relevant a beneficial
property of the WST is that truely relevant edges constitute a subset
of the oversegmented segmentation. This notion will be further
exploited in section~\ref{sec:vortest}.

Once the markers have been selected we compute the watershed transform
on the basis of an an {\it ordered queues} algorithm.  This process is
described in detail in \citep{BeuMey}, and outlined in
appendix~\ref{app:wshedimpl}. The process has a few important
advantages. It is rather efficient because each point is processed
only once while it naturally involves Watershed by Markers.

\subsection{Hierarchy Merging}
\label{sec:merging}
The WVF procedure combines two strategies to remove the artefacts generated 
by Poisson noise resulting from a density field discretely sampled by 
particles or galaxies. 
\begin{itemize}
\item the preprocessing of the image such that the insignificant 
minima are removed 
\item merging of subdivided cells into larger ones.
\end{itemize}
The first strategy involves the previously described reconstruction of
the density field by DTFE, followed by a combination of edge
preserving {\it median filtering} and smoothing with the morphological
{\it erosion} and {\it dilation} operators
(appendix~\ref{app:mathmorph}). In general, as will be argued and demonstrated 
in this study, it provides a good strategy for recovering only significant voids. 
The second strategy involves the {\it merging} of neighbouring patches via a
user-specified scheme.

Amongst a variety of possibilities we have pursued a well known method
for merging patches, the {\it watershed hierarchy}. In its original
form it assigns to each boundary a value dependent on the difference
in density values between the minima of the neighbouring patches on
either side of the ridge. We implemented a variant of this scheme 
where the discriminating value is that of the density value integrated
over the boundary. A critical contrast threshold determines the
outcome of the procedure. For an integral density value lower than the
contrast threshold the two patches are merged. If the value is higher
the edge is recognized as a genuine segment boundary.

\begin{figure*}
\begin{center}
  \includegraphics[angle=0,width=.30\textwidth]{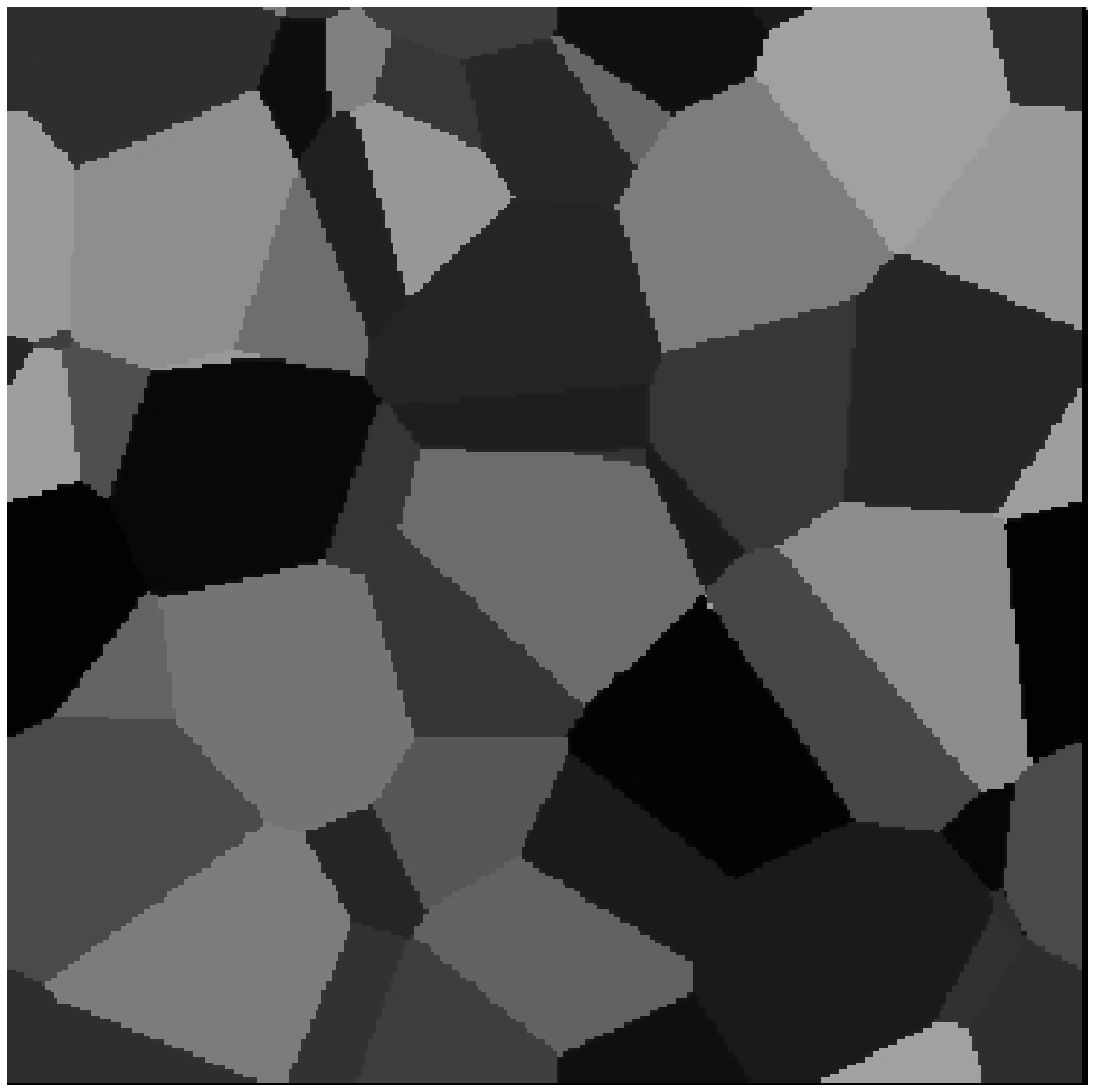}~(a)~
  \includegraphics[angle=0,width=.30\textwidth]{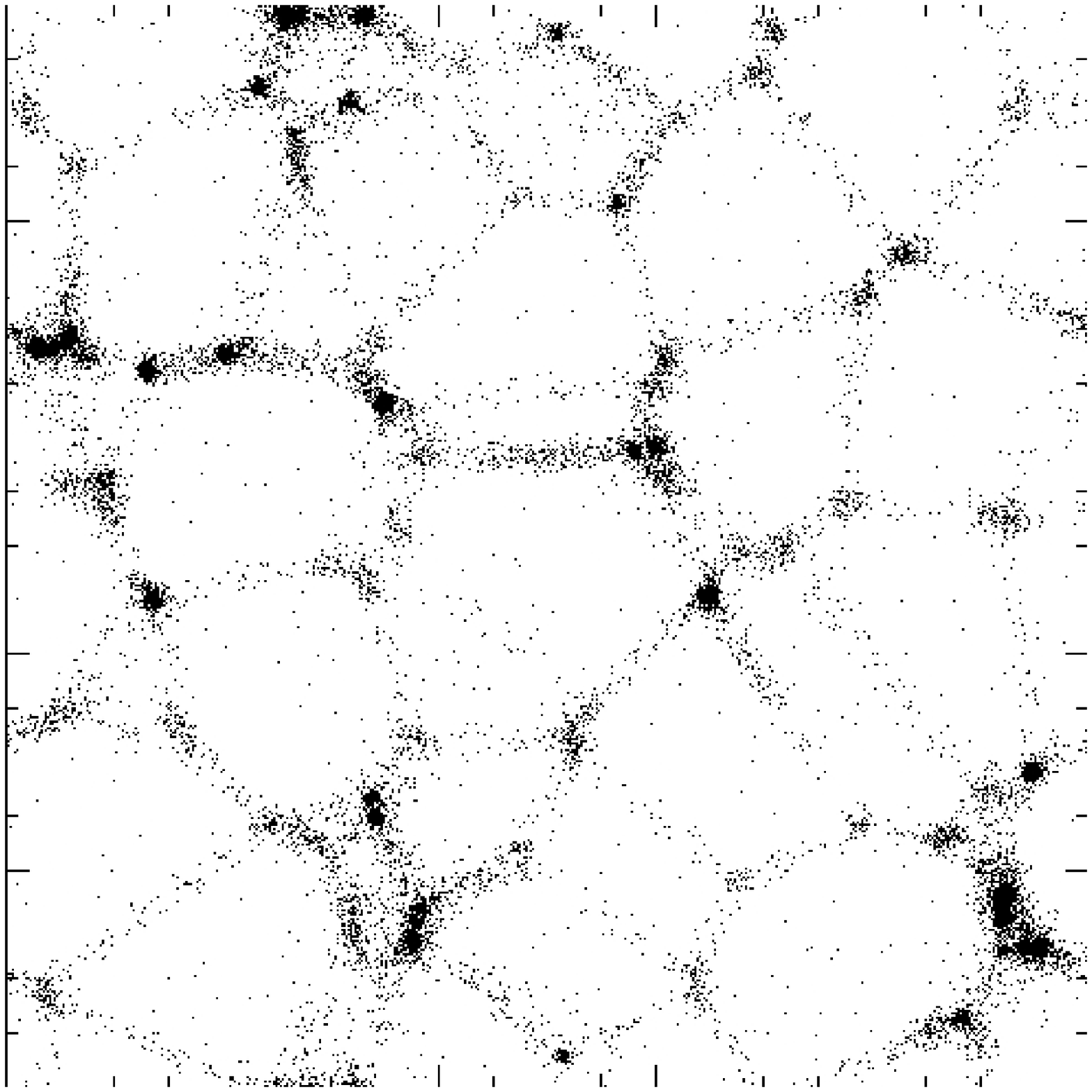}~(b)~
  \includegraphics[angle=0,width=.30\textwidth]{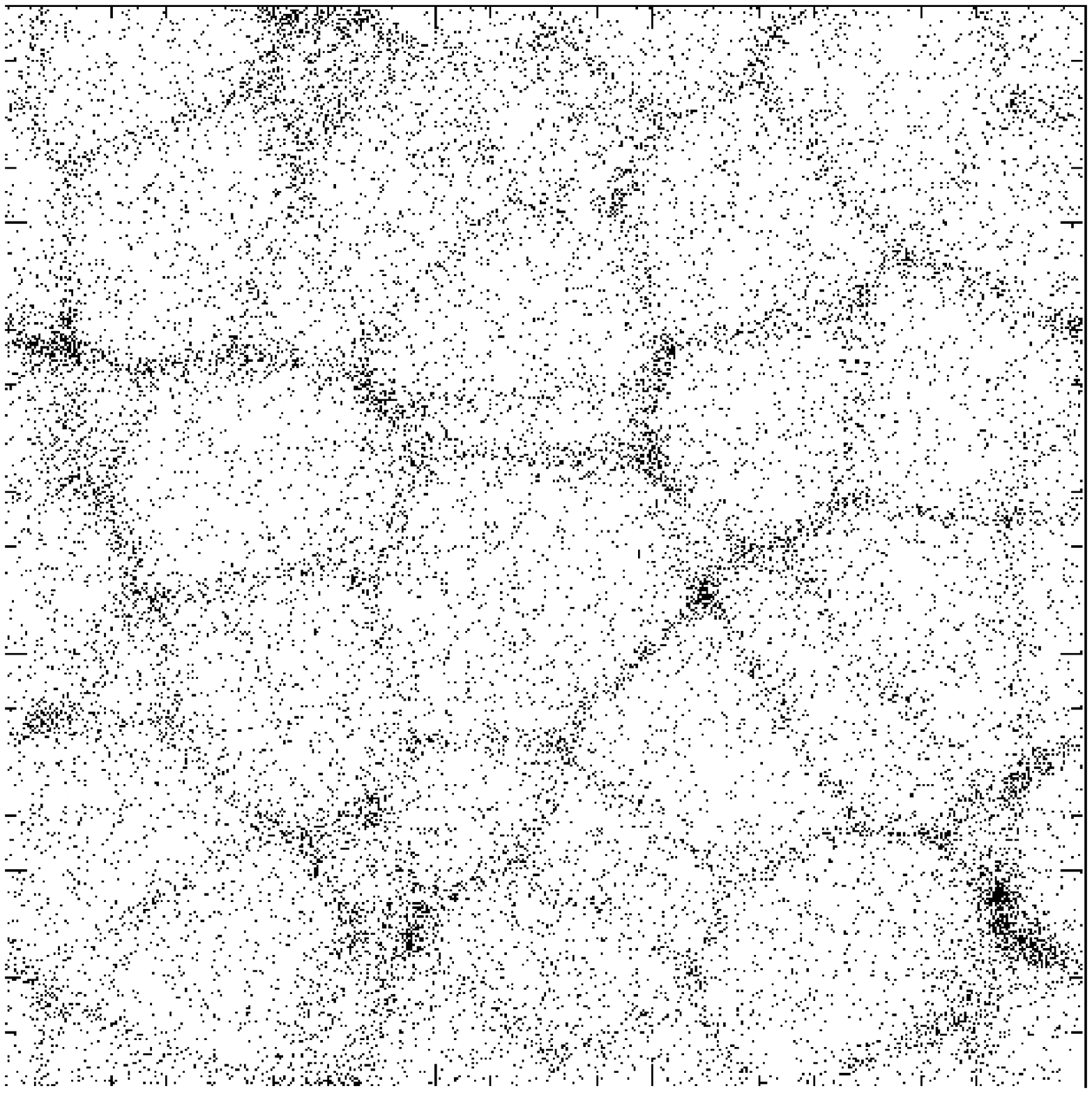}~(c)~
  \caption{Frame~(a) shows a slice through the original (geometrically 
  defined) Voronoi tessellation. For two different Voronoi clustering 
  models defined within this same tessellation, frames (b) and (c) 
  depict the particles within the same slice. Frame~(b) shows the low 
  noise case with a high density contrast between the voids and walls. 
  Frame~(c) is a high noise model with a relatively low contrast between 
  voids and walls.}
  \label{fig:segorg}
  \vspace{0.2cm}
  \includegraphics[angle=0,width=.30\textwidth]{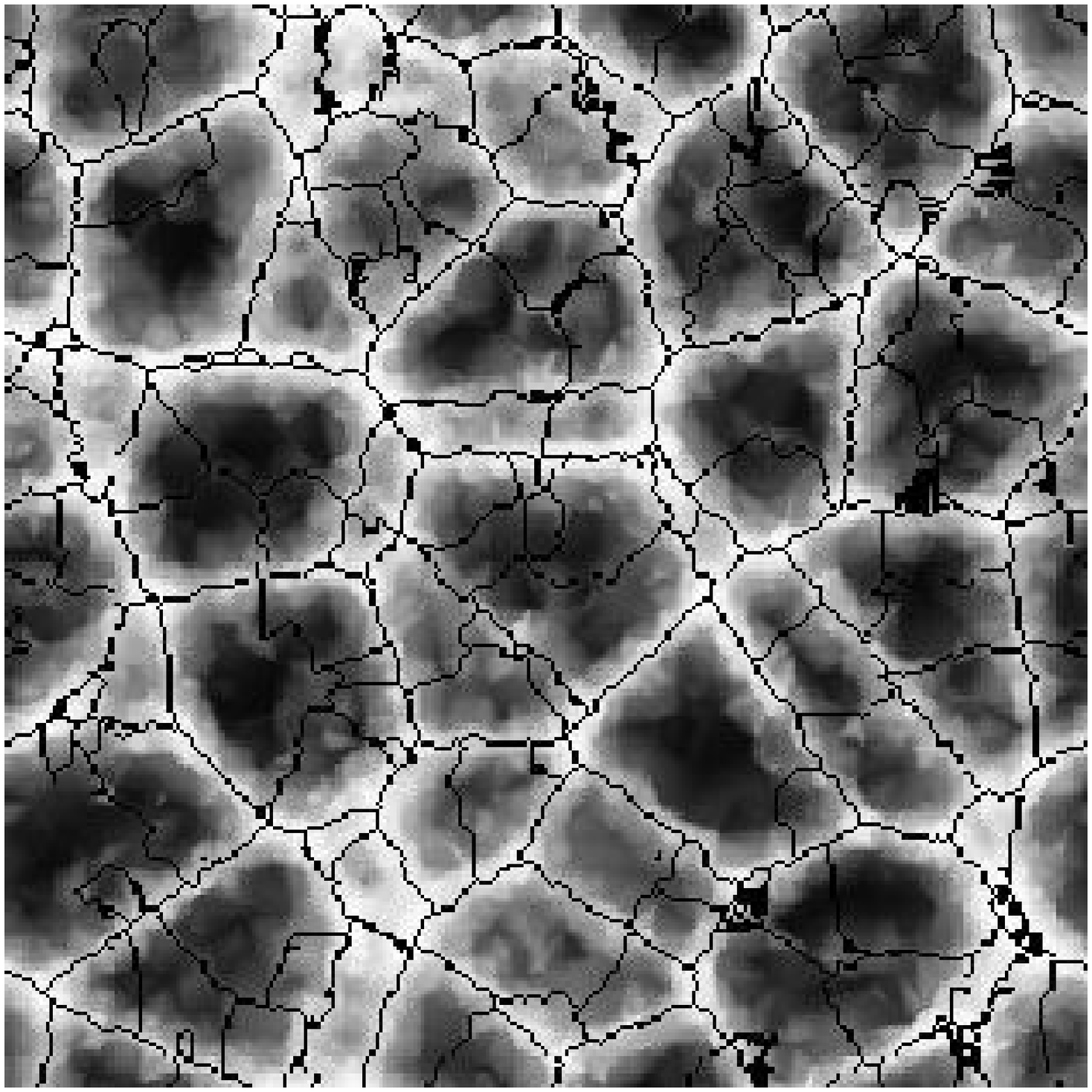}~(a)~
  \includegraphics[angle=0,width=.30\textwidth]{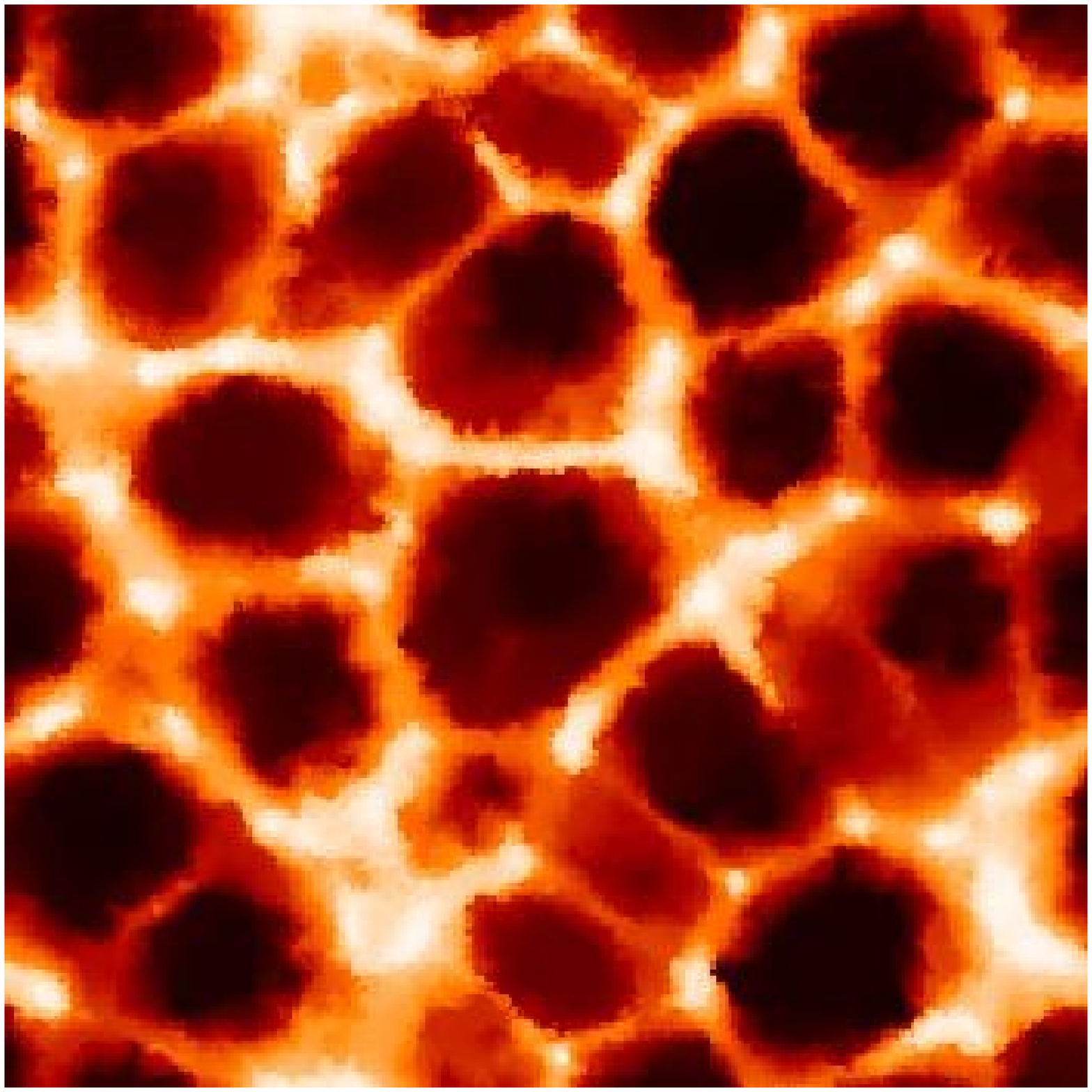}~(b)~
  \includegraphics[angle=0,width=.30\textwidth]{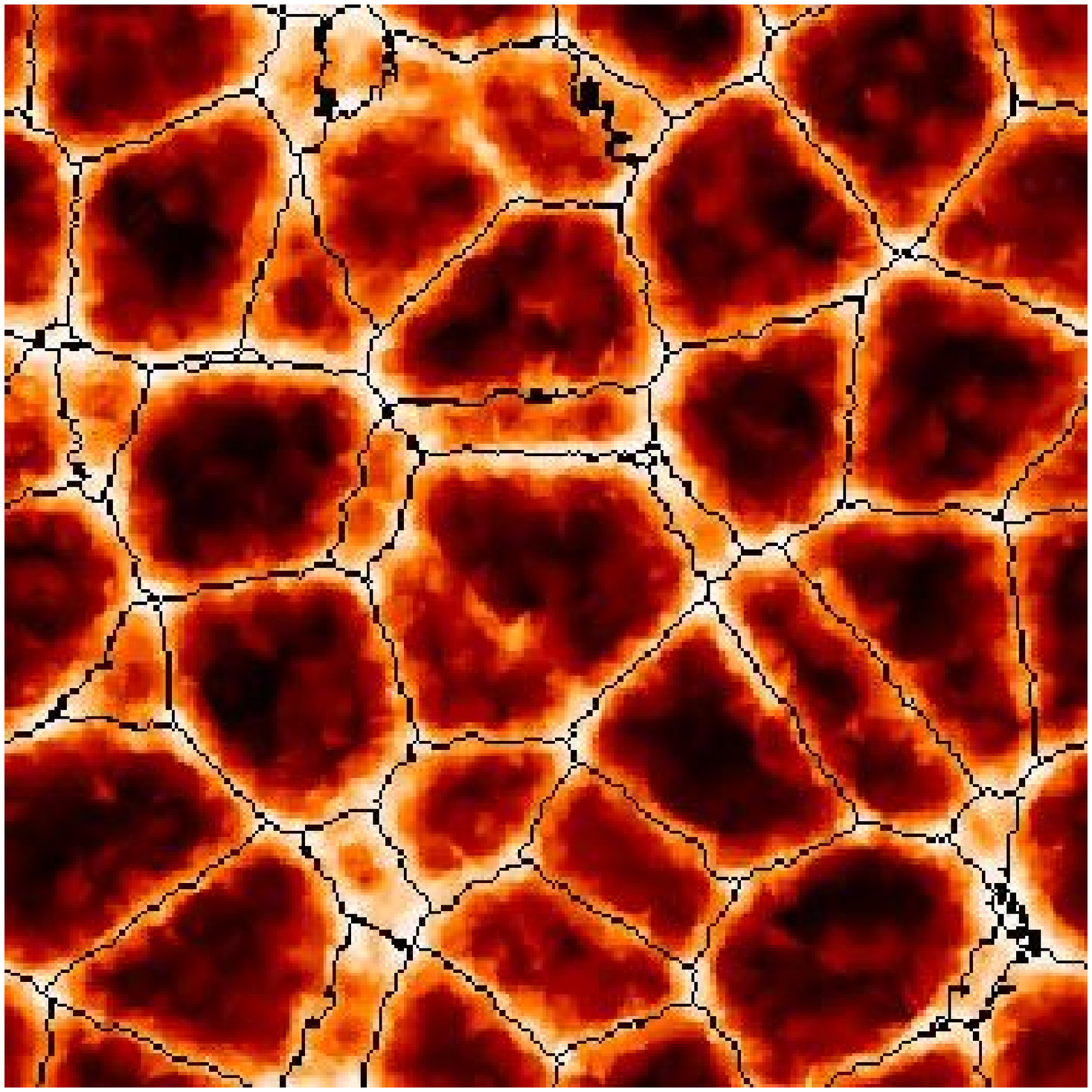}~(c)~
  \caption{The density field of the particle distribution in the
    low noise model (a). Superimposed are the WVF segmentation boundaries.
    The central frame (b) shows the resulting 5-th order median-filtered 
    density field (b). This filtered field is the input for the 
    watershed procedure whose segmentation is delineated in 
    frame (c), superimposed on top of the original density field.}
  \label{fig:vor11}
  \vspace{0.2cm}
  \includegraphics[angle=0,width=.30\textwidth]{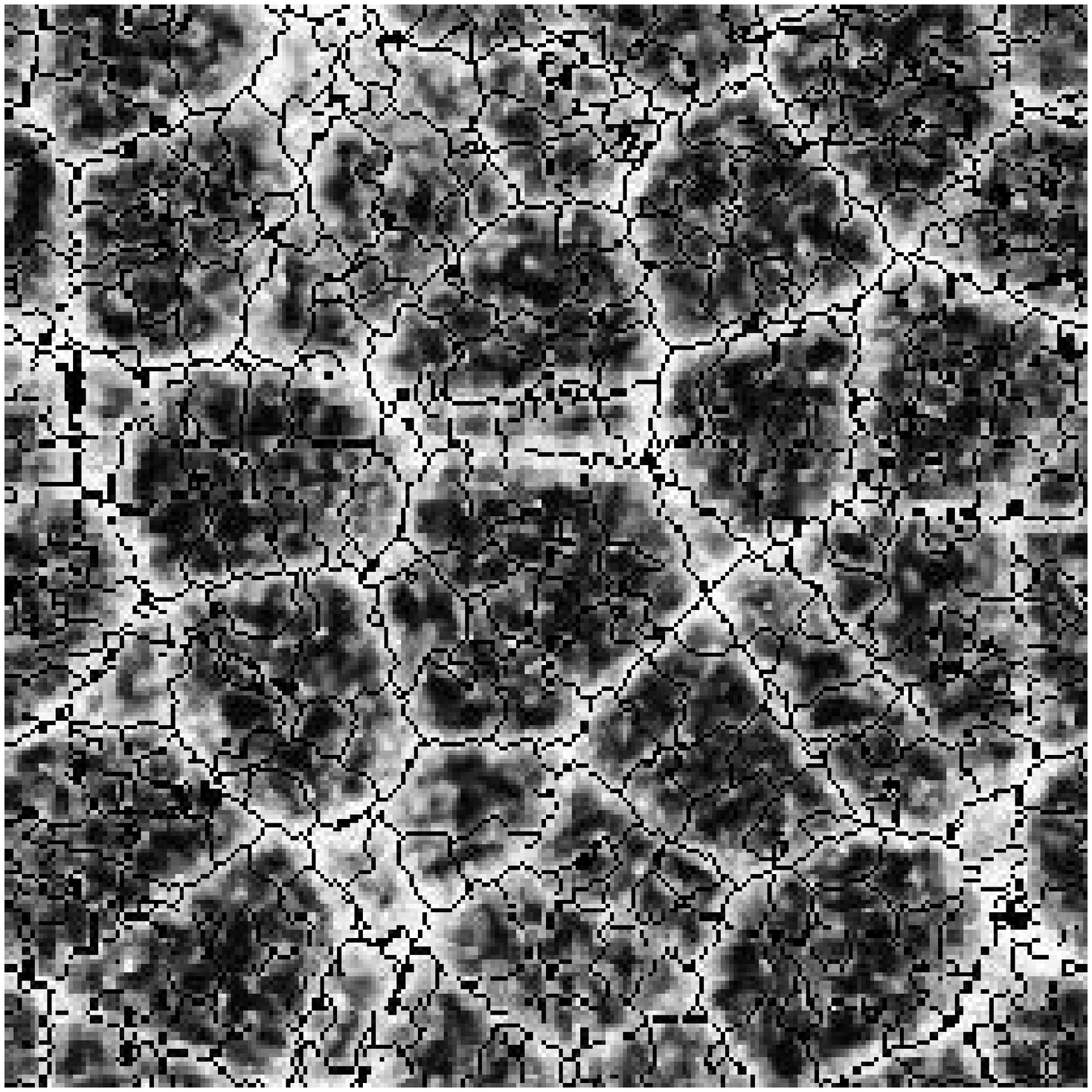}~(a)~
  \includegraphics[angle=0,width=.30\textwidth]{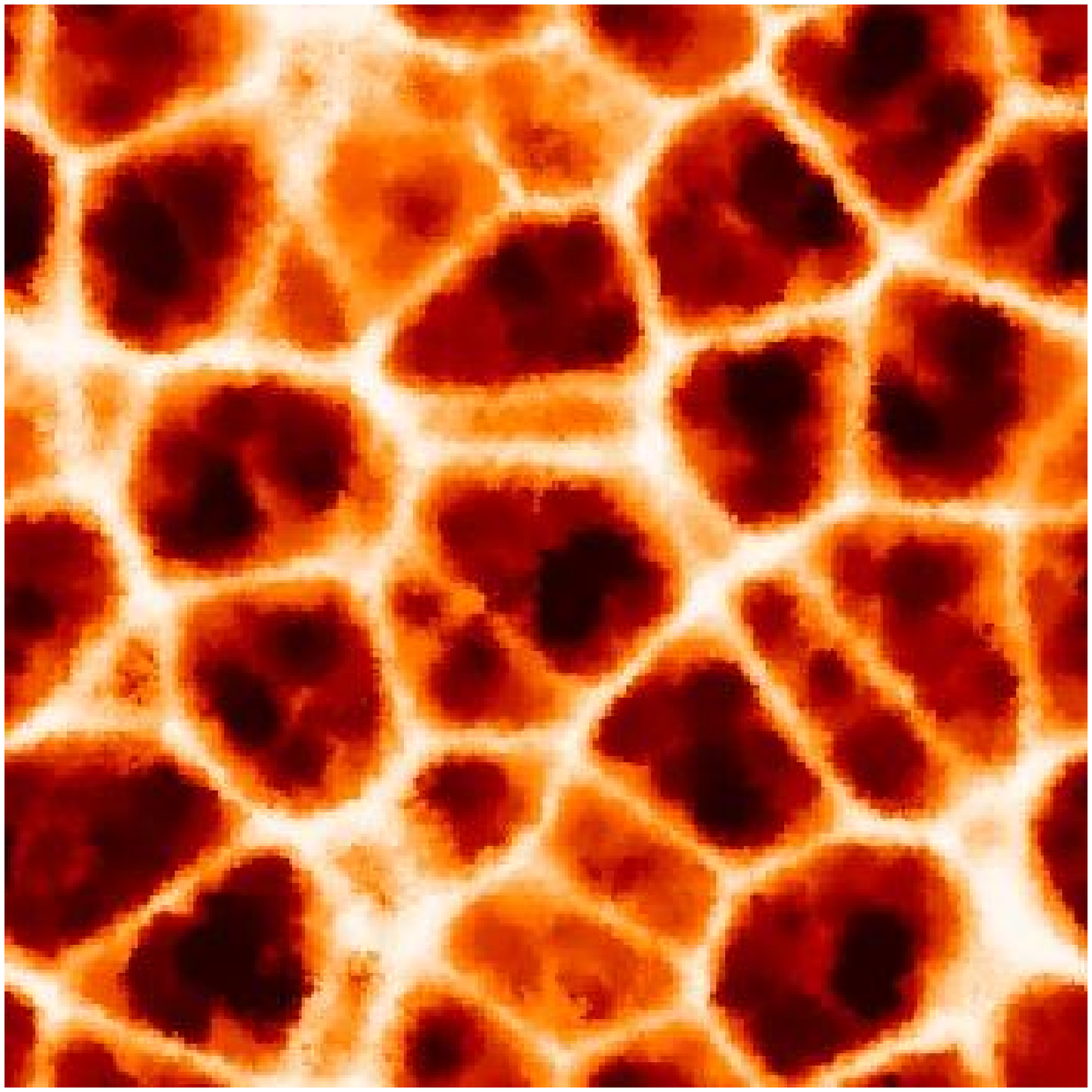}~(c)~
  \includegraphics[angle=0,width=.30\textwidth]{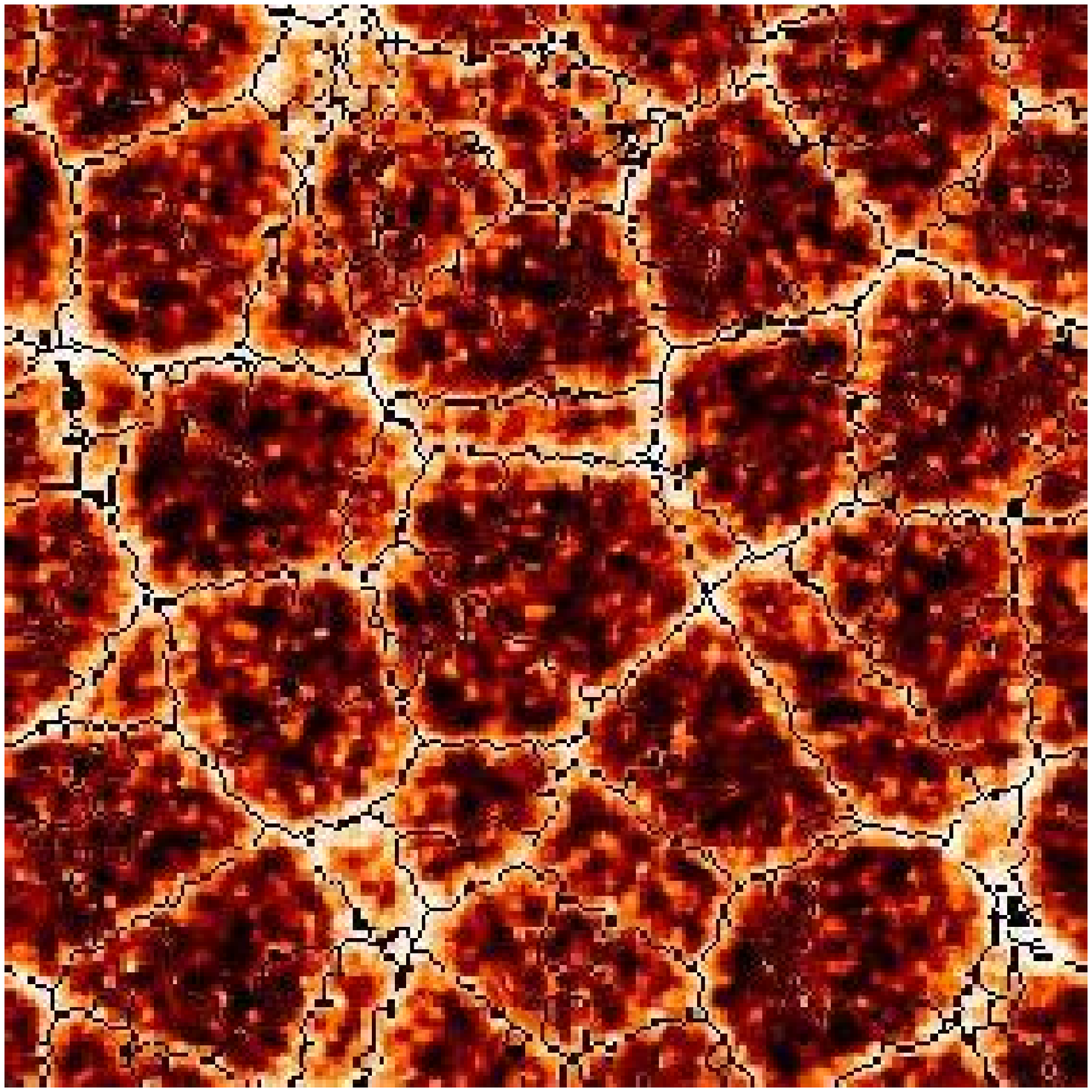}~(c)~
    \caption{The density field of the particle distribution in the
    high noise model (a). Superimposed are the WVF segmentation boundaries.
    The central frame (b) shows the resulting 20-th order median-filtered 
    density field. The WVF segmentation of the 5-th order median 
    filtered density field, followed by removal of boundaries below a 
    contrast of 0.8, is depicted in frame (c), superimposed on top of 
    the original density field.}
  \label{fig:vor15}
\end{center}
\end{figure*}

The watershed hierarchy procedure is illustrated in
Fig.~\ref{fig:pyrvor152}(a).  An example of its operation is provided
by Fig.~\ref{fig:pyrvor152}(b), one of the Voronoi clustering models
extensively analyzed in the remainder of this study. It depicts the
segmentation resulting from watershed processing of a 5 times
iteratively NN-median smoothed density field, followed by the
hierarchical removal of boundaries. The improvement compared to the 
segmentation of a merely 5 times median smoothed density field is 
remarkable (cf.  lefthand and righthand panel Fig.~\ref{fig:vor15}).

\subsubsection{Merger Threshold}
\label{sec:threshold}
In addition to the removal of features on morphological grounds, we
also have to possibility to remove features on the basis of the
involved density values.

In the case of voids we expect that they mature as they reach a
density deficit of $\Delta \approx -0.8$ \citep[see e.g][]{SheWey}.
Any structures with a lower density may be residual features, the
diminishing low density boundaries of the subvoids which have merged
\citep[see e.g][]{DubDac}. Various void finding techniques do in fact
exploit this notion and restrict their search to regions with
$\Delta<-0.8$ \citep[see e.g.][]{ColShe}. Note that in practice it may
also involve noise, of considerable significance in these diluted
regions.

\begin{figure*}
  \includegraphics[angle=0,width=.3\textwidth]{intrinsic_grep.eps}~(a)~
  \includegraphics[angle=0,width=.3\textwidth]{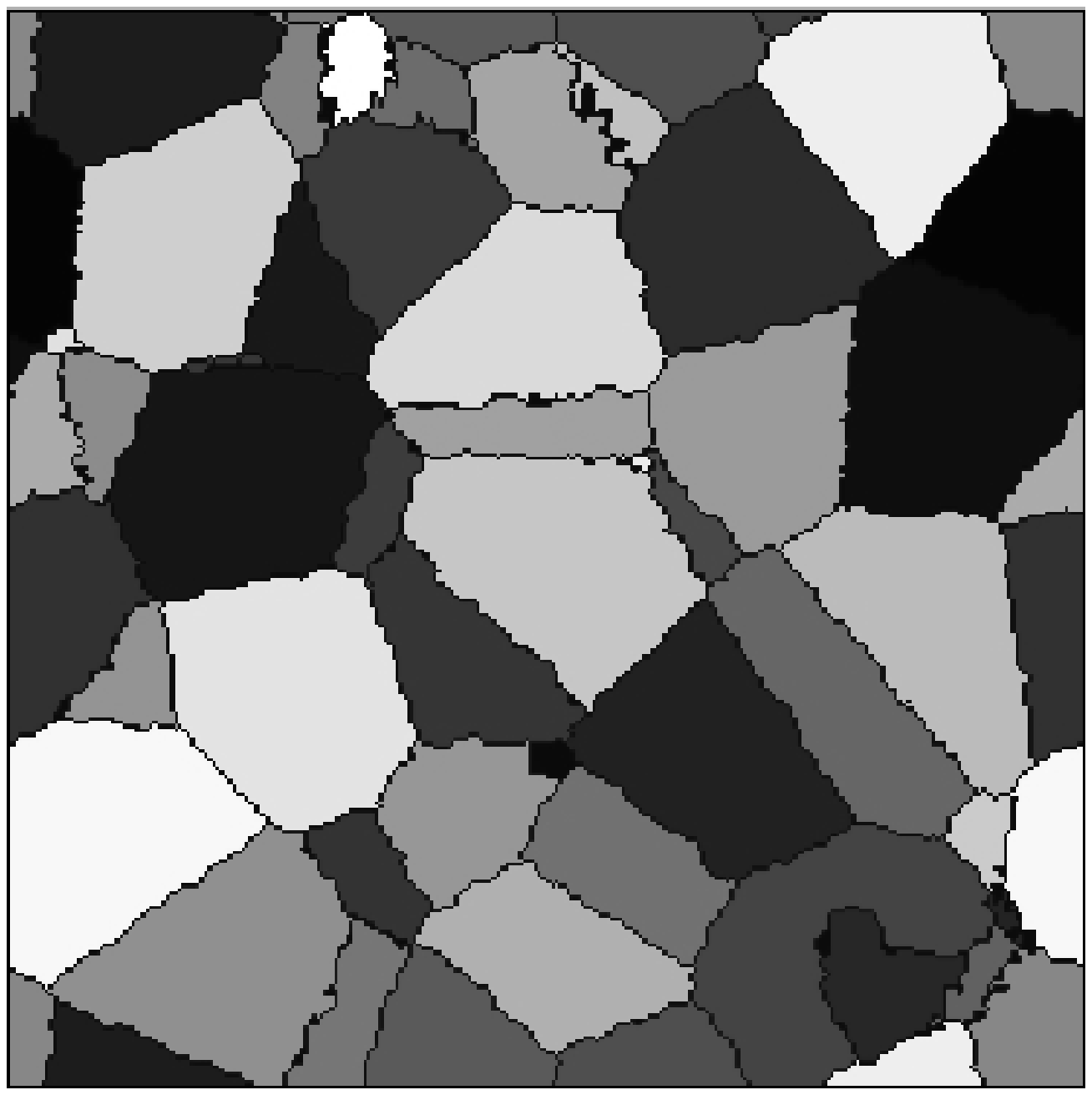}~(b)~
  \includegraphics[angle=0,width=.3\textwidth]{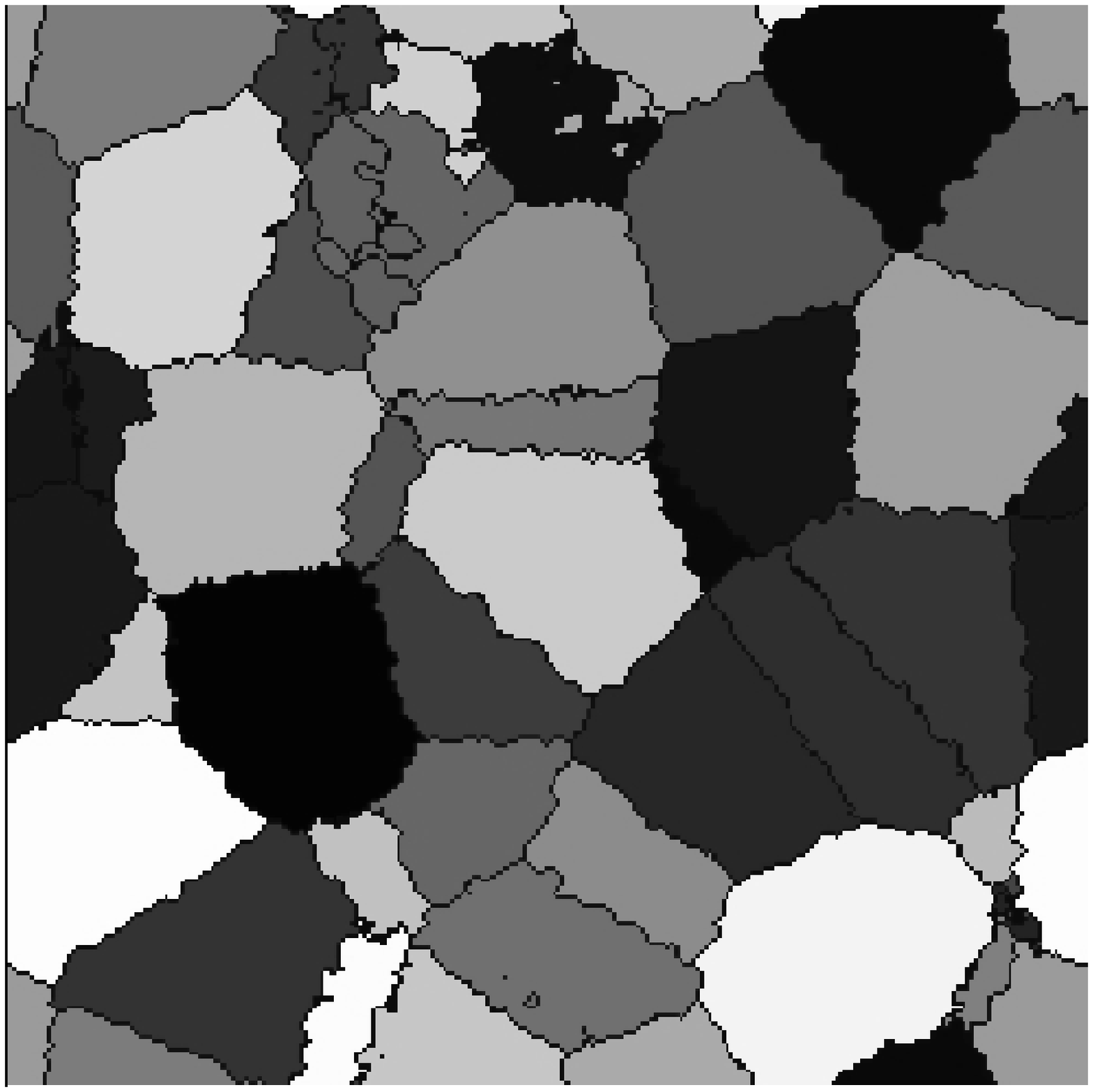}~(c)~
  \caption{Frame (a): the original (geometric) Voronoi tessellation. 
  Frames (b) and (c): the best recovered WVF segmentation of the 
  lownoise (b) and high noise (c) models.}
  \label{fig:best}
\end{figure*}

A density threshold may indeed be applied within the WVF. This
threshold is applied following the watershed transform. Any ridges and
features with a density contrast lower than a specified threshold are removed.
The threshold $\Delta=-0.8$ is a natural value of choice. The goal is
twofold: to suppress noise or spurious features within voids and to
select out subvoids.

\section{WVF Test:\ \ \\ 
\ \ \ \ \ Voronoi Clustering Model}
\label{sec:vortest}
To test and calibrate the Watershed Void Finder we have applied the
WVF to a Kinematic Voronoi Model 
\citep{WeyIck,weygaert1991,weygaert2002,weygaert2007}. In the case of
the Voronoi models we have exact quantitative information on the
location, geometry and identity of the Voronoi cells, whose interior
functions as the voids in the matter distribution, against which we
compare the outcome of the WVF analysis. These models combine the
spatial intricacies of the cosmic web with the virtues of a model that
has a priori known properties.  They are particularly suited for
studying systematic properties of spatial galaxy distributions
confined to one or more structural elements of nontrivial geometric
spatial patterns.  The Voronoi models offer flexible templates for
cellular patterns, and they are easy to tune towards a particular
spatial cellular morphology.

Kinematic Voronoi models belong to the class of Voronoi
clustering models. These are heuristic models for cellular spatial
patterns which use the Voronoi tessellation as the skeleton of the
cosmic matter distribution. The tessellation defines the structural
frame around which matter will gradually assemble during the formation
and growth of cosmic structure \citep{Vor,Oka}. The interior of
Voronoi cells correspond to voids and the Voronoi {\it planes}
with sheets of galaxies. The edges delineating the rim of each
wall are identified with the filaments in the galaxy distribution.
What is usually denoted as a flattened ``supercluster'' will consist
of an assembly of various connecting walls in the Voronoi foam, as
elongated ``superclusters'' of ``filaments'' will usually include a
few coupled edges. The most outstanding structural elements are the
vertices, corresponding to the very dense compact nodes within
the cosmic web, rich clusters of galaxies. We distinguish two
different yet complementary approaches, Voronoi Element Models 
and Kinematic Voronoi models.  The Kinematic Voronoi
models are based upon the notion that voids play a key organizational
role in the development of structure and make the Universe resemble a
soapsud of expanding bubbles \cite{Ick}. It forms an idealized and
asymptotic description of the outcome of the cosmic structure
formation process within gravitational instability scenarios with
voids forming around a dip in the primordial density field. This is
translated into a scheme for the displacement of initially randomly
distributed galaxies within the Voronoi skeleton (see
sect~\ref{app:vorclustform} for a detailed specification). Within a
void, the mean distance between galaxies increases uniformly in the
course of time. When a galaxy tries to enter an adjacent cell, the
velocity component perpendicular to the cell wall
disappears. Thereafter, the galaxy continues to move within the wall,
until it tries to enter the next cell; it then loses its velocity
component towards that cell, so that the galaxy continues along a
filament. Finally, it comes to rest in a node, as soon as it tries to
enter a fourth neighbouring void.  A detailed description of the model
construction may be found in section~\ref{app:vorclustform}.

To test and calibrate the Watershed Void Finder technique we have
applied the WVF to a high contrast/low noise Voronoi galaxy
distribution and a low contrast/high noise one. Both concern two
stages of the same Kinematic Voronoi model, the high noise one
to an early timestep with a high abundance of field galaxies and the
low noise one to an advanced stage in which most galaxies have moved
on towards filament or cluster locations. While the models differ
substantially in terms of cell filling factor, the underlying
geometric pattern remains the same: the position of the nodes, edges
and walls occupy the same location. Most importantly for our purposes:
the Voronoi cells, identified with the interior of the voids, are the
same ones, be it that the high noise cells are marked by a substantial
population of randomly distributed points.

The model has been set up in a (periodic) box with 141 $\Mpch$ size,
and is based on a Voronoi tessellation defined by 180 Voronoi
cells. In total $128^3$ particles were displaced following the
kinematic Voronoi evolution. Table~\ref{tab:vorkinmpar} specifies the
distinctive parameters defining the model realizations, and
Fig.~\ref{fig:segorg} shows the particle distribution for the two
model distributions in a central slice through the model box.

\subsection{Voronoi Model:\ \ Watershed Segmentation} The density/intensity field is
determined by DTFE, yielding a $256^3$ grid of density
values. Fig.~\ref{fig:vor11} contains an example of the outcome of the
resulting DTFE density interpolation, with the contour levels
determined according to the description in section~\ref{sec:method}.
The density map clearly reflects the filaments and nodes that were
seen in the particle distribution. The void interiors are dominated by
noise, visible as islands within a large zero density ocean.

\begin{table*}
\caption{Quantitative comparison of the original and retrieved voids}
\centering
\begin{tabular}{|l|l||c|c|c|c|c|}
\hline
\ &&&&&&\\
Model & Parameters   & Voids & Splits & Mergers & Correct & Correctness \\
\ {\hskip 2.0cm}&\ {\hskip 2.0cm}&\ {\hskip 1.5cm}&\ {\hskip 1.5cm}&\ {\hskip 1.5cm}&\ {\hskip 1.5cm}&\ {\hskip 1.5cm}\\
\hline
\hline
\ &&&&&&\\
Intrinsic &          &  180  &  -     & -  & -   & - \\
\ &&&&&&\\
\hline
\hline
\ &&&&&&\\
&original             &    847 &   -   &  - & -   & - \\
&max/min              &    259 &  82   &  3 & 118 & 66\\
Low noise &med2                 &    180 &   6   &  6 & 159 & 88\\
&med5                 &    162 &   9   & 30 & 119 & 66\\ 
&med20                &    136 &  20   & 80 &  33 & 18\\
\ &&&&&&\\
\hline
\hline
\ &&&&&&\\
&original             & 4293   &  -    & -  &  -  & - \\
&max/min              & 3540   &  -    & -  &  0  & - \\
&med5                 & 723    & 529   &  0 &  8  & 4 \\
High noise&med20                & 275    &  95   &  3 & 100 & 55\\
&hierarch            & 251    &  75   & 44 & 90  & 50\\    
&med5hr               & 172    &   6   & 12 & 144 & 80\\
&med20hr              & 175    &   1   &  6 & 160 & 89\\
\ &&&&&&\\
\hline
\end{tabular}
\label{tab:void}
\end{table*}

A direct application of the watershed transform results in a starkly
oversegmented tessellation (Fig.~\ref{fig:vor11} and
Fig.~\ref{fig:vor15}). Amongst the overabundance of mostly artificial,
noise-related, segments we may also discern real significant
watersheds. Their boundary ridges (divide lines) are defined by
filaments, walls and clusters surrounding the voids. Many of these
genuine voids are divided into small patches. They are the result of
oversegmentation induced by the noisy Poisson point distribution
within the cells. The local minima within this background noise will
act as individual watershed flood centres marking corresponding,
superfluous, watershed segments.

While for a general cosmological distribution it may be challenging to
separate genuine physical subvoids from artificial noise-generated
ones, the Voronoi kinematic models have the unique advantage of having
no intrinsic substructure. Any detected substructure has to be
artificial, rendering it straightforward to assess the action of the
various steps intent on removing the noise contributions.

\subsubsection{Smoothing and Segment Merging}
The first step in the removal of insignificant minima consists of the
application of the iterative natural neighbour median filtering
process.  This procedure, described in sect.~\ref{sec:natnghb},
removes some of the shot noise in the low density regions. At the same time it is
edge preserving.  The result of five NN-median filtering iterations on
the high noise version of the Voronoi kinematic clustering model is
shown in Fig.~\ref{fig:vor11}.  With the exception of a few
artificial edges the resulting watershed segmentation almost perfectly
matches the intrinsic Voronoi tessellation.


Figure~\ref{fig:vor15} shows the result for the high noise version of
the same Voronoi kinematic clustering model. In this case pure
NN-median filtering is not sufficient. A much more acceptable result
is achieved following the application of the watershed hierarchy 
segment merging operation and the removal of ridges with a density
contrast lower than the 0.8 contrast threshold.

\begin{figure*}
\begin{center}
  \includegraphics[width = 0.48\textwidth]{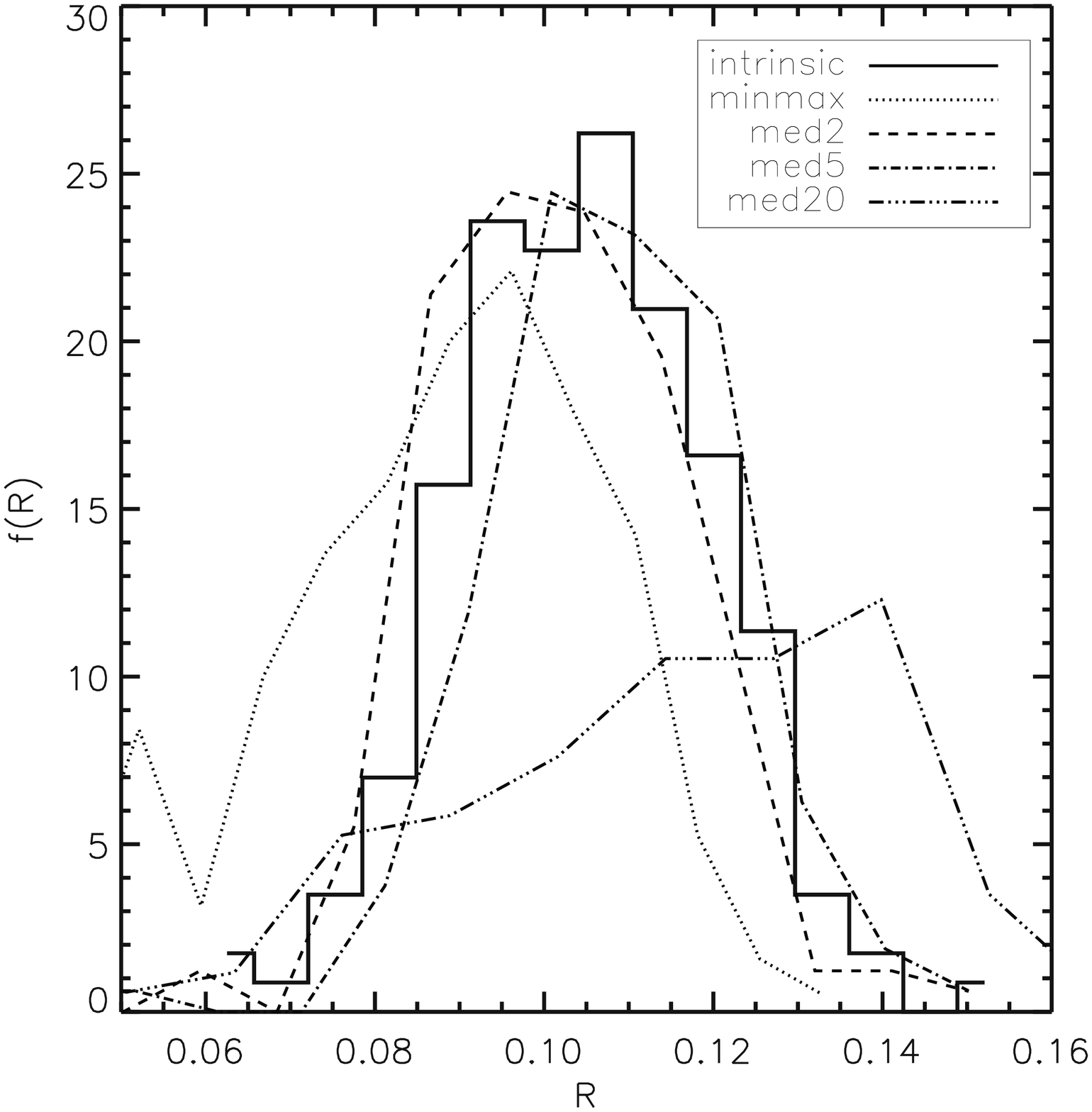}
  \includegraphics[width = 0.48\textwidth]{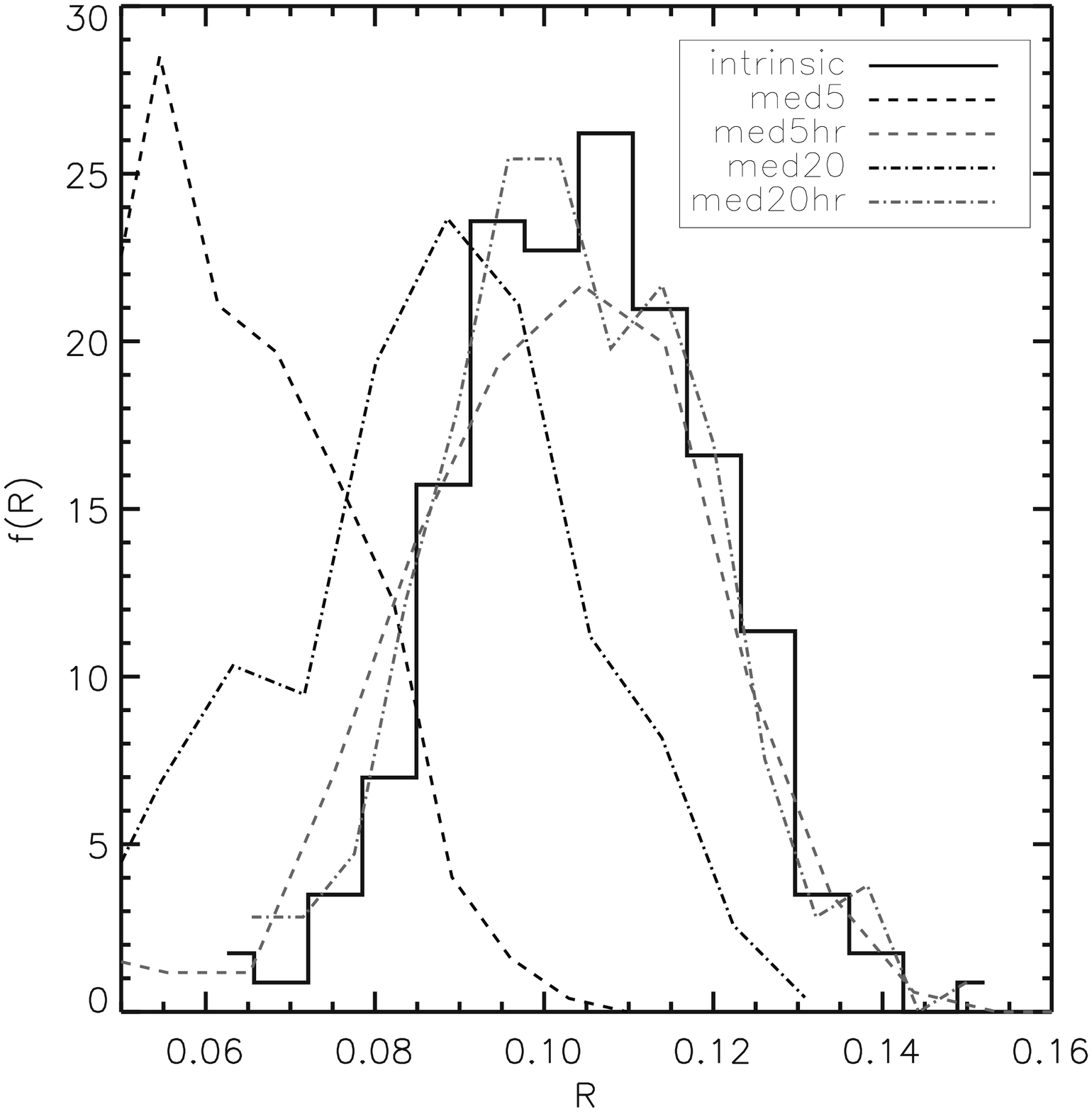}
\end{center}
  \caption{Lefthand: the volume distributions for void segments in
   low-noise models. The histogram shows the intrinsic distribution of
   the Voronoi cell volumes. Superimposed are the inferred volume 
   distribution functions for the WVF segmentations of various 
   Voronoi clustering models. The line style of each of the models 
   is indicated in the insert. Righthand: similar plot for a set  
   of noisy Voronoi clustering models.}
  \label{fig:voiddis}
\end{figure*}

For both the low-noise and high-noise realizations we find that the
intrinsic and prominent edges of the Voronoi pattern remain in
place. Nonetheless, a few shot noise induced artificial divisions survive
the filtering and noise removal operations. They mark prominent
coherent but fully artificial features in the noise. Given their rare
occurrence we accept these oversegmentations as inescapable yet
insignificant contaminations.
\section{Voronoi Clustering Model:\\
\ \ \ \ Quantitative Results Watershed} The watershed segmentation
retrieved by the watershed voidfinder is compared with the intrinsic
(geometric) Voronoi tessellation. The first test assesses the number
of false and correct WVF detections. A second test concerns the volume
distribution of the Voronoi cells and the corresponding Watershed void
segments.
\subsection{Datasets}
For our performance study we have three basic models: the intrinsic
(geometric) Voronoi tessellation, and the low noise and high noise
Voronoi clustering models (table~\ref{tab:vorkinmpar}). The Voronoi
clustering models are processed by WVF. In order to assess the various
steps in the WVF procedure the models are subjected to different
versions of the WVF.

The second column of Table~\ref{tab:void} lists the differently 
WVF processed datasets. These are:
\begin{enumerate}
\item[$\bullet$] {\it Original}: the pure DTFE density field, without 
any smoothing or boundary removal, subjected to the watershed transform.
\item[$\bullet$] {\it Minmax}: only the NN-{\it min/max} filtering 
is applied to the DTFE density field before watershed segmentation. 
\item[$\bullet$] {\it Med$n$}: $n$ iteratations of median natural-neighbour 
filtering is applied to the DTFE density field. In all situations this 
includes max/min filtering afterwards. 
\item[$\bullet$] {\it Hierarch}: following the watershed transform, 
on the pure non-filtered DTFE density, a density threshold is applied. 
The applied hierarchy threshold level is $\rho/{\rho_u}=0.8$: all segment 
boundaries with a density lower than $\delta<-0.2$ are removed as physically 
insignificant.
\item[$\bullet$] {\it Med$n$hr}: mixed process involving an $n$ 
times iterated median filtered DTFE density field, followed by the 
watershed transform, after which the segment boundaries below the 
hierarchy threshold $\delta<-0.2$ are removed. 
\end{enumerate}

Note that the physically natural threshold of $\Delta=-0.8$ is not 
really applicable to the heuristic Voronoi models. On the basis of the 
model specifications the threshold level has been set to $\Delta=-0.2$. 

\subsection{Detection Rate}
Each of the resulting segmentations is subjected to a range of
detection assessments. These are listed in the 3rd to 7th column of
Table~\ref{tab:void}.  The columns of the table contain respectively
the number of WVF void detections, the amount of false splits,
the amount of false mergers, the number of correctly identified
voids, and the correctness measure.  While the top block contains
information on the intrinsic (geometric) Voronoi tessellation, the
subsequent two blocks contain the detection evaluations for the {\it
low noise} and {\it high noise} models.

\begin{figure*}
  \begin{center}
    \includegraphics[width = 0.48\textwidth]{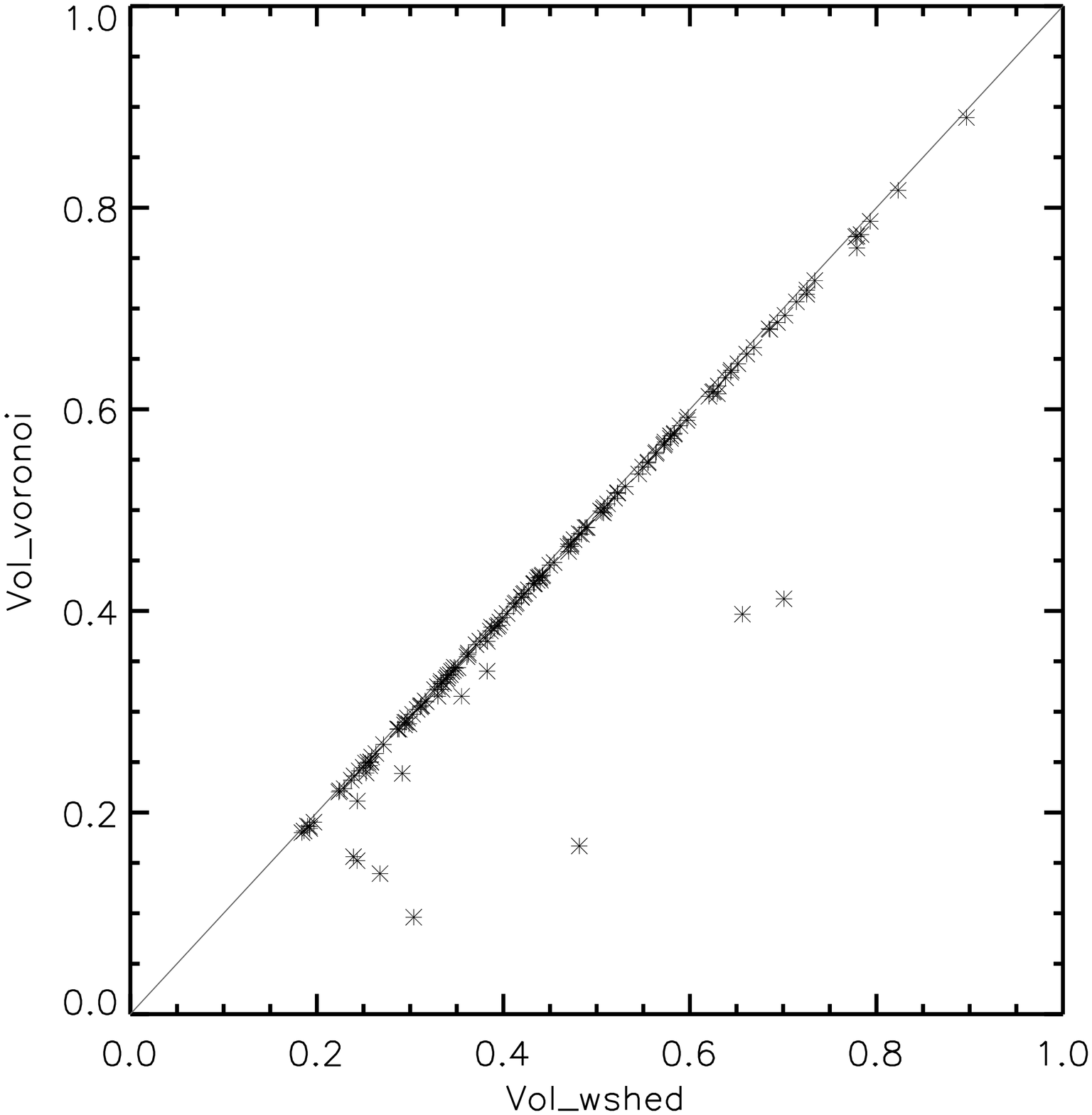}
    \includegraphics[width = 0.48\textwidth]{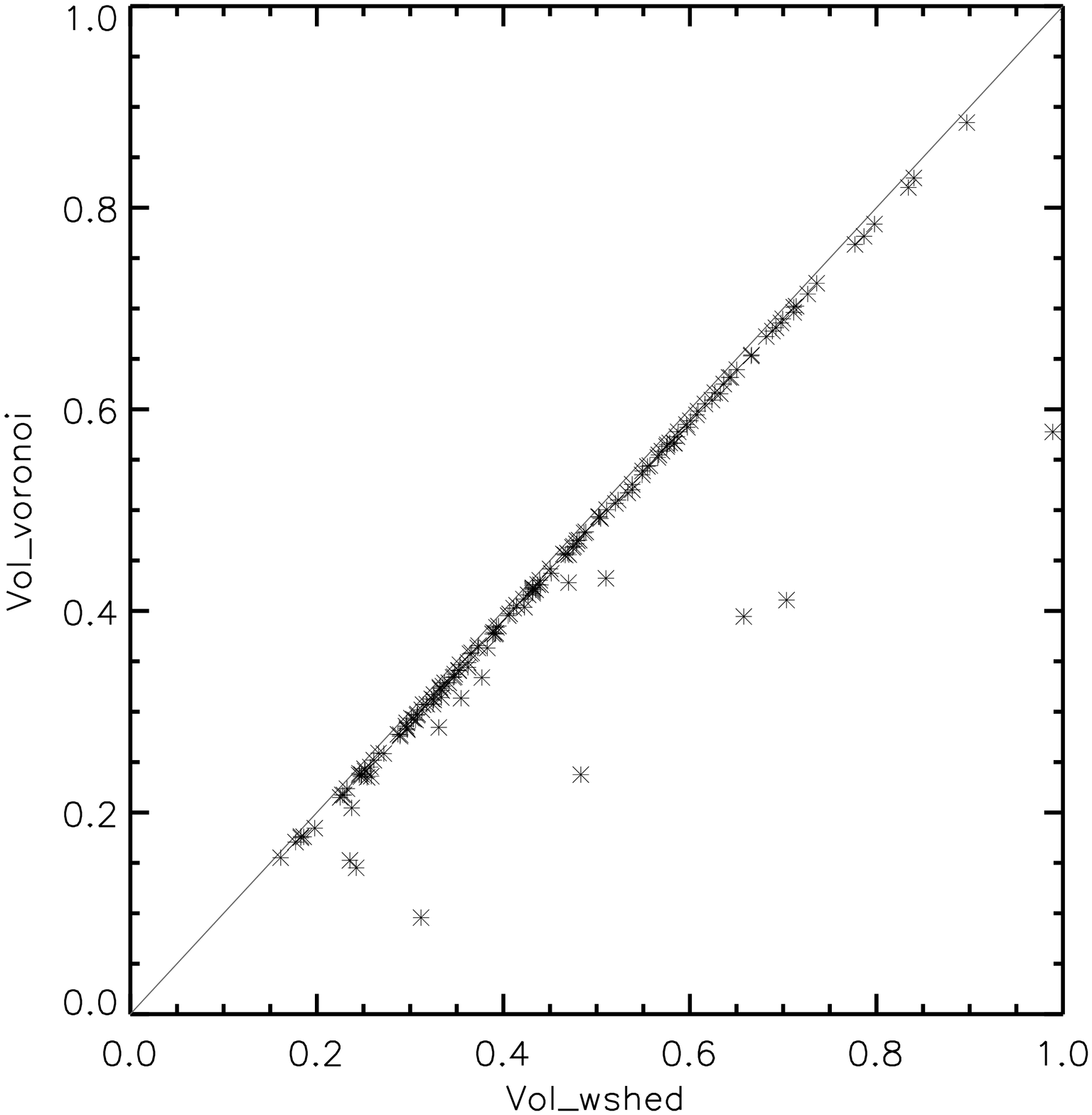}
  \end{center}
  \caption{Scatter diagram plotting the WVF void segment volumes against 
  the intrinsic geometric Voronoi cell volume. The solid line is the 
  linear 1-1 relation. Lefthand: low-noise Voronoi clustering model. 
  Righthand: noisy Voronoi clustering model.}
  \label{fig:voidvol}
  \begin{center}
    \includegraphics[width = 0.48\textwidth]{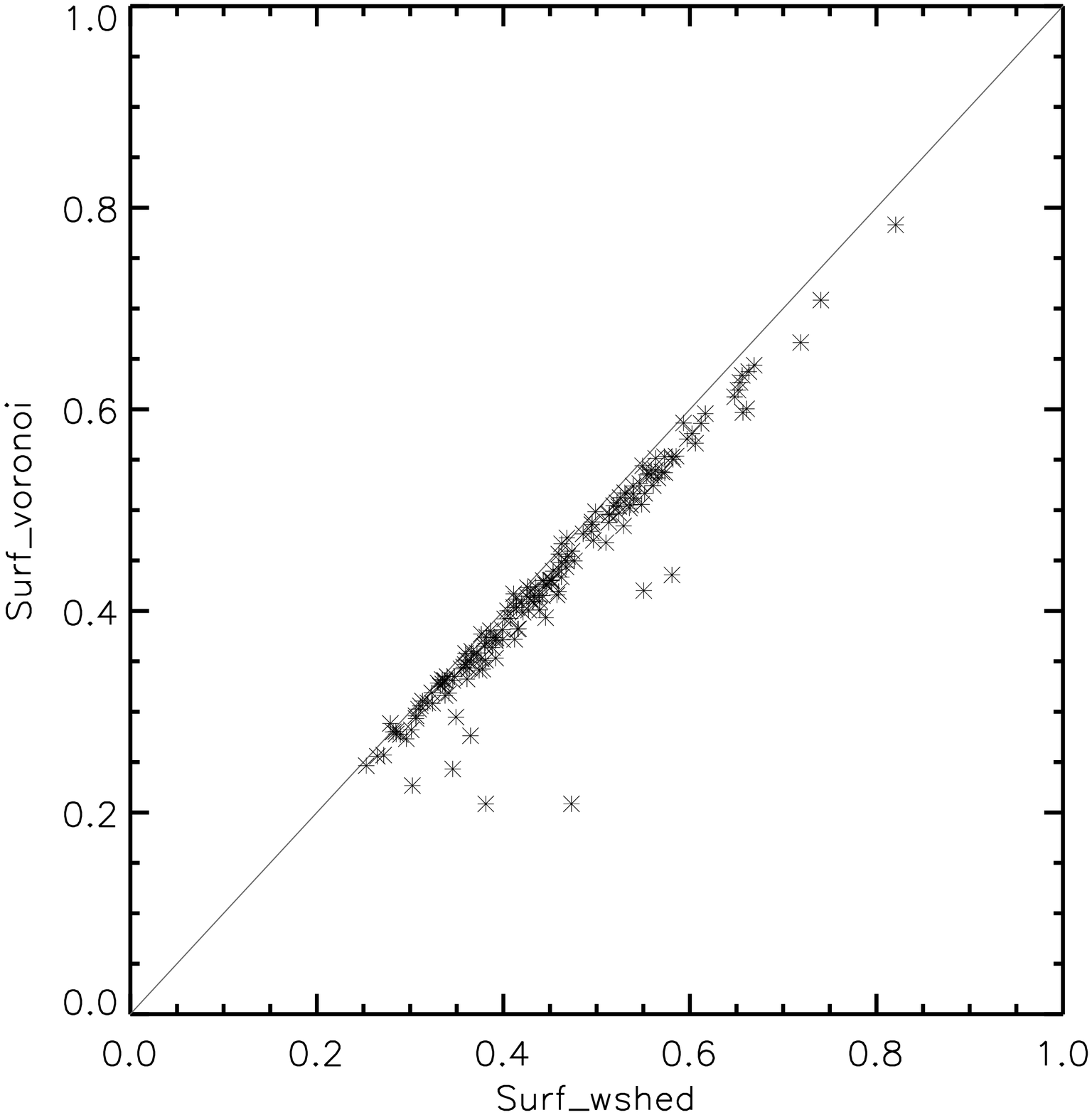}
    \includegraphics[width = 0.48\textwidth]{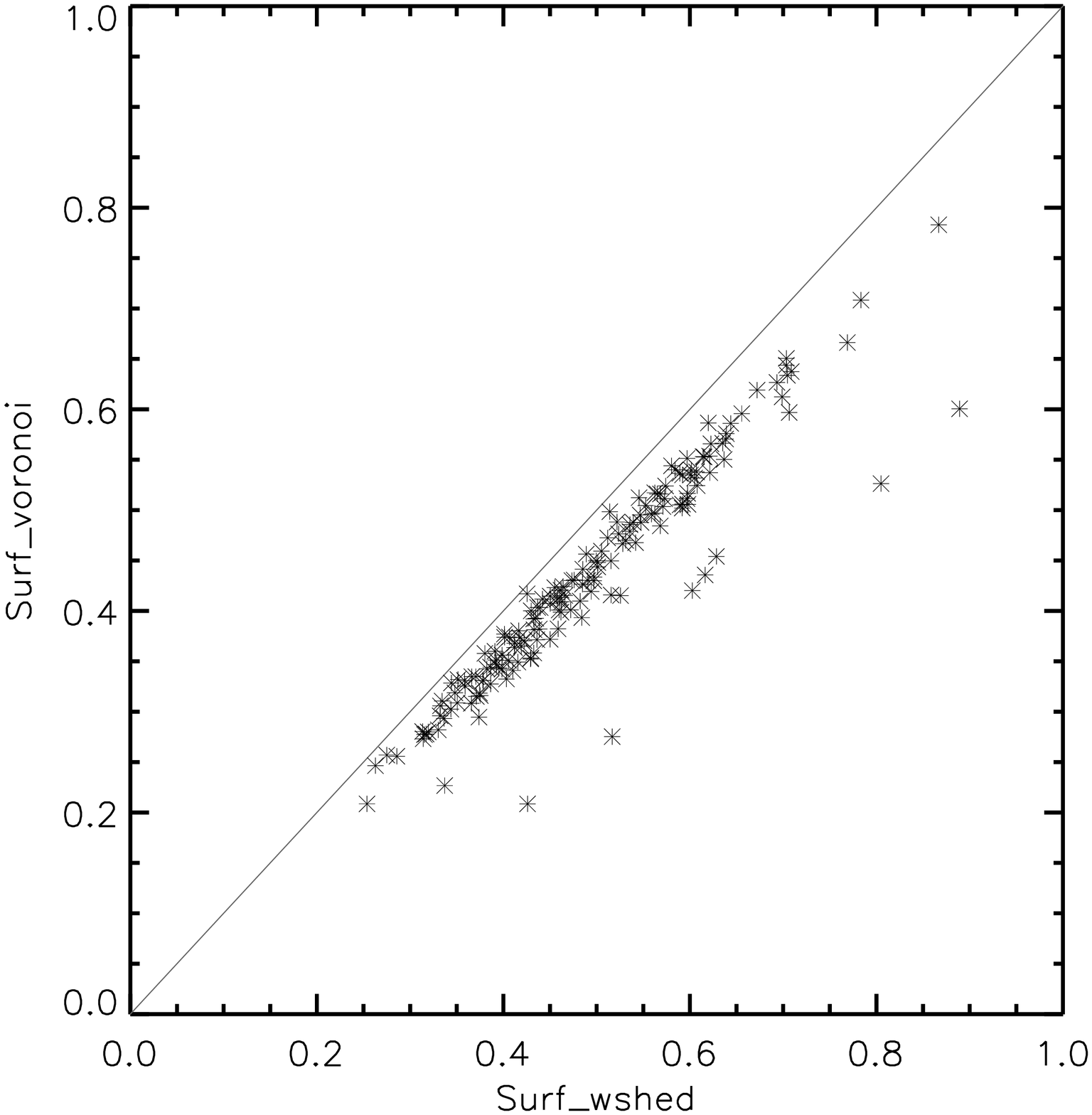}
  \end{center}
  \caption{Scatter diagram plotting the WVF void segment surface area
  against the intrinsic geometric Voronoi cell volume. The solid line
  is the linear 1-1 relation. Lefthand: low-noise Voronoi clustering
  model.  Righthand: noisy Voronoi clustering model.}
  \label{fig:voidsurf}
\end{figure*}

The false detections are split into two cases. The first case we name
{\it false splits}: a break up of a genuine cell into two or more
watershed voids. The second class is that of the {\it false mergers}:
the spurious merging of two Voronoi cells into one watershed void. The
splits, mergers and correct voids are computed by comparing the
overlap between the volume of the Voronoi cell and that of the
retrieved watershed void. A split is identified if the overlap
percentage w.r.t. the Voronoi volume is lower than a threshold of 85
percent of the overlapping volume. Along the same line, a merger
concerns an overlap deficiency with respect to the watershed void
volume. When both measures agree for at least 85 percent a void is
considered to be correct. The correctness of a certain
segmentation is the percentage of correctly identified voids with
respect the 180 intrinsic Voronoi cells.

\subsubsection{Low Noise Model}
Judging by the number of voids in the low noise model, it is clear
that smoothing or any other selection criterion remain necessary to
reduce the number of minima from 850 to a number close to the
intrinsic value 180. The second row shows the results for the case
when just the maxmin filter is applied.  This step already reduces the
number of insignificant minima by already 60 percent.  It is an
indication for the local character of the shot noise component. The
next three rows list the results for various iterations of the median
filtering. With just 2 iterations almost 90 percent of the voids are
retrieved. Most of the splits are removed at 2 iterations. This result
does not improve with more median filtering, even up to 20 iterations
this just increases the number of mergers as more walls are smoothed
away. The number of splits also increases as minima begin to merge.

\subsubsection{High noise model}
In general the same conclusion can be drawn for the high noise
model. Rank-ordered NN-median and NN-min/max filters manage to reduce
the number of insignificant minima by a factor of 80 percent (cf. the
number of voids in the second and third row). These models attain a
correctness of approximately fifty percent. Mere rank-ordered
filtering is evidently insufficient.

We also ran a {\it threshold} model which did not include median
filtering. Instead only insignificant boundaries were removed. It
achieved a recovery of fifty percent. Combining both methods ({\it med5hr}  
and {\it med20hr}) recovers 80 till 90 percent of the voids. The succes rate
may be understood by the complementarity of both methods: while the
median filtering recovers the coherent structures, the thresholding
will remove those coherent walls that are far underdense.

The translation to a cosmological density field is straightforward.
The rank-ordered filtering ensures that insignificant minima are
removed and that the watershed will pick up only coherent
boundaries. Thresholding is able to order these walls by significance
and to remove the very underdense and insignificant walls.

\subsection{Volume Comparison}
In Fig.~\ref{fig:voiddis} we compare the distribution of the void
volumes. The histogram shows the distribution of equivalent radii for
the segment cells,
\begin{equation}
R\,\equiv\,\root 3 \of {\frac{3}{4\pi}\,V}\,.
\end{equation} 
The solid line histogram shows the (geometric) volume distribution for
the intrinsic Voronoi tessellations.  On top of this we show the
distributions for the various (parameterized) watershed segmentation
models listed in Table~\ref{tab:void}.  Not surprisingly the best
segmentations have nearly equivalent volume-distributions. For the
lownoise models this is {\it med2} (lefthand), for the highnoise models
{\it med20hr} (righthand). This conclusion is in line with the detection
rates listed in Table~\ref{tab:void}.

The visual comparison of the intrinsic geometric Voronoi tessellations
and the two best segmentations - {\it med2} for the lownoise model and
{\it med20hr} for the highnoise version -- confirms that also the visual
impression between these watershed renderings and the original Voronoi
model is very much alike.

We have also assessed the cell-by-cell correspondence between the
watershed segmentations and the Voronoi model. Identifying each
watershed segment with its original Voronoi cell we have plotted the
volume of all watershed cells against the corresponding Voronoi
cell volumes. The scatter plots in Fig.~\ref{fig:voidvol} form a
convincing confirmation of the almost perfect one-to-one relation
between the volumes derived by the WVF procedure and the original
volumes.  The only deviations concern a few outliers. These are the
{\it hierarchy merger} segments for which the watershed volumes are
too large, resulting in a displacement to the right.

\subsection{Surface Comparison}
While the volumes occupied by the watershed segments in Fig.~\ref{fig:best} 
do overlap almost perfectly with that of the original Voronoi cells, their  
surfaces have a more noisy and erratic appearance. This is mostly 
a consequence of the shot noise in the (DTFE) density field, induced 
by the noise in the underlying point process. The crests in the density 
field are highly sensitive to any noise, 

In addition to assess the impact of the noise on the surfaces of the 
watershed segments we compared the watershed segement surface 
areas with the Voronoi cell surface areas. The results are 
shown in Fig.~ \ref{fig:voidsurf}. We tested the lownoise {\it med2} 
and the highnoise {\it med20hr}. In both cases we find a linear 
relationship between the watershed surface and the genuine Voronoi 
surface area. Both cases involve considerably more scatter than 
that for the volumes of the cells. In addition to an increased level 
of scatter we also find a small be it significant offset from the 
1-1 relation. The slope of the lownoise model is only slightly less 
than unity, the highnoise model slope deviates considerably more. 
These offsets do reflect the systematically larger surface areas 
of the watershed segments, a manifestation of their irregular 
surfaces. It is evident that the level of irregularity is more 
substantial for the highnoise than for the lownoise reconstructions 
(cf. Fig.~\ref{fig:best}). 

The scatter plots do also reveal several cells with huge deviations 
in surface area. Unlike expected there is no systematic trend for 
smaller cells to show larger deviations. Some of the small deviating 
cells can be recognized in Fig.~\ref{fig:best} as highly irregular 
patches. The large deviant cells correspond to watershed segments which 
as a result of noisy boundaries got wrongly merged. 

While the irregularity of the surface areas forms a good illustration of 
the noise characteristics of the watershed patches, for the purpose of void 
identification it does not pose a serious problem. Smoother contours 
may always be obtained by applying the flooding process on a 
properly smoothed field. Some suggestions for this may be 
achieved follows in the discussion.

\section{Discussion and Prospects}
The WVF void finder technique is based on the watershed transform
known from the field of image processing. Voids are identified with
the basins of the cosmological mass distribution, the filaments and
walls of the cosmic web with the ridges separating the voids from each
other. Stemming from the field of mathematical morphology, watershed
cells are considered as patches locally minimizing the ``topographic
distance''.

The WVF operates on a continuous density field. Because the
cosmological matter distribution is nearly always sampled by a
discrete distribution of galaxies or N-body particles, the first step
of the WVF is to translate this into a density field by means of the
Delaunay Tessellation Field Estimator (DTFE). Because the WVF involves
an intrinsically morphological and topological operation, the
capability of DTFE to retain the shape and morphology of the cosmic
web is essential. It guarantees that within this cosmological
application the intrinsic property of the watershed transform to act
independent of scale, shape, and structure of a segment is retained.
As a result, voids of any scale, shape and structure may be detected
by WVF.

In addition to the regular watershed transform the WVF needs to invoke
various operations to suppress (discreteness) sampling noise. In
addition, we extend the watershed formalism such that the WVF will be
capable of analyzing the hierarchy of voids in a matter distribution,
i.e. identify how and which small scale voids are embedded within a
void on larger scales \citep{PlaWey07}.  Markers indicating significant void
centers,  false segment removal by Hierarchy Merging and 
Natural Neighbour filtering all affect an efficient noise removal.
Hierarchy Merging manages to eliminate boundaries between subvoids.
Natural Neighbour median filtering, for various orders, is an essential 
new ingredient for highlighting the hierarchical embedding of the 
void population. It allows a natural selection of voids, unbiased with 
respect to the scale and shape of these structures. The voids that 
persist over a range of scales are expected to relate to the voids 
that presently dominate the cosmic matter distribution. In other 
words, WVF preserves the void hierarchy \cite{SheWey}. 

The present work includes a meticulous qualitative and quantitative
assessment of the watershed transform on the basis of a set of Voronoi 
kinematic models. These heuristic models of spatial weblike
or cellular galaxy or particle distributions facilitate the comparison
between the void detection of the WVF and that of the characteristics
of the cells in the original and intrinsically known Voronoi
tessellation. It is found that WVF is not only succesfull in
reproducing the qualititative cellular appearance of the Voronoi
models but also in reproducing quantitative aspects like an almost
perfect 1-1 match of cell size with WVF segment volume and the
corresponding void size distribution.

We foresee various possible improvements of the WVF. These concern in
particular the identification of significant edges. One possibility is
that extension proposed by \cite{NgyBo}, in which not only the
``topographic costs'' but also the lengths of the contours should be
minimized. The length minimization will result in smoother
boundaries. Additional improvements may be found in better filtering
procedures in order to facilitate studies of hierarchically structured
patterns. We expect considerable improvements by anisotropic diffusion
techniques \citep{BlaMar} and are currently implementing these in the
WVF code. 

Given the results of our study, we are confident for applying WVF to
more elaborate and realistic simulations of cosmic structure formation
and on large galaxy redshift surveys. The analysis of a set of GIF
cosmological simulations will be presented in an upcoming paper.
\section*{Acknowledgements}
We wish to thank Miguel Arag\'on-Calvo for permission to use 
Fig.~\ref{fig:vorkinmschm}. We are grateful to the referee, Mark Neyrinck, 
for the incisive, detailed and useful comments and recommendations 
for improvements. We particularly thank the participants of the KNAW 
colloquium ``Cosmic Voids'' in Amsterdam, Dec. 2006, for the many 
useful discussions and encouraging remarks.

\bigskip

\appendix
\section{\ \\ Mathematical Morphology}
\label{app:mathmorph}
\begin{figure*}
  \includegraphics[width=0.32\textwidth]{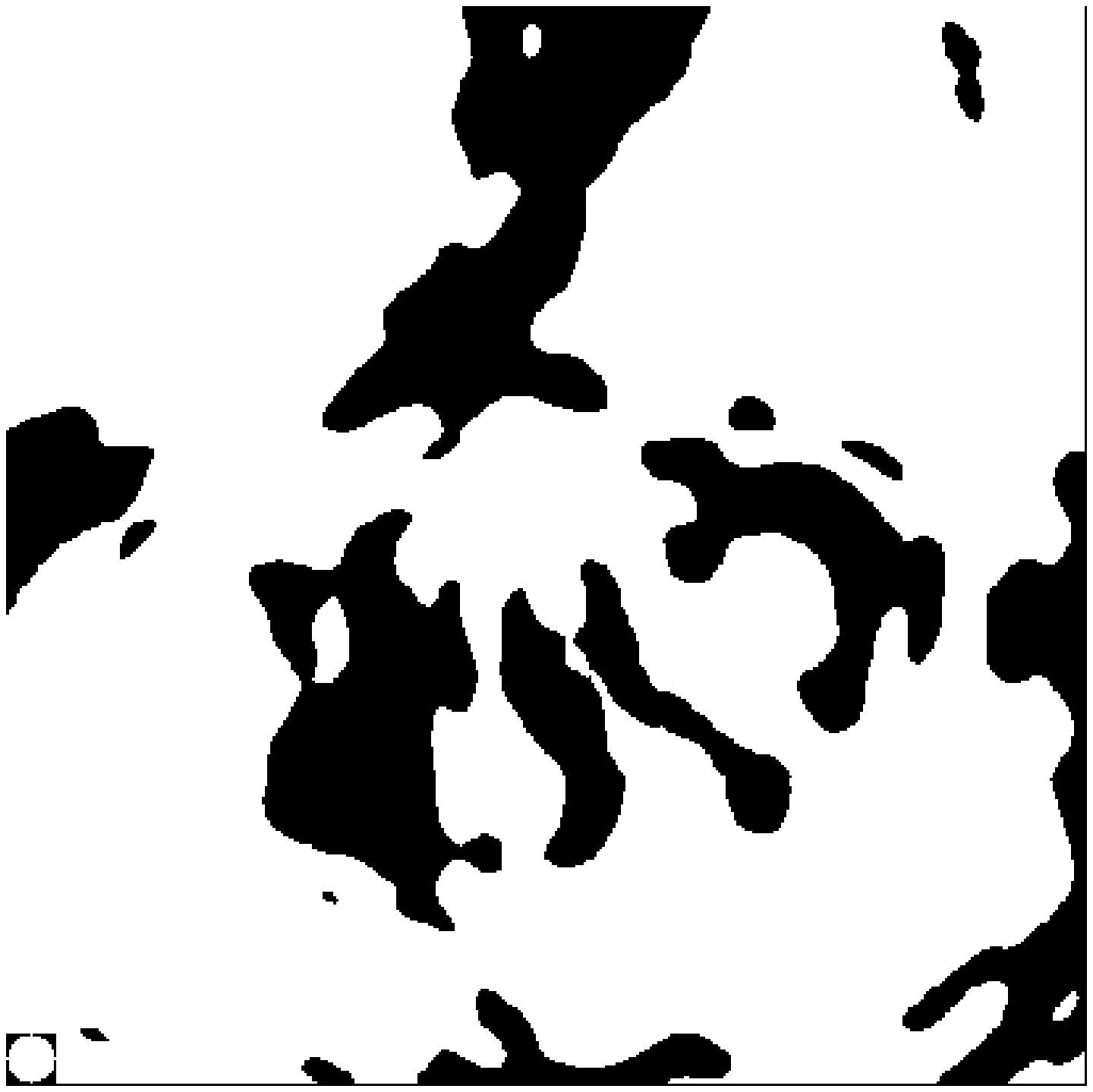}
  \includegraphics[width=0.32\textwidth]{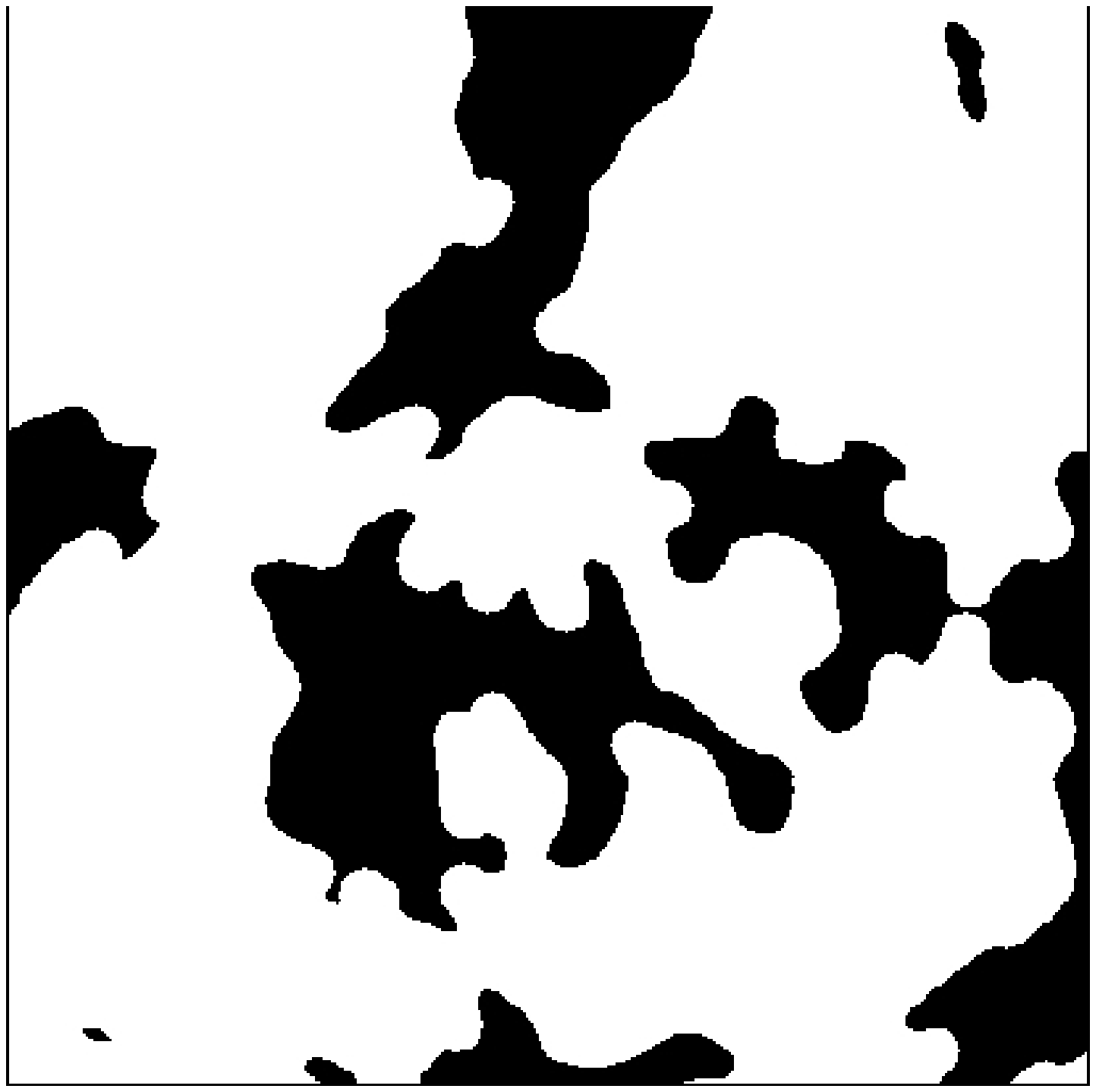}
  \includegraphics[width=0.32\textwidth]{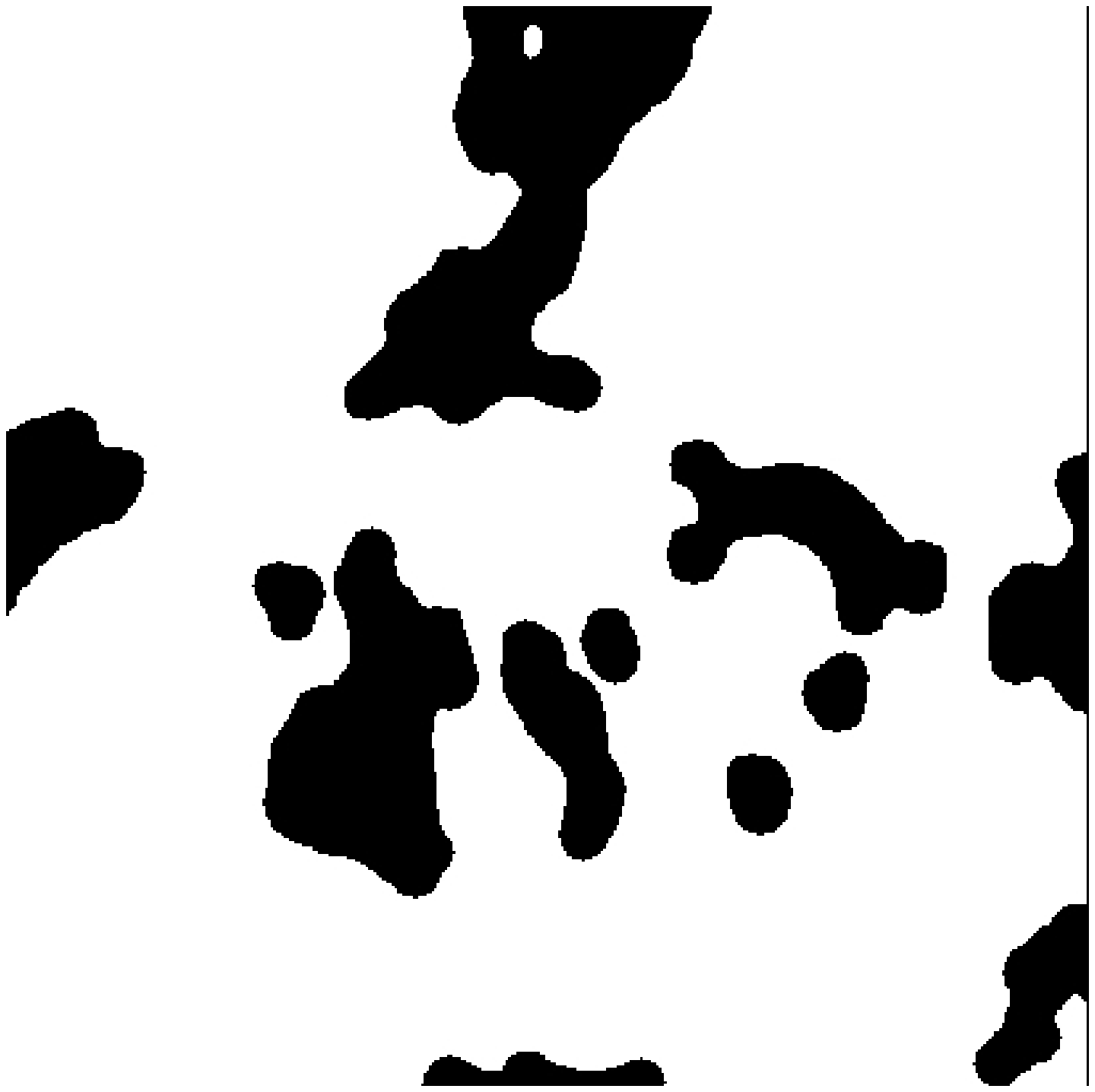}
  \vskip 0.3cm
  \includegraphics[width=0.32\textwidth]{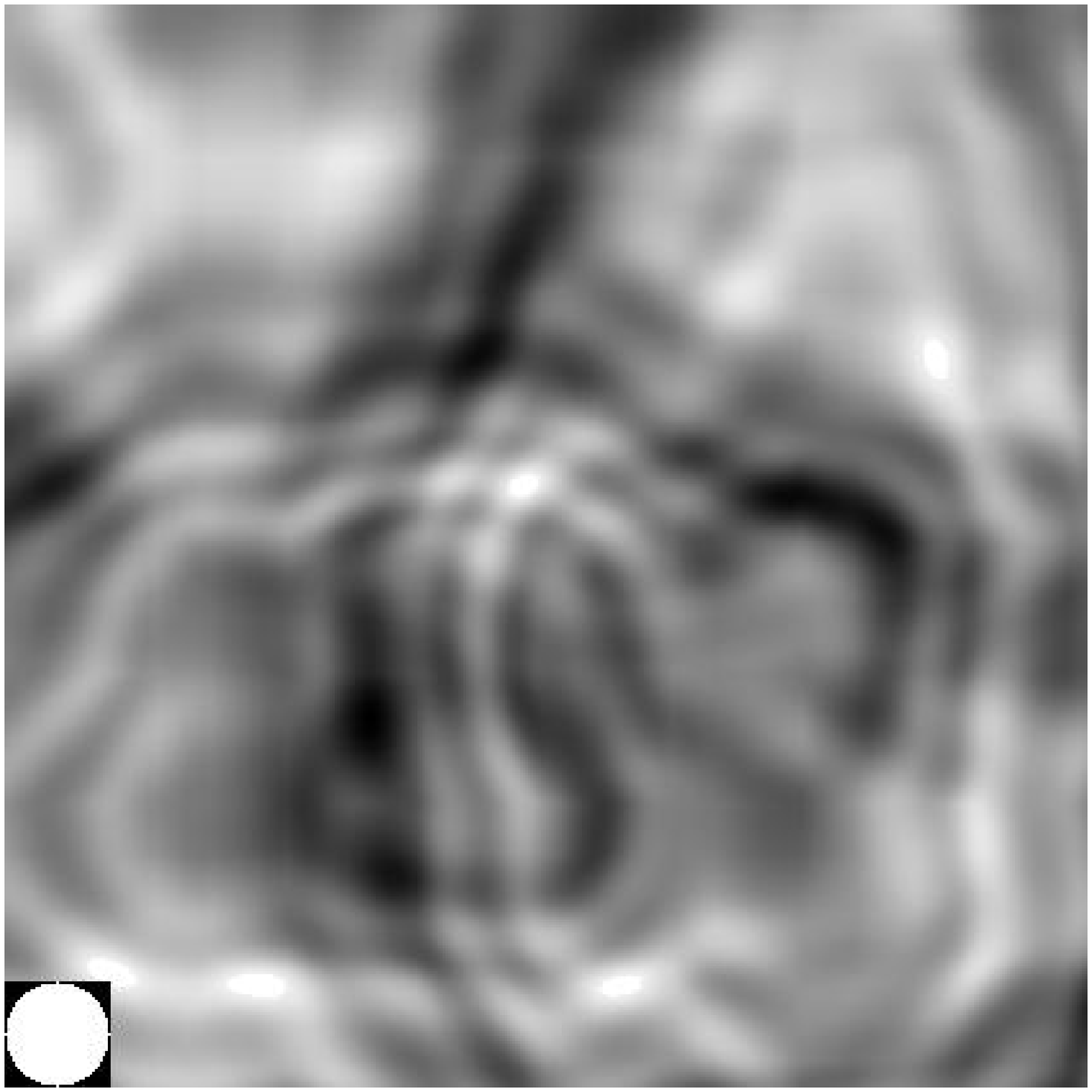}
  \includegraphics[width=0.32\textwidth]{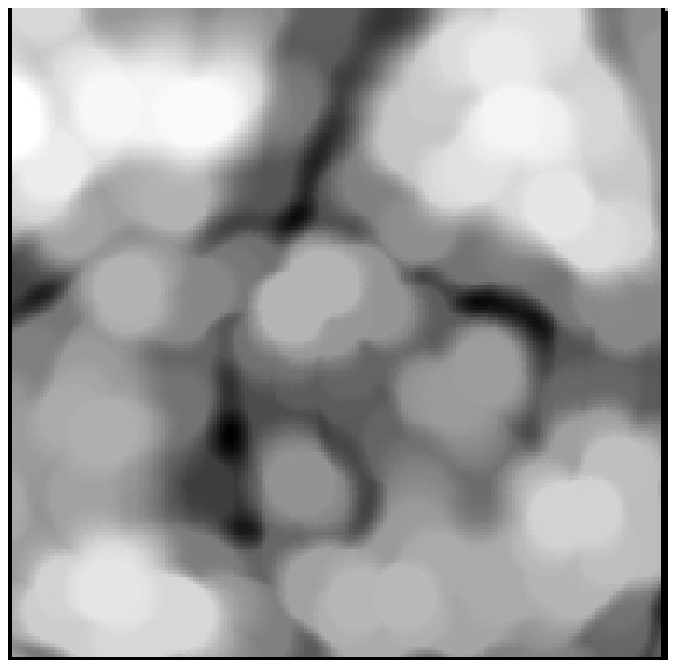}
  \includegraphics[width=0.32\textwidth]{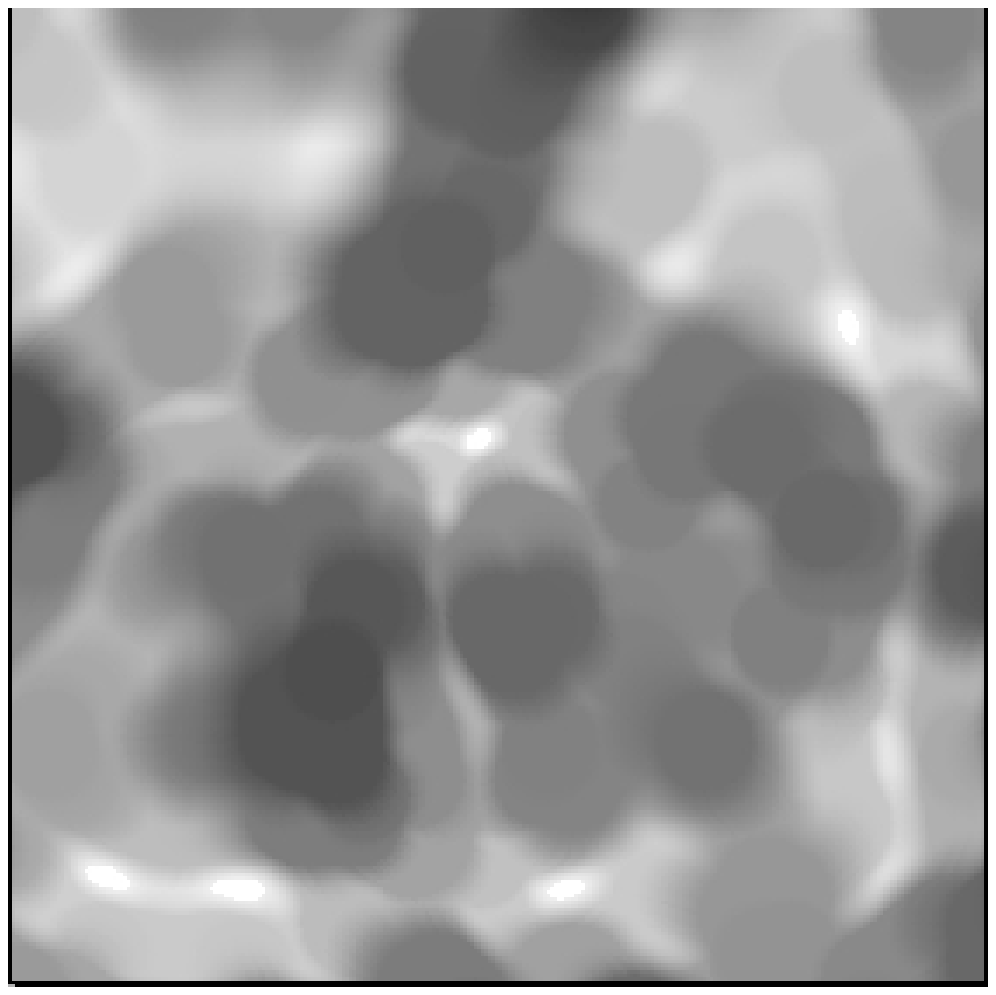}
  \caption{Illustration of the effects of a few essential operators on a binary image (top column) and 
a grayscale image (bottom row). The original image is the one in the lefthand frame. The central frame 
contains the image following a opening operation, while the righthand frame shows the effect of 
closing. The circle at the lower lefthand corner of the originals represent the circular structure 
element.}
  \label{fig:mm_oper}
\end{figure*}

Appendix~\ref{app:mathmorph} and ~\ref{app:wshedimpl} provide some
formal concepts and notations necessary for appreciating the 
watershed transform.

The WVF formalism introduced in this study is largely based upon
concepts and techniques stemming from the field of image
analysis. Although they are used within the context of the
morphological analysis of spatial patterns in cosmological density
fields the presentation is in terms of the original image analysis
nomenclature. In this we remind the cosmology reader to translate 
``image'' into ``density field'', ``basin'' into ``void interior'',
etc.

Mathematical Morphology (MM) is the field of image analysis which aims
at characterizing an image on the basis of its geometrical
structure. It provides techniques for extracting image components
which are useful for representation and description and was originally
developed by G. Matheron and J. Serra. For more details we refer to
\cite{MathSer}, also see \cite{Heij,Math,MeyBeu}. It involves a
set-theoretic method of image analysis and incorporates concepts from
algebra (set theory, lattice theory) as well as from geometry
(translation, distance, convexity).  Applications of mathematical
morphology may be found in a large variety of scientific disciplines,
including material science, medical imaging and pattern recognition.

\subsection{Images}
A cosmological density field $f({\bf x})$ may be mapped onto an image ${\mathcal F}$,  
\begin{equation}
f({\bf x})\quad\rightarrow\quad {\mathcal F}({\bf x})\,.
\end{equation}
The image ${\mathcal F}$ is a function in $n$-dimensional lattice space 
(usually $n=3$ or $n=2$) $\mathbb{Z}^n$, 
\begin{equation}
{\mathcal F} :\mathbb{Z}^n \rightarrow \mathbb{Z}\,.
\end{equation}
Although in principle images may be continuous, in practice they usually attain a finite number 
of discrete values. Two important classes are:
\begin{itemize}
\item[$\bullet$] {\it Binary image}:\ \ \ \\
\ \ \ \ image with only 2 intensity values (Fig.~\ref{fig:mm_oper}, top lefthand), 
\begin{equation}
{\mathcal F}\,=\,\begin{cases}
0\ \\
1\end{cases}
\end{equation}
We follow the convention to identify the binary image by the set $X \subseteq \mathbb{Z}^n$ for 
which ${\mathcal F}=1$.
\item[$\bullet$] {\it Grayscale image}:\ \ \ \\
\ \ \ \ image with a discrete number $N$ of values (Fig.~\ref{fig:mm_oper}, bottom lefthand), 
\begin{equation}
{\mathcal F}\,=\,\left\{{\mathcal F}_i,\ \ \ i=1,\ldots,N\right\}
\end{equation}
\end{itemize}
Mathematical morphology was originally developed for binary images, and was later extended 
to grayscale images.

\subsection{Erosion and Dilation}
The two basic operators of Mathematical morphology are {\it Erosion} and
{\it Dilation} of a binary image $X$. In order to define these 
operators we need to invoke the {\it translation} and {\it reflection} of a set $B$, 
\begin{equation}
\begin{array}{lllll}
\ \\
\textrm{translation}:&{B}_z &\,\equiv\,&\{y\,|\,y=b+z\ \ \ \forall\ \ \ b \in B\}\\
\textrm{reflection}:&{\hat B} &\,\equiv\,&\{y\,|\,y=-b\ \ \ \ \ \ \forall\ \ \ b \in B\}\\
\ \\
\end{array}
\label{eqn:bintrref}
\end{equation}
The dilation or erosion of a binary image $X$ by a structuring element $B$ identifies 
whether the translated set $B_z$ has an overlap with or is contained in a certain part of $X$, 
\begin{equation}
\begin{array}{llllll}
\ \\
\textrm{dilation}:&\quad X \oplus B &\,\equiv\,& \{z\,|\,{\hat B}_z\,\cap\,X\,\neq\,\emptyset\}\\
\textrm{erosion}:&\quad X \ominus B &\,\equiv\,& \{z\,|\,B_z\,\subseteq\,X\}\\
\ \\
\end{array}
\end{equation}
In other words, dilation consists of the Minkowski addition ($\oplus$) of the binary image $X$ with a 
structuring element $B$ while erosion is the Minkowski substraction ($\ominus$) with $B$. 
A structuring element may be any object in $\mathbb{Z}^n$. An example is the circle which functioned 
as a structuring element in Fig.~\ref{fig:mm_oper}. 
\bigskip
\noindent Erosion and dilation have a number of properties:
\begin{itemize}
\item[$\bullet$] Translation invariance
\item[$\bullet$] Global scaling invariance
\item[$\bullet$] Addition:
\begin{equation}
\hbox{\rm if}\quad X \subseteq Y\quad \rightarrow \quad X \oplus A \subseteq Y \oplus A
\end{equation}
\item[$\bullet$] Complementarity:
\item[] erosion of a set is dilation of complement
\item[] (and vice versa)
\item[$\bullet$] Adjunction relationship:
\begin{equation} 
Y \oplus A \subseteq X \iff
Y \subseteq X \ominus A.
\end{equation}
\end{itemize}
The complementarity and adjunction relationship are two aspects of the
existing duality between erosion and dilation.  In general
erosion and dilation induce a loss of information. Erosion followed by
dilation, or vice versa, will only result in a restoration of the
original image if the sets $X$, $Y$ and $A$ are convex. In fact,
various combinations of the erosion and dilation operators do result
in new operators.

\subsection{Opening and Closing}
The most straightforward combination of dilation and erosion is that
of the consecutive application of an erosion and a dilation. This
introduces two new operators, the {\it Opening} and {\it Closing}
operators. On a binary image they have the effects shown in
Fig.~\ref{fig:mm_oper} (top centre, top righthand). Opening amends
caps, removes small islands and opens isthmuses. Closing, on the other
hand, closes channels, fills small lakes and (partly) the gulfs. An
additional combination of erosion and dilation is subtraction of the
first from the latter, called a {\it morphological gradient}.

Formally, an opening is an erosion followed 
by a dilation, a closing a dilation followed by erosion. 
\begin{equation}
\begin{array}{lllllllll}
\ \\
\textrm{Opening:}&&\Lambda_{B} &\,\equiv\,&[(X \ominus B)\oplus B ] \\
\textrm{Closing:}&& \Phi_{B} &\,\equiv\,&[(X \oplus B)\ominus B ]\\
\ \\
\end{array}
\end{equation}
Characteristics of the opening and closing operators are:
\begin{itemize}
\item[$\bullet$] Increasing
\item[$\bullet$] Idempotent:
\item[] applying the operator twice yields the same output
\item[$\bullet$] Opening is anti-extensive:
\begin{equation}
\Lambda(X) \subseteq X
\end{equation}
\item[$\bullet$] Closing is extensive:
\begin{equation}
X \subseteq \Phi(X)
\end{equation}
\end{itemize}
Note that the extensivity and/or anti-extensivity of operators define the prime conditions 
for a {\it morphological operator}. 

\subsection{Grayscale Images}
\label{app:grayscale}
The morphological operators which we discussed above can be
generalized to grayscale images.

\medskip
A grayscale image is composed of subsets $S_i({\mathcal F})$, 
\begin{equation}
S_i({\mathcal F})\,=\,\{x\,|\,x \in \mathbb{Z}^n : {\mathcal F}(x)\geq i\}\,. 
\end{equation}
with $S_{i+1}({\mathcal F}) \subseteq S_{i}({\mathcal F})$ and $S_1({\mathcal F})$ 
the support of the full image. The erosion and dilation of a 
grayscale image involves their application to each individual subset $S_i({\mathcal F})$. Extension 
of the binary image definitions (eqn.~\ref{eqn:bintrref}) implies the following definition 
wrt. a grayscale image, 
\begin{equation}
\begin{array}{lll}
\ \\
{\mathcal F}\oplus B  &\,\equiv\,& {\rm sup}\{\,{\mathcal F}(x+b),\ \ \ x \in X,\ b \in {\hat B}\}\\
{\mathcal F}\ominus B &\,\equiv\,& {\rm inf}\ \{\,{\mathcal F}(x+b),\ \ \ x \in X,\ b \in {B}\}\\
\ \\
\end{array}
\end{equation}
The effect of erosion on a grayscale image is the shrinking of the
bright regions.  Bright spots smaller than the structuring element $B$
disappear completely while valleys (dark) expand. Dilation has the
opposite effect: dark regions shrink while bright regions expand. It
illustrates the duality between erosion and dilation.

For our purposes, this formal definition translates into the following practical 
implementation for 2-D grayscale images. Given a grayscale image ${\mathcal F}(A)$ with 
grid elements $a(i,j)$,
\begin{equation}
\begin{array}{lllr}
\ \\
{\mathcal F}\oplus B  &\,=\,&{\max_{(i,j) \in B}}\ \{a[r+i,s+j]\,+\,{\hat b}[i,j]\}\\
&&&\forall [r,s] \in A\\
{\mathcal F}\ominus B &\,=\,&{\min_{(i,j) \in B}}\ \{a[r+i,s+j]\,+\,b[i,j]\}\\
&&&\forall [r,s] \in A\\
\ \\
\end{array}
\end{equation}

As in the case of binary images new operators may be defined through
combinations of erosions and dilations. The closing and 
opening operators are defined in exactly the same way as that for the
binary images. Their effect is shown in the lower column of
Fig.~\ref{fig:mm_oper}. The morphological gradient 
\begin{equation}
{\cal G}\,\equiv\,({\mathcal F}\oplus B)\,\ominus\,({\mathcal F}\ominus B)
\end{equation}
is a dilation minus erosion operation. The gradient operator is often
used in object detection because an object is usually associated with
a change in grayscale with respect to the background. A variety of
additional operators involving openings and closings may be
defined. Interesting ones are the {\it granulometries}, a sequence of
erosions with increasing scale, and {\it distance transforms}.

\section{\ \\ Watershed Transform}
\label{app:wshedimpl}
The segmentation of images is defined on the basis of a distance 
criterion, referring to the concept of {\it distance} between 
subsections of an image. 

\subsection{Distance}
For appreciating the concept of watershed segmentation we consider two
distance concepts, the {\it geodesic} and {\it topographic} distance. 
Geodesic distances are used in the case of binary images while 
topographic distances form the basis for the segmentation of 
grayscale images.

Let $X\subset \mathbb{Z}^n$ be a set and x and y two points in an
$n$-dimensional lattice space $\mathbb{Z}^n$. We may define

\subsubsection{Geodesic Distance}
The geodesic distance $d_{X}(x,y)$ is the length of the shortest (geometric) 
path in $X$ connecting $x$ and $y$ (see lefthand frame, Fig.~\ref{fig:skiz}). 
Accordingly, the distance between two subsets $A$ and $B$ in $X$ 
may be defined as follows. Considering the set of all paths between 
any of the elements of $A$ and those of $B$, the distance 
$d_{X}(A,B)$ between $A$ and $B$ is defined to be the minimum length 
of any of these paths. 

Based upon the concept of $d_{X}(A,B)$ one may formulate a distance function 
of a set $Y \in \mathbb{Z}^n$. For each point $y
\in Y$, the distance $d(y,{\bar Y})$ to its complement ${\bar Y}$ is
computed. The distance function ${\mathcal D}(y)$ is the resulting
map of distance values $d_{X}(y,{\bar Y})$. Regions whose distance
$d_{X}(y,{\bar Y})$ is at least $r_i$ can be identified and equated by
erosion of $Y$ with a disk of radius $r_i$ (defining a set ${\mathcal
R}_i$). Each of these regions forms a section ${\mathcal S}_{{\mathcal
D}, i}$ (see sec.~\ref{app:grayscale}),
\begin{equation}
{\mathcal S}_{{\mathcal D}, i}\,=\,{\mathcal D}(y)\oplus {\mathcal R}_i\,,
\end{equation}
in which ${\mathcal R}_i$ is the disk of radius $r_i$. The map ${\cal D}(y)$ may 
be regarded as a stack of these sections. For illustration the distance transform 
of the the binary image Fig.~\ref{fig:mm_oper} is depicted in the central 
frame of Fig~\ref{fig:skiz}. 

\subsubsection{Topographic Distance}
The topographic distance ${\mathcal T}(x,y)$ between two points
in $\mathbb{Z}^n$ is defined with respect to the image map ${\mathcal
F}$.  Taking the limit of a continuous map ${\mathcal F}$, the
topographic distance from $x$ to $y$ is the path which attains the
minimum length through the ``image landscape'',
\begin{equation}
{\mathcal T}(x,y)\,\equiv\,{\rm inf} \int_{\Gamma} |\nabla {\mathcal F}(\gamma(s))| ds\,.
\label{eq:topdist}
\end{equation} 
In this definition, the integral denotes the {\it image pathlength}
${\mathcal F}(\gamma)$ along all paths $\gamma(s)$ in the set of all
possible paths, $\Gamma$. This concept of distance is related to the
geodesics of the surface ${\mathcal F}$: the path of steepest descent,
specifing the track a droplet of water would follow as it flows down a
mountain surface.

\subsection{Segmentation}
\label{app:segment}
Based on the specific definition of distance, we may segment a
binary image ${\mathcal F}$ through the identification of the 
zones of influence of well-defined subsets of ${\mathcal F}$. In
general the binary image ${\mathcal F}$ contains $n$ connected subsets
$Y_i$ ($i=1,\ldots,n$) (the black regions of Fig.~\ref{fig:mm_oper})
and a set X (white region) which contains all points $x$ which do not
belong to any of the subsets $Y_i$, i.e.
\begin{equation}
X\,=\,\left\{x \in {\cal F}|x\,\notin\,\bigcup_l Y_l\right\}\,.
\end{equation}

\subsubsection{Zone of Influence}
The geodesic zone of influence ${Z}_{\mathcal{F}}(Y_i)$ of a
subset $Y_i \in \mathcal{F}$ is the set of all the points $x \in X$
that are stricktly closer to $Y_i$ than to any other subset $Y_j,\ j \neq i$,
\begin{equation}
{Z}_{\mathcal{F}}(Y_{i})\,\equiv\,\left\{x\in
X|\,d_{X}(x,Y_{i}) < d_{X}(x,Y_{j}),\ \forall j \neq i\,\right\}
\end{equation}

The {\it Zone of Influence} ${\mathcal Z}$ of $\mathcal{F}$ is the union of
all influence zones of ${Z}_{\mathcal{F}}(Y_{i})$,
\begin{equation}
{\mathcal Z}_{\mathcal{F}}\,\equiv\,\bigcup_n\,{Z}_{\mathcal{F}}(Y_{i})
\end{equation}

\subsubsection{Skeleton}
\begin{figure*}
  \includegraphics[width=0.32\textwidth]{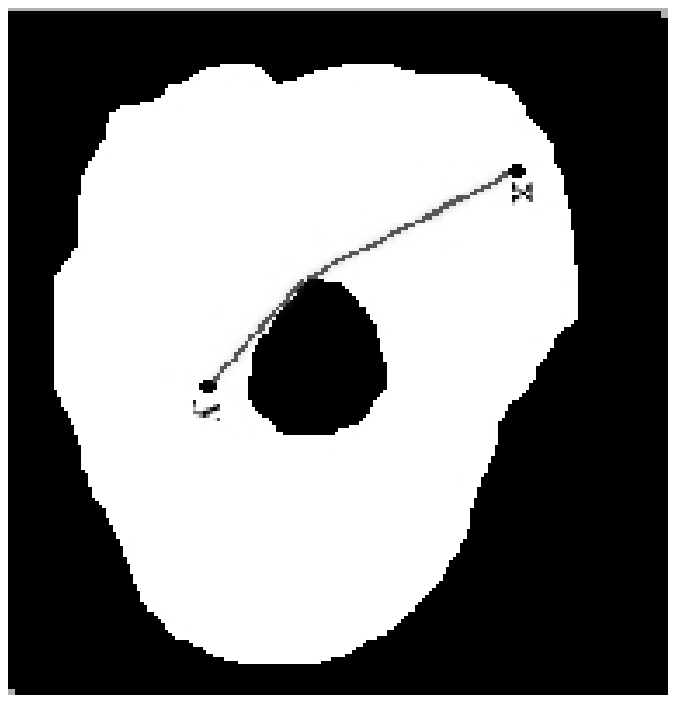}
  \includegraphics[width=0.32\textwidth]{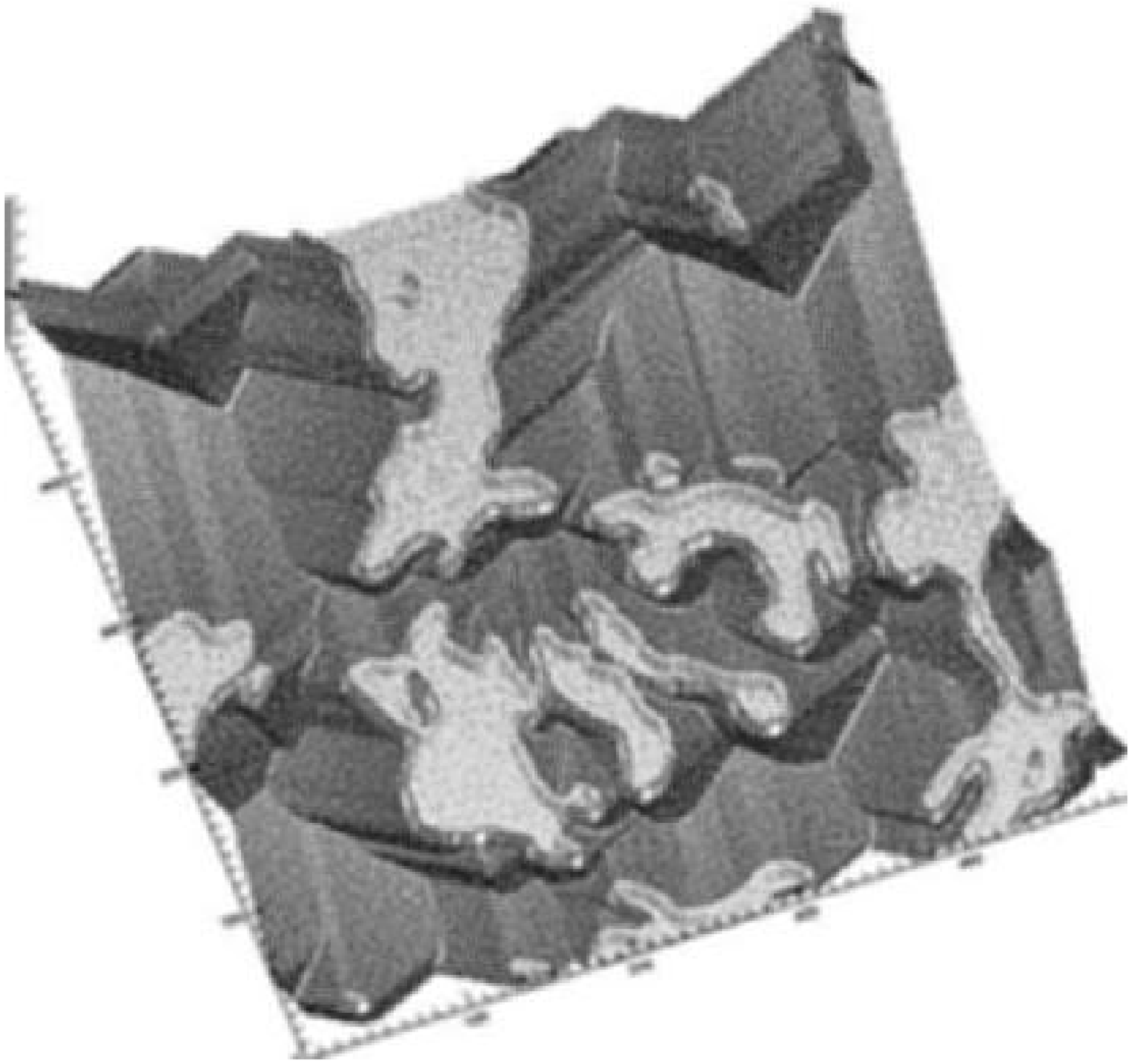}
  \includegraphics[width=0.32\textwidth]{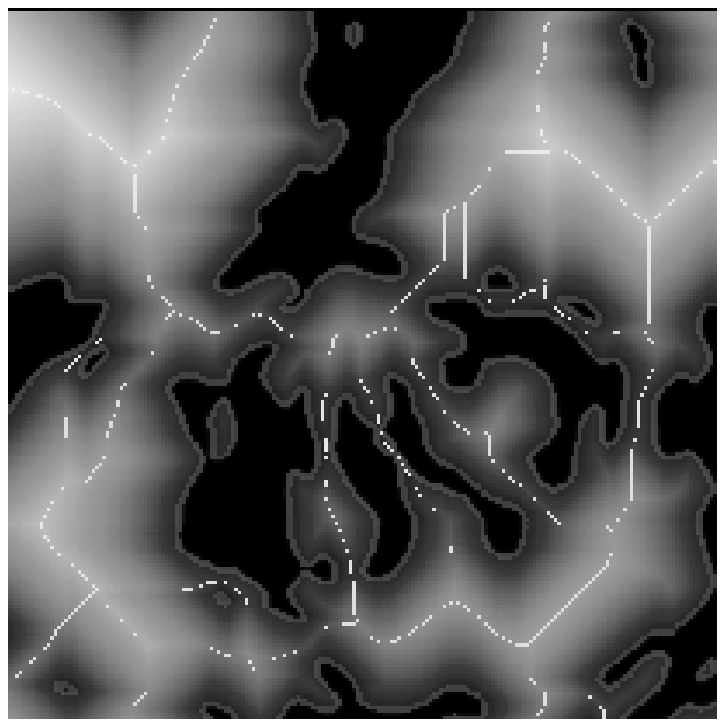}
  \caption{These three images illustrate the concept of {\it distance} and 
{\it segmentation}. Left: the {\it geodesic distance} between points $x$ and $y$ 
is depicted within a binary image. Centre/Right: a landscape defined by a 
distance function (centre) and the resulting segmentation and skeleton 
(white line, righthand frame).}
  \label{fig:skiz}
\end{figure*}

The boundary set in $X$ consists of those points which do belong to
$X$ yet are not contained in any of the zones of influence
${Z}_{\mathcal{F}}(Y_i)$. These boundary points define the geodesic
{\it Skeleton} ${\mathcal K}$ of $\mathcal{F}$,
\begin{equation}
{\mathcal K}_{\mathcal{F}}\,\equiv\,X \backslash {\mathcal Z}_{\mathcal{F}}\,.
\end{equation}
In Fig.~\ref{fig:skiz} (righthand frame) the skeleton in set $X$ is
outlined by white lines. The skeleton is superimposed on its defining
distance function landscape, its values indicated by a red colour
gradient scheme (the corresponding landscape profile is depicted in
the central frame). We should point out that here we follow the 
definition of mathematical morphology although the name of 
{\it skeleton} of the large scale cosmic matter distribution has 
been used for different be it related concepts \citep[see e.g.][]{weygaert1991,NovCol}.

It is interesting to note that if we restrict the subsets $Y_i$ to
single points the skeleton naturally evolves into a {\it (first-order)
Voronoi Tessellation}. It is the definition of the concept of skeleton
${\mathcal K}$ within the specific context of grayscale images which
brings us to the definition of the {\it Watershed Transform} (see next
section).

\subsection{The watershed transform:\ \ Algorithms}
Grayscale images consist of a finite number of discrete levels. This
results in a slightly more complicated situation for its segmentation.
In the case of a binary image an image is segmented on the basis 
of a {\it geodesic distance}. For the segmentation of grayscale 
images the distance concept needs to be generalized to that of 
{\it topographic distance}.  

\bigskip
\noindent
{\bf Definition:} Each {\it watershed basin} is the collection of
points which are closer in topographic distance to the defining
minimum then to any other minimum.
\bigskip

The literature is replete with algorithms for the construction of the
{\it watershed transform} of an image into its constituting {\it
watershed segments}. They may be divided into two classes.  One class
simulates the {\it watershed basin} immersion process. The second aims
at detecting the watershed skeleton on the basis of the distance.

\subsubsection{Watershed by Immersion}
{\it Watershed by immersion} was introduced and defined by
\cite{VinSoi}.  The first step of the procedure concerns the
identification of the minima. Formally, a minimum is a plateau at
altitude $h$ from which it is impossible to reach a point of lower
height. Starting with the lowest grayscale level $h_{min}$ and
recursively proceeding to the highest level $h_{max}$ the algorithm
allocates the zone of influence of each minimum by gradually filling
up the surrounding catchment basin.

At a particular grayscale level $i$, with altitude $h_i$, the
algorithm has hypothetically inundated a landscape region with an
altitude ${\mathcal F}({\bf x})\leq h_i$. The total of inundated area,
\begin{equation}
{\tilde {\mathcal S}}_i({\mathcal F})\,=\,\{x \in \mathbf{Z}^n |\, {\mathcal F}(x) \leq i\,\}\,,
\end{equation}
is the complement of the section ${\mathcal S}_i$.  Having arrived at
level $i$ and proceeding to level $i+1$ the algorithm has three 
possibilities,
\begin{itemize}
\item[1] encounter a new minimum (at level $i+1$)
\item[2] add new points to existing catchment basins 
(condition: points connected to only one existing basin). 
\item[3] encounter new points that belong to several basins
\end{itemize}
Situation (1) signals the event that a new minimum becomes active in
the image. The second option concerns the extension of an existing basin
by an additional collection of points. These are points identified at 
level $(i+1)$ which find themselves within the realm of a single basin 
existent at level $i$. They belong to the zone of influence of 
${\tilde {\mathcal S}}_{i}$ and find themselves embedded in its 
extended counterpart ${\tilde {\mathcal S}}_{i+1}$ at level $(i+1)$. 
In situation (3) more than one basin may be connected to the 
stack ${\tilde {\mathcal S}}_{i+1}$. Its correct subdivision 
is determined by computing the influence zones of all connected 
basins. Defining ${\mathcal B}_i({\cal F})$ to be the union of all catchment 
basins at level $i$ the union of catchment basins at level $(i+1)$ becomes 
\begin{equation}
  {\mathcal B}_{i+1}({\mathcal F})\,=\,{Z}_{{\tilde {\mathcal S}_{i+1}}}({\tilde {\mathcal S_{i}}})\,\cup\,{\mathcal B}_{i}
\end{equation}
The watershed procedure may be viewed as iteratively computing the
zone of influence at each new grey scale level. 

Following the rationale above the  final ``immersion'' definition for the 
watershed ${\mathcal W}$ of the image ${\mathcal F}$ within a domain $X$ is 
\begin{equation}
  {\mathcal W}({\mathcal F})\,=\,X\,\backslash\,{\mathcal B}_{h_{max}}
\end{equation}
On completion of the procedure the union of points attached to every
minimum $m$ in $X$ is equal to the union of catchment basins,
${\mathcal B}_{h_{max}}$. The skeleton remains as the watershed
segmentation.

\subsubsection{Watershed by Topographic Distance}
The alternative strategy for determining the watershed transform is
that of following the strict definition of segmentation by minimum
topographic distance. \cite{RoeMeij} give a summary of the most
notable schemes. These algorithms seek to find all points (pixels)
whose topographic distance to a particular marker - ie. a significant
minimum in the density field -- is the shortest amongst that to all
other markers in the image.

The formalism bears some resemblance to Dijkstra's graph theoretical
problem of tracing the shortest path forest in a point
distribution. Based on this similarity an image is seen as a connected (di)graph 
in which the pixels of the image function as the
nodes of the graph. Each point $p$ is reachable from each other point
$p'$ via the graph's edges. The latter usually define a network on the
basis of $4-$connectivities or $8$-connectivities.

The shortest path between two points (nodes) $p$ and $p'$ is found by
traversing the graph and keeping track of the walking cost. Critical
for the procedure is the assignment of a proper measure of cost 
to each path. By definition it should be a non-negative increasing
function and be related to the definition of topographic distance
(eq.~\ref{eq:topdist}). This suggests the use of the quantity
\begin{equation} 
{\mathcal F'}(p,p')\,=\,{\rm max}\left\{\frac{{\mathcal F}(p)-{\mathcal F}(p')}{d(p,p')}\right\},
\end{equation}
the maximum slope linking the two pixels $p$ and $p'$. This leads to the following 
{\it cost function} for the link between two neighbouring pixel $p$ and pixel $p'$, 
\begin{eqnarray*}
{\mathcal C}(p,p')\,=\,
\begin{cases}
{\mathcal F'}(p,p')\, d(p,p')\qquad\qquad\qquad\quad {\mathcal F}(p)>{\mathcal F}(p') \\
\ \\
{\mathcal F'}(p',p) \, d(p,p')\qquad\qquad\qquad\quad {\mathcal F}(p)<{\mathcal F}(p') \\
\ \\
{\displaystyle {\mathcal F'}(p,p')+{\mathcal F'}(p',p) \over \displaystyle 2}\,\,\,d(p,p')\qquad 
{\mathcal F}(p)={\mathcal F}(p')\\
\end{cases}
\end{eqnarray*}
\begin{equation}
\end{equation}
The total cost ${\mathcal C}$ for a path $\gamma(p_1,p_2,...p_n)$ connecting any two 
points $p_1$ and $p_n$ via the points $\{p_2,\ldots,p_{n-1}\}$ will then simply be the sum
\begin{equation}
{\cal C}^{\gamma}(p_1,p_n)\,=\,\sum_{i\leq n} {\mathcal C}(p_i,p_{i-1})\,,
\end{equation}
The topographic distance ${\mathcal T}(p_1,p_n)$ is the infimum of 
${\mathcal C}^{\gamma}(p_1,p_n)$ over all paths connecting $p_1$ and $p_n$. 
\begin{equation}
{\mathcal T}(p_1,p_n)\,=\,\inf_{\Gamma}\,{\mathcal C}^{\gamma}(p_1,p_n)
\end{equation}
Given this definition for the topographic distance within a grayscale image 
we can pursue the segmentation process described in section~\ref{app:segment}, 
ultimately yielding the watershed segmentation.  

\subsection{Ordered Queues Algorithm}
\label{app:queue}
We follow the watershed transform algorithm by \cite{BeuMey}. Their 
method implicitly incorporates the concept of Markers. These 
markers are the minima used as sources of the watershed flooding 
procedure. As such they form a select subgroup amongst all 
minima of an image ${\mathcal F}$. 

The code for the watershed proceudre involves the following steps:
\begin{itemize}
\item {\bf Initialization} \ \ \ All pixels of the cube are initialized 
and tagged to indicate they have not yet been processed. Each 
grayscale level is allocated a queue and all pixels are 
attached to the queue corresponding to their level. 
\item {\bf Minima} \ \ \ Each minimum plateau is tagged by a unique 
``minimum tag''. The pixels corresponding to 
a minimum are inserted into the corresponding queue. 
\item {\bf Flooding} \ \ \ All pixels in the grayscale level queues are 
processed, starting at the lowest grayscale level. Unless a pixel is 
surrounded by a complex of unprocessed neighbours it will be assigned  
to the queue of the corresponding minimum. Pixels which also border 
another minimum obtain a boundary tag.
\item {\bf Final Stage} \ \ \ For any grayscale level the flooding stops when 
the queue has emptied. The procedure continues with processing the pixels 
in the queue for the next grayscale level. The process is finished once 
all level queues have been emptied. 
\end{itemize}

\section{\ \\ Kinematic Voronoi models}
Voronoi Clustering Models are a class of heuristic models for
cellular distributions of matter
\cite{WeyIck,weygaert1991,weygaert2002,weygaert2007}. They use the
Voronoi tessellation as the skeleton of the cosmic matter
distribution, identifying the structural frame around which matter
will gradually assemble during the emergence of cosmic structure. The
interior of Voronoi cells correspond to voids and the Voronoi
planes with sheets of galaxies. The edges delineating the
rim of each wall are identified with the filaments in the galaxy
distribution.  The most outstanding structural elements are the vertices, 
corresponding to the very dense compact nodes within the
cosmic web, the rich clusters of galaxies.

We distinguish two different yet complementary approaches. One is the
fully heuristic approach of Voronoi Element models. They are
particularly apt for studying systematic properties of spatial galaxy
distributions confined to one or more structural elements of
nontrivial geometric spatial patterns. The second, supplementary,
approach is that of the Voronoi Kinematic models, which attempts
to ``simulate'' foamlike galaxy distributions on the basis of
simplified models of the evolution of the megaparsec scale
distribution.

The Voronoi Kinematic Model is based upon the notion
that voids play a key organizational role in the development of
structure and make the Universe resemble a soapsud of expanding
bubbles \cite{Ick}. It forms an idealized and asymptotic description
of the outcome of the cosmic structure formation process within
gravitational instability scenarios with voids forming around a dip in
the primordial density field.  For plausible structure formation
scenarios, most notably the concordance $\Lambda$CDM cosmology, this
evolution will proceed hierarchically. A detailed assessment of the
resulting void hierarchy by \citep{SheWey} demonstrated that this
leads to a selfsimilarly evolving peaked void size distribution. 
By implication, most voids have comparable sizes and excess 
expansion rates. The geometrically interesting
implication is that the asymptotic limit of the ``peaked'' void
distribution degenerating into one of only one characteristic void
size. It yields a cosmic matter distribution consisting of {\em
equally sized} and expanding spherical voids, a geometrical
configuration which is precisely that of a Voronoi Tessellation.
This is translated into a scheme for the displacement of initially
randomly distributed galaxies within the Voronoi skeleton (see
sect.~\ref{app:vorclustform} for a detailed specification). Within a
void, the mean distance between galaxies increases uniformly in the
course of time. When a galaxy tries to enter an adjacent cell, the
velocity component perpendicular to the cell wall
disappears. Thereafter, the galaxy continues to move within the wall,
until it tries to enter the next cell; it then loses its velocity
component towards that cell, so that the galaxy continues along a
filament. Finally, it comes to rest in a node, as soon as it tries to
enter a fourth neighbouring void.

\begin{figure}
  \centering
  \vskip 0.5truecm
  \mbox{\hskip 0.0truecm\includegraphics[height=7.0cm]{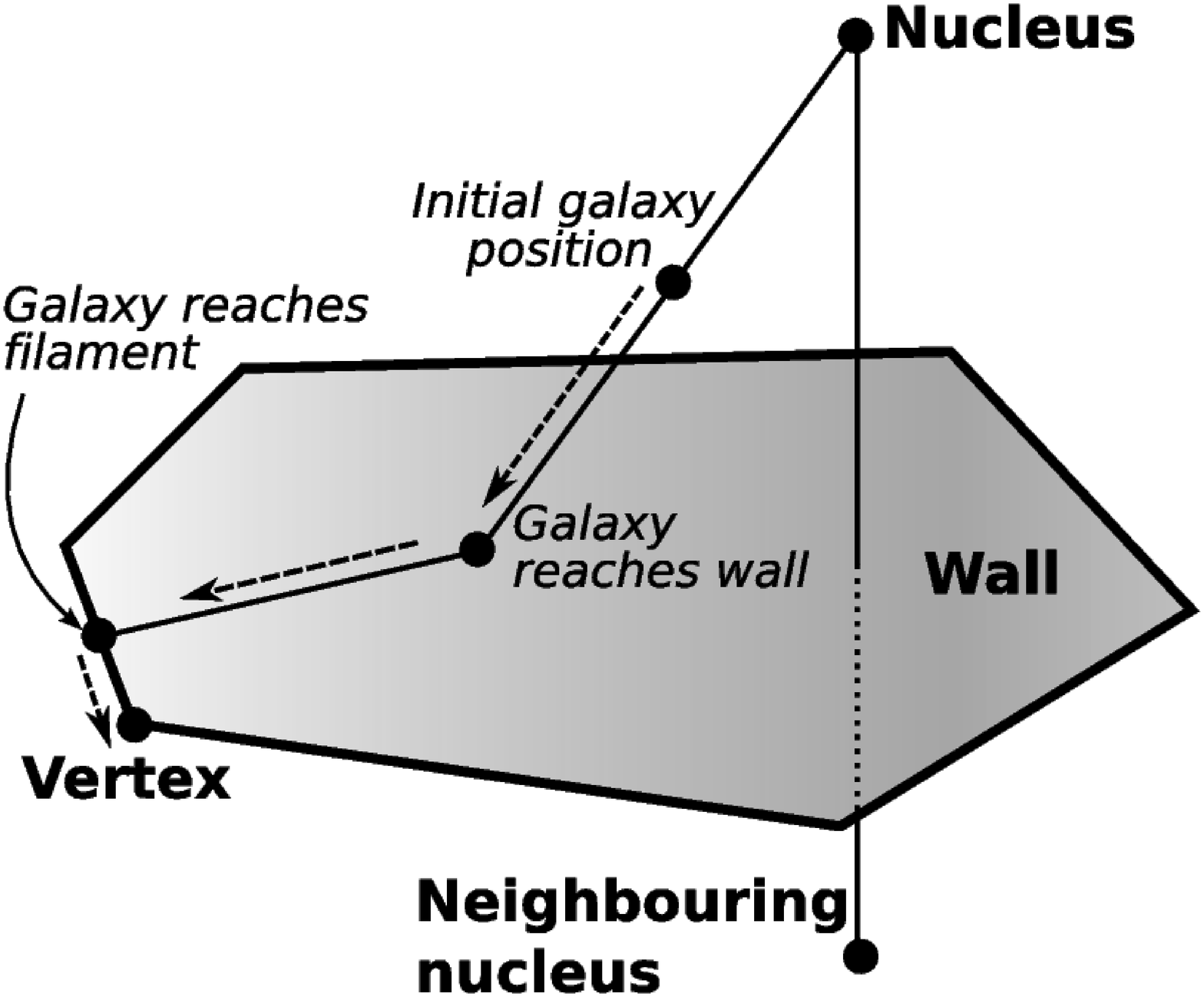}}
  \vskip 0.2cm
  \caption{Schematic illustration of the Voronoi kinematic model. Courtesy: Miguel Arag\'on.}
  \label{fig:vorkinmschm}
\end{figure} 
\subsection{Initial Conditions}
\label{app:vorclustform}
The initial conditions for the Voronoi galaxy distribution are:
\begin{itemize}
\item[$\bullet$] Distribution of $M$ nuclei, {\it expansion centres}, within the simulation volume $V$. The location 
of nucleus $m$ is ${\bf y}_m$.
\item[$\bullet$] Generate $N$ model galaxies whose initial locations, ${\bf x}_{n0}$ $(n=1,\ldots,N)$, are randomly 
distributed throughout the sample volume $V$. 
\item[$\bullet$] Of each model galaxy $n$ determine the Voronoi cell ${\cal V}_{\alpha}$ in which it is located, ie. 
determine the closest nucleus $j_{\alpha}$.
\end{itemize}
\begin{figure*}
  \centering
  \includegraphics[width=19.0cm]{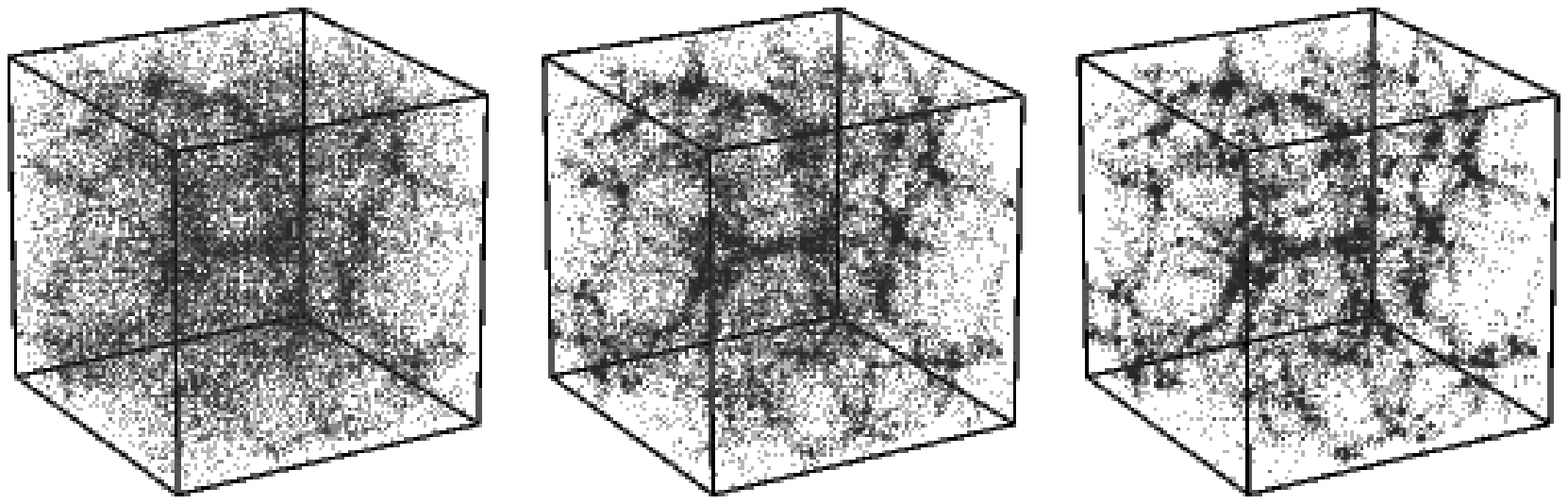}
  \caption{Evolution of galaxy distribution in the Voronoi kinematic model. A sequel 
           of 3 consecutive timesteps within the kinematic Voronoi cell formation process,  
           proceeding from left to right, and from top to bottom. The depicted boxes  
           have a size of $100h^{-1} \hbox{Mpc}$. Within these cubic volumes some $64$ Voronoi 
           cells with a typical size of $25h^{-1}\hbox{Mpc}$ delineate the cosmic framework 
           around which some 32000 galaxies have aggregated. Taken from a total (periodic) 
           cubic ``simulation'' volume of $200h^{-1}\hbox{Mpc}$ containing 268,235 
           ``galaxies''.}
  \label{fig:vorkinm3cube}
\end{figure*} 
\noindent All different Voronoi models are based upon the displacement
of a sample of $N$ ``model galaxies''. The initial spatial
distribution of these $N$ galaxies within the sample volume $V$ is
purely random, their initial locations ${\bf x}_{n0}$ ($n=1,\ldots,N)$
defined by a homogeneous Poisson process. A set of $M$ nuclei within the volume 
$V$ corresponds to the cell centres, or {\it expansion centres} driving the 
evolving matter distribution. The nuclei have locations ${\bf y}_m$ $(m=1,\ldots,M)$.

Following the specification of the initial positions of all galaxies,
the second stage of the procedure consists of the calculation of the
complete Voronoi track for each galaxy $n=1,\ldots,N$
(sec.~\ref{sec:vortrack}). Once the Voronoi track has been
determined, for any cosmic epoch $t$ one may determine the displacement 
${\bf x}_n$ that each galaxy has traversed along its
path in the Voronoi tessellation (sec.~\ref{sec:vortrack}).

\subsection{Voronoi Tracks}
\label{sec:vortrack}
The first step of the formalism is the determination for each galaxy $n$ the Voronoi cell ${\cal V}_{\alpha}$ in which it is initially located, 
ie. finding the nucleus $j_{\alpha}$ which is closest to the galaxies' initial position ${\bf x}_{n0}$. 

In the second step the galaxy $n$ is moved from its initial position ${\bf x}_{n0}$ along the radial path emanating form its expansion 
centre $j_\alpha$, ie. along the direction defined by the unity vector ${\hat {\bf e}}_{n\alpha}$. Dependent on how far the galaxy is 
moved away from its initial location ${\bf x}_{n0}$ - set by the {\it radius of expansion} $R_n$ to be specified later -- the galaxies' 
path ${\bf x}_n$ (see Fig.~\ref{fig:vorkinmschm}) may be codified as 
\begin{eqnarray}
{\bf x}_n&\,=\,&{\bf y}_{\alpha}\,+\,{\bf s}_{n\alpha}\,+\,{\bf s}_{n\alpha \beta}\,+\,{\bf s}_{n\alpha \beta \gamma} \nonumber\\
\ \\
&\,=\,&{\bf y}_{\alpha}\,+\,s_{n\alpha}{\hat {\bf e}}_{n\alpha}\,+\,s_{n\alpha\beta}{\hat {\bf e}}_{n\alpha\beta}\,+\,s_{n\alpha\beta\gamma}
{\hat {\bf e}}_{n\alpha\beta\gamma}\nonumber
\label{eq:galpath}
\end{eqnarray}
\noindent in which the four different components are:
\begin{enumerate}
\item[$\bullet$] ${\hat {\bf e}}_{n\alpha}$: \ \ \ \ unity vector path within Voronoi cell ${\cal V}_{\alpha}$
\item[$\bullet$] ${\hat {\bf e}}_{n\alpha\beta}$:\ \ \ \,unity vector path within Voronoi wall $\Sigma_{\alpha\beta}$
\item[$\bullet$] ${\hat {\bf e}}_{n\alpha\beta\gamma}$:\ \ unity vector path along Voronoi edge $\Lambda_{\alpha\beta\gamma}$ 
\item[$\bullet$] Vertex $\Xi_{\alpha\beta\gamma\delta}$
\end{enumerate}
The identity of the neighbouring nuclei $j_{\alpha}$, $j_{\beta}$, $j_{\gamma}$ and $j_{\delta}$, and therefore the identity 
of the cell ${\cal V}_{\alpha}$, the wall $\Sigma_{\alpha\beta}$, the edge $\Lambda_{\alpha\beta\gamma}$ and the vertex 
$\Xi_{\alpha\beta\gamma\delta}$, depends on the initial location ${\bf x}_{n0}$ of the galaxy, the position 
${\bf y}_{\alpha}$ of its closest nucleus and the definition of the galaxies' path within the Voronoi skeleton. 

\bigskip
The cosmic matter distribution at a particular cosmic epoch is obtained by calculating the individual displacement factors 
$(s_{n\alpha}(t),s_{n\alpha\beta}(t),s_{n\alpha\beta\gamma}(t))$ for each model galaxy. These are to be derived from the 
global ``void'' expansion factor $R(t)$. This factor parameterizes the cosmic epoch and specifies the (virtual) radial 
path of the galaxy from its expansion centre $j_{\alpha}$. 

At first, while still within the cell's interior, the galaxy proceeds according to 
\begin{eqnarray}
{\bf s}_{n\alpha}(t)\,=\,{\bf x}_n(t)\,-\,{\bf y}_{\alpha}\,=\,R(t)\ |{\bf x}_{n0}-{\bf y}_{\alpha}|\,{\hat {\bf e}}_{n\alpha}\,. 
\end{eqnarray}
As a result within a void the mean distance between galaxies increases uniformly in the course 
of time. Once the galaxy tries to enter an adjacent cell $j_{\beta}$ and reaches a Voronoi wall, i.e. when 
$R(t)\,|{\bf x}_{n0}-{\bf y}_{\alpha}|\,>\,\upsilon_n$, the galaxy's motion will be constrained to the 
radial path's component within the wall $\Sigma_{\alpha\beta}$. The galaxy moves along the wall until the displacement 
supersedes the extent of the path within the wall and it tries to enter a third cell $j_{\gamma}$, i.e. 
when $s_{n\alpha\beta}(t)\,>\,\sigma_n\,$. Subsequently, it moves along $\Lambda_{\alpha\beta\gamma}$ until 
it comes to rest at the node $\Xi_{\alpha\beta\gamma\delta}$ as soon as it tries to enter a fourth neighbouring 
void $j_{\delta}$ when $s_{n\alpha\beta\gamma}\,>\,\lambda_n$. 

A finite thickness is assigned to all Voronoi structural elements. The walls, filaments and vertices are assumed to have a 
Gaussian radial density distribution specified by the widths $R_{\rm W}$ of the walls, $R_{\rm F}$ of the filaments 
and $R_{\rm V}$ of the vertices. Voronoi wall galaxies are displaced according to the specified Gaussian density profile in
the direction perpendicular to their wall. A similar procedure is followed for the Voronoi filament galaxies and
the Voronoi vertex galaxies. As a result the vertices stand out as three-dimensional Gaussian peaks.


\section{\ \\ DTFE Reconstruction Procedure}
\label{app:dtfe}
For a detailed specification of the DTFE density field procedure we
refer to \cite{Sch}. In summary, the DTFE procedure for density
field reconstruction from a discrete set of points consists of the
following steps:
\begin{enumerate}
\item[$\bullet$] {\bf Point sample}\\ Given that the point sample is
supposed to represent an unbiased reflection of the underlying density
field, it needs to be a general Poisson process of the (supposed)
underlying density field.
\medskip
\item[$\bullet$] {\bf Boundary Conditions}\\ The boundary conditions
will determine the Delaunay and Voronoi cells that overlap the
boundary of the sample volume. Dependent on the sample at hand, a
variety of options exists:
\begin{enumerate}
\item[+] {\it Empty boundary conditions:}\\ outside the sample volume
there are no points.
\item[+] {\it Periodic boundary conditions:}\\ the point sample is
supposed to be repeated periodically in boundary boxes, defining a
toroidal topology for the sample volume. 
\item[+] {\it Buffered boundary conditions:}\\ the sample volume box is
surrounded by a bufferzone filled with a synthetic point sample.
\end{enumerate}
\medskip
\item[$\bullet$] {\bf Delaunay Tessellation}\\ Construction of the
Delaunay tessellation from the point sample.  While we also still use the
Voronoi-Delaunay code of \citep{weygaert1991,weygaert1994}, at present there is a 
number of efficient library routines available. Particularly noteworthy
is the \cgal initiative, a large library of computational geometry
routines\footnote{\cgal is a \texttt{C++} library of algorithms and
data structures for Computational Geometry, see \url{www.cgal.org}.}\\
\medskip
\item[$\bullet$] {\bf Field values point sample}\\ The estimate of the
density at each sample point is the normalized inverse of the volume
of its contiguous Voronoi cell ${\cal W}_i$ of each point
$i$. The {\it contiguous Voronoi cell} of a point $i$ is the union of
all Delaunay tetrahedra of which point $i$ forms one of the four
vertices. We recognize two applicable situations:
\\ \itemitem{-} {\it uniform sampling process}:\\ 
the point sample is an unbiased sample of
the underlying density field. Typical example is that of $N$-body
simulation particles. For $D$-dimensional space the density estimate
is,
\begin{equation}
{\widehat \rho}({\bf x}_i)\,=\,(1+D)\,\frac{w_i}{V({\cal W}_i)} \,.
\label{eq:densvor}
\end{equation}
\noindent with $w_i$ the weight of sample point $i$, usually we assume
the same ``mass'' for each point. \\ 
\itemitem{-} {\it systematic non-uniform sampling process}:\\
 sampling density according to specified
selection process.  The non-uniform sampling process is quantified by
an a priori known selection function $\psi({\bf x})$. This situation
is typical for galaxy surveys, $\psi({\bf x})$ may encapsulate
differences in sampling density $\psi(\alpha,\delta)$ as function of
sky position $(\alpha,\delta)$, as well as the radial redshift
selection function $\psi(r)$ for magnitude- or flux-limited
surveys. For $D$-dimensional space the density estimate is ,
\begin{equation}
{\widehat \rho}({\bf x}_i)\,=\,(1+D)\,\frac{w_i}{\psi({\bf x}_i)\,V({\cal W}_i)} \,.
\label{eq:densvornu}
\end{equation}

\medskip
\item[$\bullet$] {\bf Field Gradient}\\ Calculation of the field
gradient estimate $\widehat{\nabla f}|_m$ in each $D$-dimensional
Delaunay simplex $m$ ($D=3$: tetrahedron; $D=2$: triangle) by solving
the set of linear equations for the field values $f_i$ at the positions 
${\bf r}_i$ of the $(D+1)$ tetrahedron vertices,\\
\begin{eqnarray}
\widehat{\nabla f}|_m \ \ \Longleftarrow\ \ 
\begin{cases}
f_0 \ \ \ \ f_1 \ \ \ \ f_2 \ \ \ \ f_3 \\
\ \\
{\bf r}_0 \ \ \ \ {\bf r}_1 \ \ \ \ {\bf r}_2 \ \ \ \ {\bf r}_3 \\
\end{cases}\,
\label{eq:dtfegrad}
\end{eqnarray}
Evidently, linear interpolation for a field $f$ is only meaningful
when the field does not fluctuate strongly.

\medskip
\item[$\bullet$] {\bf Interpolation}.\\ The final basic step of the
DTFE procedure is the field interpolation. The processing and
postprocessing steps involve numerous interpolation calculations, for
each of the involved locations ${\bf x}$. Given a location ${\bf x}$, the 
Delaunay tetrahedron $m$ in which it
is embedded is determined. On the basis of the field gradient
$\widehat{\nabla f}|_m$ the field value is computed by (linear)
interpolation,
\begin{equation}
{\widehat f}({\bf x})\,=\,{\widehat f}({\bf x}_{i})\,+\,{\widehat {\nabla f}} \bigl|_m \,\cdot\,({\bf x}-{\bf x}_{i}) \,.
\label{eq:fieldval}
\end{equation}
In principle, higher-order interpolation procedures are also
possible. Two relevant procedures are: 
\itemitem{-} {\it Spline Interpolation} 
\itemitem{-} {\it Natural Neighbour Interpolation}\\ 
For NN-interpolation see \cite{Wat, BraSam, Suk} and \cite{Oka}. 
Implementation of Natural neighbour interpolations is presently in progress.
\medskip
\item[$\bullet$] {\bf Processing}.\\ Though basically of the same
character, for practical purposes we make a distinction between
straightforward processing steps concerning the production of images
and simple smoothing filtering operations and more
complex postprocessing.  The latter are treated in the next item. 
Basic to the processing steps is the determination of
field values following the interpolation procedure(s) outlined
above. Straightforward ``first line'' field operations are {\it
Image reconstruction} and {\it Smoothing/Filtering}.
\begin{enumerate}
\item[+] {\bf Image reconstruction}.\\ For a set of image
points, usually grid points, determine the image value. 
formally the average field value within the corresponding gridcell.
In practice a few different strategies may be followed
\itemitem{-} {\it Formal geometric approach} 
\itemitem{-} {\it Monte Carlo approach}
\itemitem{-} {\it Singular interpolation approach}\\
The choice of strategy is mainly dictated by accuracy requirements. For WVF we 
use the Monte Carlo approach in which the grid density value is the average 
of the DTFE field values at a number of randomly sampled points within the grid 
cell. \\
\item[+] {\bf Smoothing} and {\bf Filtering}:\\
A range of filtering operations is conceivable. Two of relevance to WVF are:
\itemitem{-} {\it Linear filtering} of the field ${\widehat f}$\\ 
Convolution of the field ${\widehat f}$ with a filter function $W_s({\bf x},{\bf y})$, usually user-specified, 
   \begin{equation}
     f_s({\bf x})\,=\,\int\,{\widehat f}({\bf x'})\, W_s({\bf x'},{\bf y})\,d{\bf x'}     
   \end{equation}
\itemitem{-} {\it Natural Neighbour Rank-Ordered filtering}\\ (see sec.~\ref{sec:natnghb}).  
\end{enumerate}
\medskip
\item[$\bullet$] {\bf Post-processing}.\\ The real potential of DTFE
fields may be found in sophisticated applications, tuned towards
uncovering characteristics of the reconstructed fields.  An important
aspect of this involves the analysis of structures in the density
field. The WVF formalism developed in this study is an obvious
example.
\end{enumerate}

\label{lastpage}

\end{document}